\pdfoutput=1
\documentclass[10pt,twocolumn,letterpaper]{article}
\usepackage[T1]{fontenc}
\usepackage{mathptmx}
\usepackage[letterpaper,textwidth=516pt,textheight=696pt,columnsep=12pt,top=54pt,hcentering,headheight=12pt,headsep=12pt,footskip=24pt]{geometry}
\usepackage{titlesec}
\usepackage[font=footnotesize,labelsep=period]{caption}
\renewcommand\thesection{\Roman{section}}
\renewcommand\thesubsection{\Alph{subsection}}
\titleformat{\section}{\centering\normalfont\scshape}{\thesection.}{0.5em}{}
\titlespacing*{\section}{0pt}{1.2ex plus .4ex}{0.6ex}
\titleformat{\subsection}{\normalfont\itshape}{\thesubsection.}{0.5em}{}
\titlespacing*{\subsection}{0pt}{0.9ex plus .3ex}{0.4ex}
\makeatletter
\def\@maketitle{\newpage\null\vskip -2.5em
  \begin{center}{\fontsize{22}{26}\selectfont \@title\par}\vskip 0.9em
  {\fontsize{11}{13}\selectfont \@author\par}\end{center}\par\vskip 1.2em}
\makeatother
\renewenvironment{abstract}{\noindent\small\bfseries\textit{Abstract}---}{\par\vspace{0.6ex}}
\newenvironment{IEEEkeywords}{\noindent\small\bfseries\textit{Index Terms}---}{\par\vspace{0.6ex}}

\makeatletter
\def\@fnsymbol#1{}
\makeatother
\renewcommand\footnoterule{\kern-3pt\hrule width 0.4\columnwidth\kern 2.6pt}
\usepackage[utf8]{inputenc}
\usepackage{cite}
\usepackage{graphicx}
\usepackage{amsmath}
\usepackage{booktabs,tabularx}
\usepackage{url}
\usepackage[hidelinks]{hyperref}
\usepackage{microtype}
\graphicspath{{figs/}}
\renewcommand\thesubsection{\arabic{subsection}}

\providecommand{\theHfigure}{\arabic{figure}}
\extrafloats{100}
\renewcommand{\dblfloatpagefraction}{0.75}
\DeclareUnicodeCharacter{00D7}{\ensuremath{\times}}
\DeclareUnicodeCharacter{2264}{\ensuremath{\le}}
\DeclareUnicodeCharacter{2265}{\ensuremath{\ge}}
\DeclareUnicodeCharacter{2212}{\ensuremath{-}}
\DeclareUnicodeCharacter{207B}{\textsuperscript{\ensuremath{-}}}
\DeclareUnicodeCharacter{2074}{\textsuperscript{4}}
\DeclareUnicodeCharacter{2075}{\textsuperscript{5}}
\DeclareUnicodeCharacter{2076}{\textsuperscript{6}}
\DeclareUnicodeCharacter{2070}{\textsuperscript{0}}
\DeclareUnicodeCharacter{2009}{\,}
\DeclareUnicodeCharacter{03BA}{\ensuremath{\kappa}}
\DeclareUnicodeCharacter{03C1}{\ensuremath{\rho}}
\DeclareUnicodeCharacter{2010}{-}
\DeclareUnicodeCharacter{00B5}{\ensuremath{\mu}}
\DeclareUnicodeCharacter{2248}{\ensuremath{\approx}}
\DeclareUnicodeCharacter{2192}{\ensuremath{\rightarrow}}
\DeclareUnicodeCharacter{222A}{\ensuremath{\cup}}
\begin{document}
\title{Quality-Control Policies Shape Model Behaviour in Computational Pathology}

\author{Mario Parreño Centeno, Yu Guan, Pedro Montero Pavón, Sanna Palosaari, and Juha Väyrynen%
\thanks{M. Parreño Centeno, S. Palosaari and J. Väyrynen are with the Research Unit of Translational Medicine, Faculty of Medicine, University of Oulu, 90014 Oulu, Finland. Y. Guan is with the Department of Computer Science, University of Warwick, Coventry CV4 7AL, U.K. P. Montero Pavón is with the Department of Pathology, Virgen Macarena University Hospital, 41009 Seville, Spain. Corresponding author: Mario Parreño Centeno (e-mail: mario.parreno-centeno@oulu.fi).}}

\maketitle

\begin{abstract}
Quality-control policies determine which image regions are used by computational pathology models, but their effects are difficult to isolate when quality estimation and input selection are coupled. We make these policies explicit, configurable and testable. We introduce PathQC-3C, comprising 1,661 densely annotated regions from 1,194 haematoxylin and eosin whole-slide images across 32 cancer/tissue categories, and TileNet, a framework that separates dense tissue/background/artefact estimation from task-specific input-selection policies. We evaluated the default predictor across three held-out cohorts and against harmonised quality-control comparators, assessed policy decisions in a blinded single-pathologist audit, and tested alternative policies in three downstream settings. In retrospective CAMELYON16 multiple-instance learning, background-aware selection improved the pooled out-of-fold area under the receiver operating characteristic curve from 0.758 to 0.839 compared with random sampling from the same exhaustive candidate pool under a fixed tile budget. In weak-label lung classification, strict artefact exclusion reduced the pre-matching candidate pool by approximately 43\%, with small metric-dependent differences after matching training-set size and patient contribution. PathQCAug, which composites real artefact components during tumour--stroma model adaptation, improved macro Dice on high-artefact tiles under severe held-out composited artefacts across three training seeds and outperformed random-erasing and clean-tissue controls. Separating quality estimation from policy execution provides an auditable way to control which image content enters computational pathology models; the appropriate policy is task specific, and clinical utility remains to be established.
\end{abstract}

\begin{IEEEkeywords}
Computational pathology, quality control, image segmentation, multiple-instance learning, data augmentation.
\end{IEEEkeywords}

\section{Introduction}
Computational pathology increasingly relies on tiling whole-slide images (WSIs) for feature extraction, self-supervised pretraining and downstream learning~\cite{r1,r2,r3,r4}. Routine haematoxylin and eosin (H\&E) WSIs contain non-tissue background and technical artefacts introduced during slide preparation and scanning~\cite{r5,r6,r7,r8}. Including these regions consumes computation and can expose models to structured nuisance variation, particularly when slide- or region-level labels are assigned to all retained tiles~\cite{r3,r7,r9}. Decisions about which regions to retain therefore affect model inputs, yet quality control (QC) is often implemented as fixed preprocessing rather than as an explicit, auditable policy.

Most tile-based workflows use binary foreground detection~\cite{r1,r2,r4,r10}. Thresholding and colour heuristics remove blank regions but cannot reliably distinguish interpretable tissue from tissue affected by folds, blur, pen marks or scanning defects~\cite{r7,r8,r10,r11}, and they may discard legitimate structures: in a recent study, saturation-based tissue detection omitted regions containing adipose tissue and alveolar parenchyma from the training pool of a diffusion-based artefact detector, contributing to false-positive detections on normal tissue~\cite{r12}. Beyond detection, QC requires deciding how to act on the estimated quality state: empty background can generally be discarded, severely compromised regions may require exclusion or review, and mildly affected tissue may remain informative. These decisions depend on the downstream objective and cannot be inferred from binary foreground masks alone.

Artefact-aware QC approaches address aspects of this problem. GrandQC provides multi-class artefact segmentation with institutional and scanner benchmarking~\cite{r13}; WSI-SmartTiling identifies folds and blur for content-aware tiling~\cite{r14}; SlideInspect combines tissue and artefact assessment with slide-level actions~\cite{r15}; and patch-based approaches, including those of Kahaki et al. and Kanwal et al., support artefact classification, quality reporting and artefact-guided filtering~\cite{r16,r17}. These approaches differ in annotation taxonomy, spatial resolution and intended application (Table~\ref{tab:resources}), while the consequences of mapping quality estimates to downstream selection rules remain difficult to isolate when estimation and policy are coupled within a particular implementation.

This distinction also matters for attention-based multiple-instance learning (MIL), which identifies regions informative for a target label but whose attention is not a measure of image quality~\cite{r3}, and for foundation encoders, which process the image regions supplied to them and do not by themselves define which regions should enter training~\cite{r18}. Artefact-aware control can therefore act at complementary stages of the computational pipeline: before learning, by determining which regions enter computation and how finite tile budgets are allocated, and during learning, by controlling how technical nuisance variation is represented in the training data.

We therefore separate quality estimation from policy execution. Our contributions are threefold: 1)~PathQC-3C, a dense three-class tissue/background/artefact resource comprising 1,661 regions of interest (ROIs) from 1,194 WSIs across 32 cancer/tissue categories, in which the artefact class intentionally aggregates heterogeneous technical artefacts to support operational QC decisions rather than fine-grained artefact classification; 2)~TileNet, a predictor-agnostic framework that aggregates dense QC estimates over candidate tile footprints and applies explicit, configurable selection policies (Fig.~1), here with a single-scale 20× Swin--UPerNet-Base model~\cite{r19,r20} as the default dense QC predictor; and 3)~PathQCAug, a real-artefact compositing strategy for foundation-model adaptation, building on the use of annotated real artefacts to augment the training of artefact detectors~\cite{r21}. The predictor is an implementation choice; the defining component of TileNet is the policy interface.

We evaluate held-out segmentation performance, harmonised comparators, tile-level policy fidelity and agreement with blinded pathologist usability assessments. Downstream consequences are examined through fixed-budget MIL on CAMELYON16~\cite{r22}, a matched-data weak-label lung stress test and tumour--stroma model adaptation on the Breast Cancer Semantic Segmentation (BCSS) dataset~\cite{r23}, in which real-artefact compositing is compared with geometry-matched random erasing and mean-coverage-matched clean-tissue replacement. Across these analyses, artefact delineation was the main technical challenge, and the appropriate response to detected artefacts depended on the downstream task.

\begin{figure}[!tp]
\centering
\includegraphics[width=\columnwidth]{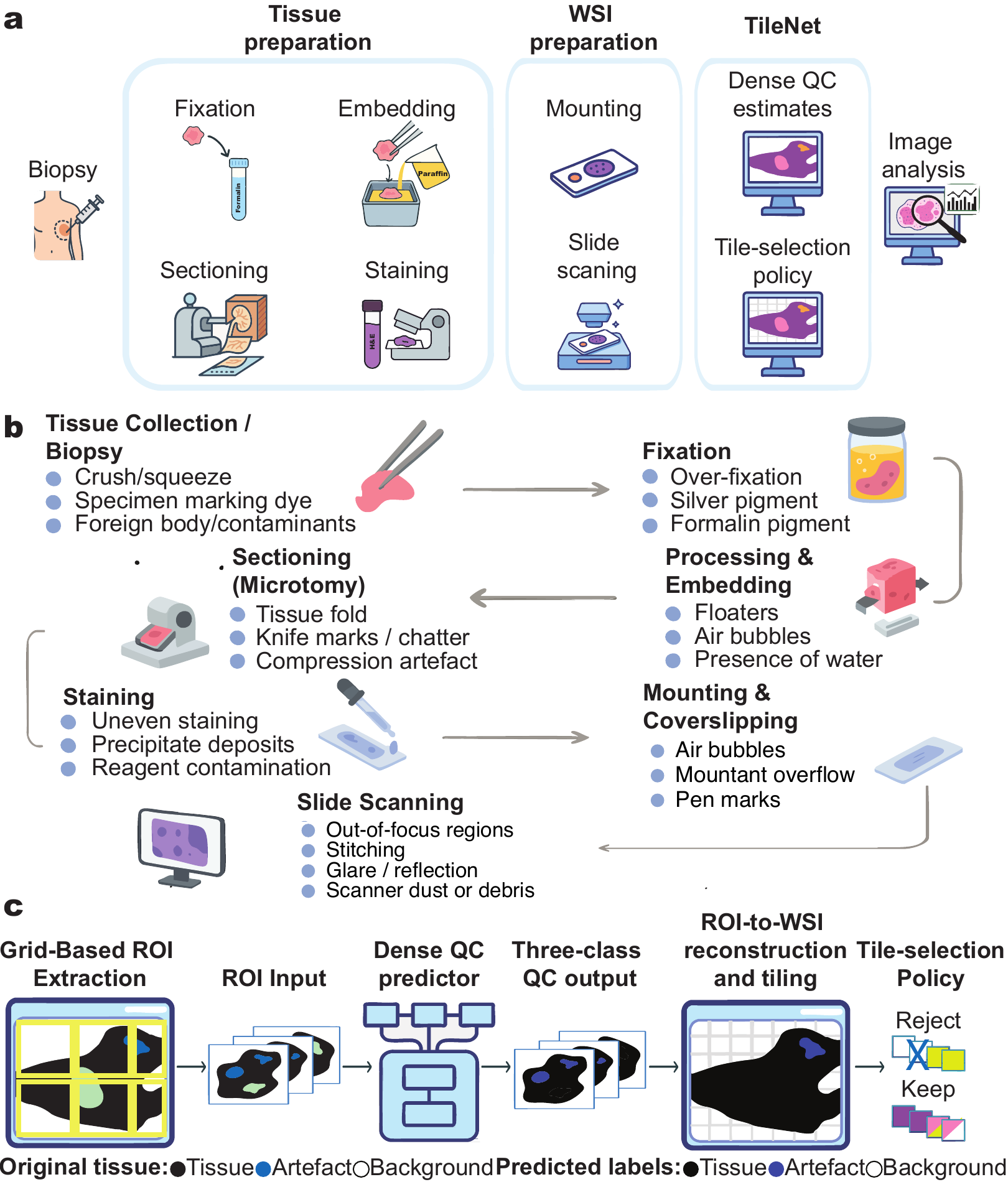}
\setcounter{figure}{0}\caption{\textbf{Artefact-aware quality control as an explicit policy layer in computational pathology.} a, Overview of a tile-based computational pathology workflow. Technical variation may be introduced during tissue preparation, whole-slide image (WSI) preparation and digitisation. TileNet is inserted upstream of downstream image analysis as an artefact-aware quality-control framework that generates dense tissue/background/artefact maps and converts them into explicit tile-selection policies. b, Representative sources of technical artefacts and variability across the histopathology workflow, including tissue collection, fixation, processing, embedding, sectioning, staining, mounting, coverslipping and slide scanning. c, TileNet execution pipeline. WSIs are partitioned into regions of interest (ROIs), processed by a dense QC predictor to generate three-class tissue/background/artefact outputs, and reconstructed into WSI-aligned QC maps. Tile-level tissue, background and artefact fractions are then aggregated over each tile footprint and converted into keep/reject decisions using user-defined policy thresholds.}
\label{fig:1}
\end{figure}

\begin{table*}[!tp]
\caption{Comparison of selected whole-slide image artefact and quality-control annotation resources}
\label{tab:resources}
\centering\scriptsize\setlength{\tabcolsep}{3pt}\renewcommand{\arraystretch}{1.12}
\begin{tabularx}{\textwidth}{@{}>{\raggedright\arraybackslash}p{2.1cm}>{\raggedright\arraybackslash}X>{\raggedright\arraybackslash}X>{\raggedright\arraybackslash}X>{\raggedright\arraybackslash}X@{}}
\toprule
\textbf{Feature} & \textbf{PathQC-3C} & \textbf{GrandQC released test dataset} ~\cite{r13} & \textbf{WSI-SmartTiling dataset} ~\cite{r14} & \textbf{SlideInspect released test set} ~\cite{r15}\\
\midrule
\textbf{Artefact representation} & One operational artefact mask class covering a broad set of workflow-related artefact appearances, including floaters, out-of-focus regions, pen marks, scanner dust/debris, specimen marking dye, formalin pigment, tissue folds, knife marks/chatter, air bubbles and staining or mounting defects & Multiple artefact-related classes, including air bubble/slide edge, out-of-focus regions, pen markings, tissue folds and dark spot/foreign object, together with tissue without artefacts & Two artefact categories: fold and blurring & Three artefact categories: tissue folds, out-of-focus regions and mounting artefacts, with mounting artefacts grouping bubbles, dust, pen marks and coverslip-related issues\\
\textbf{Dense pixel-wise annotation} & Yes & Yes, with caveat: dense annotations are provided for most regions, while regions with numerous small artefacts may include sparse annotations and unannotated/ignore areas & Yes & Yes; manual annotations are provided for tissue regions and artefacts\\
\textbf{Annotation / evaluation scale} & 20× equivalent resolution; 0.5 µm per pixel & Artefact detection evaluated at 5×, 7× and 10×; tissue detection at approximately 1× & 20× and 40× & Tissue segmentation at 1.25× and artefact detection at 5×; public test images are provided at 5×\\
\textbf{Source-site / acquisition diversity} & Yes; 166 distinct recorded GDC source-site names in TCGA, with 12 separately reported TCIA identifiers. Institutional aliases were not canonicalised. Scanner metadata were not systematically available, and scanner diversity was not explicitly controlled. & Released manually annotated test dataset drawn from a single pathology department (318 slides); the full GrandQC study additionally includes multi-institutional and multi-scanner benchmarking. & Yes; four dataset sources overall, with the exact number of individual source sites not clearly specified & Yes in the full study; five centres overall, and the released test set contains slides from centres not represented in training and validation\\
\textbf{Cancer/tissue-category diversity} & Yes; 32 cancer/tissue categories & Yes; four organs/tumour types in the released artefact test dataset & Yes; six tissues/organs in external validation & Yes; eight tissue types overall: adrenal, breast, colon, kidney, liver, lung, myocardial and prostate; four staining methods: H\&E, IHC, PAS and trichrome\\
\textbf{Downstream or actionable QC evaluation} & Yes; the associated framework was evaluated for tile selection, weakly supervised learning and artefact-aware foundation-model adaptation. & Yes; GrandQC was evaluated as a QC benchmark and shown to improve selected downstream image-analysis tasks, but the released manually annotated test dataset is primarily a segmentation/QC benchmark & Yes; evaluated for WSI tile selection & Yes; provides slide-level traffic-light quality assessment and recommended corrective actions: no action, re-scan, re-mount or re-cut; not a downstream computational pathology tile-retention benchmark\\
\textbf{Annotated slides/images in compared dataset} & 1,194 & 318 & 917 (available upon request) & 441\\
\bottomrule
\end{tabularx}
\par\smallskip
\parbox{\textwidth}{\scriptsize PathQC-3C is compared with the GrandQC released manually annotated test dataset, WSI-SmartTiling dataset and SlideInspect released test set with respect to artefact representation, annotation granularity, imaging scale, source composition, cancer/tissue-category diversity and downstream or actionable quality-control evaluation. The comparison focuses on resources providing spatially resolved reference annotations suitable for artefact segmentation or region-level quality-control evaluation, rather than datasets limited to patch-level classification labels. Counts refer to the annotated datasets or subsets compared in the table; for GrandQC and SlideInspect, these correspond to the released manually annotated test sets rather than the larger cohorts used elsewhere in the respective studies. PathQC-3C includes 166 exact recorded GDC source-site names for TCGA and 12 separately reported TCIA source identifiers. Registry names were not canonicalised to physical institutions and do not identify distinct scanning sites or scanners; TCIA identifiers cannot be mapped directly to TCGA names. Cancer/tissue categories in PathQC-3C correspond to TCGA study/disease categories and TCIA diagnostic categories rather than strictly anatomical organ sites. This is a selected resource comparison rather than a ranking of all current QC systems.}
\end{table*}

\section{Materials and Methods}
\subsection{Study design and quality-control framework}

This study evaluated artefact-aware quality control as an explicit representation and policy layer for computational pathology. The work comprised four linked components: construction of PathQC-3C, a dense tissue/background/artefact annotation resource; development and held-out evaluation of dense quality-control predictors; evaluation of TileNet-guided tile-selection policies in fixed-budget multiple-instance learning and matched weak-label classification; and development of PathQCAug for training-time augmentation with real histological artefacts.

TileNet separated quality estimation from policy execution. A dense quality-control predictor generated spatially aligned tissue, background and artefact outputs. For each downstream tile, these outputs were aggregated over the corresponding tile footprint to obtain predicted tissue, background and artefact fractions. TileNet then converted these estimates into explicit keep, reject or ranking decisions using task-specific eligibility rules. The default predictor used in this study was a single-scale 20× Swin--UPerNet-Base model, but the policy interface was not restricted to this architecture. All downstream quality-control estimates, including CAMELYON16 patch fractions, TCGA-LUSC tile filtering and BCSS artefact-burden stratification, were generated with a single checkpoint of this predictor, the fold-0 development model, which is the checkpoint released with TileNet (Code availability).

Implementation details for data units and leakage control, ROI extraction, label encoding, predictor training, the exploratory SAM3 experiment, runtime profiling, optional hole filling, CLAM training, TCGA-LUSC replicate sensitivity analyses, artefact-library construction and the PathQCAug software are given in the Supplementary Methods (Section S1).

\subsection{Ethics and secondary use of public images}

This study used publicly available, de-identified human histopathology images from The Cancer Genome Atlas (TCGA), the Clinical Proteomic Tumor Analysis Consortium (CPTAC) collections hosted by The Cancer Imaging Archive (TCIA), CAMELYON16 and the Breast Cancer Semantic Segmentation (BCSS) dataset. The original data were collected and released under the ethical approvals and consent or waiver arrangements applicable to the respective source studies. No new human participants were recruited or tissue specimens collected. As the study involved secondary analysis of existing de-identified images, no additional institutional ethics approval was required. The pathologist reviewed de-identified image crops as an expert rater, not as a research participant.

\subsection{PathQC-3C cohorts and annotation}

PathQC-3C comprised 1,194 haematoxylin and eosin whole-slide images (WSIs) and 1,661 manually annotated regions of interest (ROIs), covering 32 cancer/tissue categories. Each ROI represented a 6,000 × 6,000-pixel field of view at 0.5 µm per pixel, corresponding to 20×-equivalent resolution, and was paired with a dense three-class quality-control mask comprising mutually exclusive tissue, background and artefact labels.

The TCGA-derived component contained 1,144 WSIs and 1,611 ROIs from 510 barcode Tissue Source Site (TSS) codes mapped to 166 distinct recorded GDC source-site names and 32 cancer/tissue categories. All TCGA-derived WSIs were diagnostic slides identified by DX-designated TCGA slide-type codes, corresponding to formalin-fixed, paraffin-embedded (FFPE) tissue preparation. It was divided without WSI overlap into a development cohort of 947 WSIs and 1,350 ROIs and a held-out internal evaluation cohort of 197 WSIs and 261 ROIs. The evaluation cohort comprised TCGA-common-source, containing 152 WSIs and 197 ROIs from 54 recorded GDC source-site names represented during development, and TCGA-unique-source, containing 45 WSIs and 64 ROIs from 19 recorded GDC source-site names absent from development. External evaluation used 50 WSIs and 50 ROIs from 12 deidentified source-site identifiers in the CPTAC-SAR, CPTAC-HNSCC and CPTAC-PDA collections hosted by The Cancer Imaging Archive. TCGA and TCIA source-site identifiers could not be reliably cross-mapped; therefore, external validation denotes evaluation on separate CPTAC collections rather than verified institutional disjointness. Source-site overlap and cancer/tissue-category coverage across these cohorts are summarised in Supplementary Figs. 20--22 at WSI and ROI levels. For each ROI, metrics from the three fold-specific prediction outputs were averaged before cohort-level summarisation. TCIA provides the original collection metadata and source images~\cite{r33}.

TCGA barcode TSS codes were mapped to exact `Source Site' strings in the GDC code table (lookup on 11 November 2026; Source Data, worksheet Fig2\_site\_mapping). The analysed TCGA resource contains 510 distinct TSS codes mapping to 166 recorded names. Development contains 395 codes and 147 names, TCGA-common-source 92 codes and 54 names, and TCGA-unique-source 24 codes and 19 names; one code (FA) occurs in both development and TCGA-common-source, so the cohort-level code counts sum to 511. All 54 common-source names occur in development; none of the 19 unique-source names does. Names were used as registry labels without canonicalising institutional aliases. Thus, `unique-source' denotes registry-name disjointness and does not establish institutional or scanner independence. TCGA-common-source includes source-site--category combinations also represented during development and should not be interpreted as uniformly unseen combinations.

Each ROI was labelled using three mutually exclusive classes. Tissue denoted histologically interpretable biological material, background denoted non-tissue regions, and artefact denoted technical alterations that degraded, obscured or distorted biological content. The artefact class intentionally aggregated heterogeneous preparation, staining, mounting and scanning defects into one operational category suitable for exclusion, ranking, down-weighting or review. Initial masks were manually generated using Adobe Photoshop 26.8.0 by a digital pathology expert and subsequently visualised using QuPath 0.5.1 during iterative consensus review with a board-certified pathologist and an expert histologist. Label encoding, supervision integrity and annotation conventions are described below.

\subsection{Dense quality-control predictor development and ablation design}

Candidate three-class predictors included EfficientNet-B0 U-Net++ at 10× and 20×, SegFormer-B5 at 20×, Swin--UPerNet-Base at 20× and Swin--UPerNet-Large at 20×. The baseline architecture comparison used 512 × 512-pixel source crops, presented unchanged as 512 × 512-pixel network inputs, and the Dice--IoU objective. A two-class Swin--UPerNet-Base control was additionally trained after collapsing background and artefact into a composite non-tissue class. This control used the same patient-level partitions and fitting procedure as the corresponding three-class configuration and was intended to determine whether accurate foreground separation alone provided the representation required for artefact-specific policies.

Subsequent Swin--UPerNet-Base ablations evaluated source crops of 256 × 256, 512 × 512 and 1,024 × 1,024 pixels sampled from the 20× ROIs. All source crops were processed to 512 × 512-pixel tensors before network input. The source-crop comparison therefore altered the physical context and image-resampling transformation but not the dimensions presented to the neural network.

The reported optimisation grid included Dice--IoU and Dice--cross-entropy objectives, unweighted and fold-specific weighted cross-entropy, and minimum-foreground fractions of 0.0 and 0.1 where applicable. Once the source-crop, loss and sampling configuration had been selected, per-device batch sizes of 2, 4 and 6, per-epoch sampling budgets of 15,000 and 30,000 crops, and initial learning rates of 1.2 × 10⁻⁵, 2.0 × 10⁻⁵ and 3.0 × 10⁻⁵ were evaluated. The complete shared and run-specific configurations are reported in Supplementary Data 4, Tables 23 and 24.

The selected single-scale configuration used 1,024 × 1,024-pixel source crops at 20×, processed to 512 × 512-pixel network inputs, unweighted Dice--cross-entropy, a minimum-foreground fraction of 0.0, a per-device batch size of 2, gradient accumulation over six steps, 15,000 sampled crops per epoch and an initial learning rate of 2.0 × 10⁻⁵. Configuration selection was based descriptively on artefact Dice, macro Dice, fold-to-fold stability, computational cost and implementation simplicity rather than on a formal comparison among hyperparameter settings.

\subsection{Held-out ROI inference and metric computation}

Held-out evaluation used complete annotated ROIs rather than the deterministic validation crops used during development. Each 6,000 × 6,000-pixel ROI was processed at 20×-equivalent resolution using spatial inference windows. Window-level outputs were reconstructed into an ROI-aligned tissue/background/artefact map. Multi-magnification models received paired 10×-equivalent inputs corresponding to the same physical locations as the 20× windows.

Each fold-specific checkpoint was applied independently to every held-out ROI. ROI-level metrics were first calculated separately for each checkpoint and subsequently averaged within ROI, so that each ROI contributed one value per predictor configuration to the cohort-level summaries. Mean ± s.d. across ROIs was then reported.

Prediction and reference masks were mapped to the shared tissue/background/artefact encoding before evaluation. Reference pixels labelled 255 were excluded. Prediction pixels that could not be mapped to a valid evaluation class were treated as incorrect predictions for the corresponding reference class rather than removed. Macro Dice and macro intersection-over-union were calculated within each ROI as the unweighted mean over classes present in that ROI's reference mask. Class-specific metrics were summarised only over ROIs containing the corresponding reference class; artefact Dice was therefore calculated over artefact-containing ROIs only.

Fold-specific ROI--predictor pairs were retained only for explicitly paired visualisations. In those plots, the difference for a fusion model was calculated against the single-scale prediction for the same ROI and checkpoint fold. Cohort-level performance tables used the fold-averaged ROI values and did not treat repeated predictions for one ROI as independent observations.

\subsection{Comparator harmonisation}

The default dense predictor was compared with Otsu~\cite{r10}, H\&E-Otsu~\cite{r7}, WSI-SmartTiling~\cite{r14} and GrandQC~\cite{r13} on the same held-out ROIs. Comparator predictions were generated once per ROI and harmonised to the shared tissue/background/artefact encoding before evaluation (Supplementary Data 4, Table 25). Predictions were mapped to the dimensions of the reference mask using nearest-neighbour interpolation.

Otsu and H\&E-Otsu were evaluated as binary foreground/background methods. Foreground was mapped to tissue and non-foreground to background; neither method generated artefact predictions. Their artefact Dice was therefore zero on artefact-containing ROIs by construction and was not interpreted as a conventional artefact-detection estimate.

WSI-SmartTiling was evaluated from its native thumbnail-scale output. Qualified-tissue and background labels were preserved, and its fold- and blur-related defect labels were collapsed into the common artefact class.

GrandQC was evaluated using the pretrained GrandQC\_MPP15 quality-control checkpoint at its native 1.5 µm-per-pixel operating resolution with sliding-window inference. The separate Tissue\_Detection\_MPP10 model was used only as a gate to avoid applying the artefact model to predominantly background windows. Final comparator labels were derived from the GrandQC quality-control output: class 1 was mapped to tissue, class 7 to background and classes 2--6 to artefact.

As for the default predictor, reference ignore pixels were excluded, while invalid or unmapped prediction pixels were counted as errors. No comparator was retrained or threshold-tuned using the held-out cohorts. Comparisons therefore measured agreement after harmonisation rather than performance on the comparators' native tasks or at matched resolution. The ROI was the inferential unit. Predictor--comparator differences were calculated within ROI using the fold-averaged default-predictor value and the corresponding comparator result.

For WSI-scale use, dense predictor outputs were reconstructed into slide-aligned tissue/background/artefact maps at approximately 0.5 µm per pixel. Candidate downstream-tile coordinates were mapped to these outputs, and the predicted tissue, background and artefact fractions were calculated within each footprint. These terms refer to quantities estimated by the dense QC predictor; TileNet consumed these estimates to execute the specified selection policy.

\subsection{Tile-level burden and policy-fidelity analysis}

Dense reference and prediction masks were divided into non-overlapping 512 × 512- and 1,024 × 1,024-pixel footprints. Only footprints containing valid annotated pixels over the complete evaluated footprint were retained. Within each footprint, reference-derived and prediction-derived tissue, background and artefact fractions were calculated as the proportions of valid pixels assigned to the corresponding class.

Continuous agreement was quantified using absolute error, signed error and Spearman rank correlation. Pooled Spearman correlation was calculated over all footprints in a cohort. Spearman correlation was also calculated separately within each ROI where sufficient variation was present and summarised across ROIs. Mean absolute error and signed bias were calculated over footprints, with uncertainty estimated using 10,000 ROI-cluster bootstrap resamples that retained all footprints from a sampled ROI together.

Policy fidelity was evaluated by applying identical eligibility rules to reference-derived and predicted fractions. Artefact-only analyses evaluated maximum artefact fractions of 0, 0.01, 0.02, 0.05, 0.10, 0.15, 0.20 and 0.25 over 512 × 512-pixel footprints. A combined rule retained a footprint when artefact fraction was ≤0.01 and tissue fraction was ≥0.50. The reference-derived decision was treated as the comparison target, and rejection as the positive class. Overall agreement, balanced accuracy and the proportions retained by the reference- and prediction-derived policies were calculated. Confidence intervals were obtained from 10,000 ROI-cluster bootstrap resamples.

For the combined rule, component-gate agreement was decomposed into concordant decisions, artefact-gate-only disagreement, tissue-gate-only disagreement, disagreement involving both gates and masked gate disagreement. Masked disagreement denoted footprints for which one or both component gates differed between reference and prediction but the final combined decision remained concordant because the other gate independently determined rejection.

Larger artefact-underestimation events were defined descriptively as footprints with reference artefact burden \textgreater0.05 and predicted artefact burden ≤0.01. Events were counted by ROI, ROIs were ranked by event count, and the cumulative fractions attributable to the highest-contributing one, three and five ROIs were calculated separately for each cohort and footprint size. These analyses characterised the concentration of policy-relevant errors and were not used to define a new operating threshold.

\subsection{Blinded pathologist audit}

The blinded pathologist audit used 100 512 × 512-pixel crops sampled from 41 ROIs in the TCIA external-validation cohort at 20×-equivalent resolution. Crop selection was completed before review using only reference-mask-derived artefact fractions and ROI metadata and without visual inspection of the H\&E images.

Reference artefact burden was calculated as the fraction of valid pixels labelled artefact in the reference mask. Reference binary artefact presence was defined as any non-zero \textbf{reference-derived} artefact fraction. For reporting and sampling, crops were divided into six reference-burden strata: none (0\%), very low (\textgreater5\% to \textless7\%), low (7\% to \textless10\%), moderate (10\% to \textless20\%), high (20\% to \textless50\%) and very high (≥50\%). The final set comprised 22 artefact-negative, 23 very-low, 18 low, 15 moderate, 14 high and 8 very-high crops. No ROI contributed more than four artefact-positive crops or one artefact-negative control, giving a maximum of five crops from one ROI. Crop coordinates were unique, although crops from the same ROI could overlap partially.

A board-certified pathologist from a separate institution, who had not participated in reference annotation, crop selection or model development, reviewed the 100 crops in a single blinded batch. The reviewer was unaware of the reference masks, reference burden values, sampling strata, predictor outputs and policy decisions. Artefact presence was recorded as yes or no; severity as none, mild, moderate or severe; usability as usable, reject or uncertain; and dominant artefact type using a predefined category list. Label variants arising during recording or translation were harmonised to the original category set before summarisation.

A reference-derived usability proxy classified a crop as reject when background occupied \textgreater70\% of valid pixels or artefact occupied \textgreater5\%; otherwise, the crop was classified as usable. For severity comparison, the six burden strata were mapped to four categories: none to none, very low and low to mild, moderate to moderate, and high and very high to severe.

Artefact-presence and usability agreement were quantified using percentage agreement and Cohen's κ. Crops rated uncertain were excluded from the primary binary usability analysis; a secondary analysis retained uncertain as a separate category. Severity agreement was assessed using exact agreement and quadratically weighted Cohen's κ. Percentile confidence intervals were obtained from 10,000 ROI-cluster bootstrap resamples.

Prediction-derived tissue, background and artefact fractions were calculated from the dense predictor output over the same crop footprints. Among crops with definite usability calls, discrimination of pathologist-rejected versus usable crops was quantified using AUROC. Confidence intervals for reference- and prediction-derived AUROC and their paired difference were estimated using 10,000 case-cluster bootstrap resamples, retaining all crops from a selected TCIA case together and applying identical resamples to both burden sources. The combined rule (artefact fraction ≤0.01; tissue fraction ≥0.50) was additionally applied to the reference- and prediction-derived fractions; in this audit sample, it yielded the same reference-derived decisions as the usability proxy. Usable retention, rejected-crop exclusion, balanced accuracy and overall agreement with the pathologist were calculated for each burden source. The audit was not designed to optimise thresholds below 5\%, because artefact-positive crops were sampled with reference burdens of approximately 5\% or greater.

\subsection{CAMELYON16 candidate universes and fixed-budget policies}

The CAMELYON16 experiment used the 270 labelled training WSIs represented in the processed patch dataset, comprising 159 normal and 111 tumour slides. Two candidate-patch universes were evaluated under a fixed budget of 4,000 patches per slide.

G0 used the candidate universe produced by the original CLAM foreground-selection procedure. U0--U5 used an exhaustive candidate universe comprising all in-bounds 256 × 256-pixel tiles over each WSI. The common-support cohort comprised 244 slides---134 normal and 110 tumour---for which both G0 and U5c, the most restrictive combined QC-based eligibility rule, could provide the full 4,000-patch budget. Because every other U policy imposed less restrictive criteria on the same exhaustive universe, all evaluated arms contained exactly 4,000 patches per slide on these 244 slides.

One fixed set of five stratified slide-level cross-validation splits was used for every condition. Development, validation and test partitions were identical across arms within each fold. The five outer test folds formed an exact partition of the common-support cohort, so every slide contributed one held-out prediction per arm and training seed. Three independent training seeds were used, yielding 15 fold--seed evaluations per arm.

G0 applied the original fixed-budget random-sampling procedure to the original CLAM candidate universe. Within the exhaustive universe, each patch was assigned a deterministic slide-specific random priority using a fixed seeding procedure. This priority was frozen and reused across all U policies. U0 imposed no QC criterion and selected the first 4,000 patches in the frozen random order, corresponding to reproducible uniform sampling without replacement.

U1 policies imposed maximum predicted artefact fractions of 0.20, 0.10 or 0.001. U2 policies imposed maximum predicted background fractions of 0.50, 0.30 or 0.10. U3, U4 and U5 combined background thresholds of 0.50, 0.30 and 0.10, respectively, with each of the three artefact thresholds. For each filtered policy, only eligible patches were considered. When more than 4,000 patches were eligible, the first 4,000 eligible patches in the frozen priority order were retained. Policy-failing patches were never used to backfill a bag.

Because G0 and the U arms sampled from different candidate universes, G0-versus-U contrasts were treated as descriptive. U0 was the matched reference for QC-policy inference because U0 and U1--U5 shared the same exhaustive universe, fixed random priority and patch budget.

\subsection{CAMELYON16 attention-localisation analysis}

Attention localisation was evaluated on all 110 tumour slides in the common-support cohort, each of which had a CAMELYON16 lesion annotation. For each slide, arm and training seed, attention weights were reconstructed from the out-of-fold CLAM model corresponding to the fold in which that slide was held out. Retained patches were ranked by normalised attention weight.

For k = 1, 5, 10, 50 and 100, the level-0 centre coordinate of each of the k highest-attention patches was tested for overlap with the lesion annotation. Tumour-hit fraction was the proportion of the top-k patch centres falling within annotated tumour. The any-hit endpoint indicated whether one or more of the top-k centres overlapped tumour.

Because the selection policy altered tumour prevalence in the retained bag, localisation was also evaluated relative to a slide- and arm-specific random-ranking expectation. Expected tumour-hit fraction was the proportion of all retained patches whose centres overlapped tumour. Hit-fraction enrichment was calculated as observed top-k hit fraction minus retained-bag tumour prevalence.

Metrics were calculated separately for each of the three trained models and averaged across seeds within slide. Policy-level estimates and confidence intervals were obtained from 10,000 slide-level bootstrap resamples. Pairwise comparisons preserved policy pairing within slide. U4c-versus-U0 was the controlled comparison because both arms operated on the same exhaustive candidate universe; comparisons involving G0 were descriptive cross-universe comparisons.

\subsection{Scanner-mark audit and same-model counterfactual analysis}

Scanner-mark-containing regions admitted by the original CLAM foreground-selection procedure were identified through blinded manual review. Candidate regions were reviewed without slide identity and subsequently linked to their WSIs using a blinding key. Twenty-eight slides contained at least one reviewed scanner-mark region; 27 were members of the 244-slide common-support cohort and were eligible for out-of-fold analysis. Reviewer-confidence levels were assigned to each region. High-confidence regions defined the primary QC-channel analysis, and broader confidence sets were retained as sensitivity analyses.

For each manually delineated scanner-mark region, the mean predicted tissue, background and artefact fractions were calculated from the dense QC-predictor output. The dominant QC channel was defined as the channel with the largest mean fraction. Scanner marks were treated as manually identified non-biological technical regions; no assumption was made that they should map specifically to the predicted artefact channel. Background and artefact denote model output channels, not mutually exclusive physical explanations for the presence of technical marks. A scanner mark assigned to background can escape artefact-only filtering while being removed by a background-aware policy.

Predicted QC fractions were then mapped to the patch coordinates of the original G0 bags. High-background patches were defined by predicted background fraction \textgreater0.50.

For each scanner-mark-positive slide and each out-of-fold G0 model, CLAM attention weights and attention-weighted instance contributions to the tumour-versus-normal classifier margin were recomputed from the stored patch embeddings. Attention mass was the sum of normalised attention weights assigned to high-background patches. Attention enrichment was this mass divided by the proportion of the bag classified as high background.

Absolute-logit-contribution fraction was the proportion of the total absolute attention-weighted instance contribution assigned to high-background patches. Absolute-logit-contribution enrichment was this fraction divided by the proportion of patches classified as high background. These quantities were treated as components of the trained CLAM decision function and not as independent causal attribution measures.

For same-model counterfactual evaluation, patches exceeding the specified predicted-background threshold were removed and the remaining bag was passed through the unchanged out-of-fold G0 model. Attention and slide-level probability were recomputed without retraining. For each seed, counterfactual tumour probability, decision-threshold crossing and change in correctness were recorded. Positive values of P$_{\mathrm{original}}$ - P$_{\mathrm{counterfactual}}$ indicated that the removed patches had provided net support for the tumour class; negative values indicated support for the normal class. Because removal was based on predicted background fraction rather than the manually delineated scanner-mark mask, this analysis measured dependence on QC-defined high-background instances and was not interpreted as a causal test of scanner marks.

\subsection{G0--U4c prediction-disagreement analysis}

Out-of-fold G0 and U4c predictions were aligned by slide, fold and training seed for descriptive prediction-disagreement analysis. U4c was the combined background-and-artefact policy with the highest fold--seed mean AUROC and retained patches with predicted background fraction ≤0.30 and predicted artefact fraction ≤0.001.

For each slide and seed, the signed U4c-minus-G0 tumour-probability difference, its absolute magnitude and hard-classification agreement were calculated. Seed-level values were aggregated by slide to obtain mean G0 and U4c probabilities, mean signed difference, mean and maximum absolute differences, number of seed-level classification disagreements and correctness under each arm.

Slides were defined as reproducibly discordant when G0 and U4c produced different hard classifications in all three independently trained models. Because subset membership was determined from the observed predictions, this group was treated as a post hoc exploratory mechanistic subset rather than an independently sampled inferential cohort. Its class and correctness composition were therefore reported explicitly.

The same-model G0 counterfactual procedure was applied separately to each slide and seed in this subset after removing patches with predicted background fraction \textgreater0.50. High-background patch fraction, attention enrichment, absolute-logit-contribution enrichment, mean absolute probability change, prediction flips and correctness changes were summarised across seeds. This analysis distinguished immediate sensitivity of a fixed G0 model to high-background-patch removal from the broader differences between independently trained G0 and U4c models arising from different candidate universes and selection procedures.

\subsection{Matched weakly supervised lung classification}

The weakly supervised lung experiment used 112 H\&E image regions (56 squamous-cell-carcinoma (SCC) and 56 normal regions), with two regions from each of 56 TCGA frozen-section slides (TS/BS), representing 41 tissue samples from 40 patients (Supplementary Data 3). Regions were partitioned before tile generation into 72 fitting, 16 validation and 24 test regions, with balanced class counts. Partitions were disjoint at the patient, sample, slide and source-region levels and used non-overlapping TCGA Tissue Source Site (TSS) codes (fitting: 34, 37, 38 and 43; validation: 39; test: 51 and 56). This experiment was a weak-supervision stress test rather than a diagnostic benchmark. Unlike the FFPE diagnostic slides used to develop the dense QC predictor, these source images were frozen sections. Source-slide and patient identifiers were cross-referenced against the complete TCGA-derived PathQC-3C cohort, confirming no WSI- or patient-level overlap.

Each image region was divided into non-overlapping 224 × 224-pixel RGB tiles at 20×-equivalent resolution. The region-level normal or SCC label was inherited by all retained tiles from that region. Tiles were therefore weakly labelled and were not interpreted as independent diagnostic specimens.

Predicted background fraction was restricted to \textless0.30 in every condition. Four predicted-artefact policies were evaluated: strict exclusion of tiles containing any predicted artefact pixels, maximum predicted artefact fractions of 0.01 and 0.05, and unrestricted artefact inclusion.

The strict-exclusion fitting pool contained 18,183 eligible tiles and defined the matched training budget. For each more permissive policy, tiles were sampled without replacement to reproduce the strict-exclusion count in every patient-by-class stratum. The resulting fitting sets had identical total size, class counts, patient identities and patient-by-class contributions. Equality of the matching strata was verified directly from the generated manifests before fitting.

All conditions used the same validation and primary test sets, defined under strict artefact exclusion and containing 2,523 and 5,042 tiles, respectively. The primary test set comprised 1,325 normal and 3,717 SCC tiles from ten patients, with five patients per class.

The ≤1\%, ≤5\% and unrestricted policies each used three independently sampled matched fitting sets generated with sampling seeds 42, 43 and 44. Each set was trained with optimisation seeds 42, 43 and 44, yielding a fully crossed 3 × 3 design and nine fitted models per artefact-permitting policy. Strict exclusion produced one deterministic fitting set, which was trained with the same three optimisation seeds. Thirty models were fitted in total.

Each condition used the ViT-Base classifier~\cite{r35} google/vit-base-patch16-224-in21k, pretrained on ImageNet-21K. Training augmentation comprised random horizontal and vertical flips and colour jitter with brightness, contrast and saturation parameters of 0.1 and hue of 0.03. Models were trained using unweighted cross-entropy, AdamW~\cite{r36} with learning rate 5 × 10⁻⁵ and weight decay 0.01, batch size 16 and a constant learning-rate schedule with 10\% warm-up. Training continued for up to 50 epochs. The checkpoint with the highest validation macro-F1 was retained, and early stopping was applied after seven consecutive epochs without improvement.

Performance was quantified using accuracy, macro-F1, balanced accuracy, class-specific recall, class-specific precision and patient-weighted accuracy. Patient-weighted accuracy was the unweighted mean of the ten patient-specific tile accuracies and was treated as a coarse secondary measure because only ten patients contributed to the test set.

\subsection{Tumour--stroma segmentation model and optimisation}

Dense downstream adaptation used a binary tumour--stroma segmentation task derived from BCSS annotations. Images and reference masks were resized to 384 × 384 pixels. Pixels outside the annotated target support were assigned ignore label 255 and excluded from losses and metrics.

The model combined the CONCH v1.5 histopathology image encoder (ViT-L/16), obtained through the MahmoodLab/TITAN distribution, with a lightweight dense-prediction head. Patch tokens were reshaped to their spatial grid, processed by two convolutional layers and bilinearly upsampled to the input resolution. The CONCH v1.5 backbone contained approximately 303 million parameters and the segmentation head 12.2 million parameters, corresponding to 3.87\% of the complete 315.5-million-parameter model.

Optimisation proceeded in two stages for each random seed. During stage 1, the CONCH v1.5 backbone was frozen and only the segmentation head was trained. Stage 2 was initialised from the seed-specific stage-1 checkpoint with the highest validation macro Dice. In the standard stage-2 condition, the trunk and head were both trainable. In the frozen-backbone analysis, the trunk remained frozen and only the head was updated.

AdamW was used with a head learning rate of 1 × 10⁻⁴, a backbone learning rate of 1 × 10⁻⁵ when trainable and weight decay of 1 × 10⁻⁴. Training used batch size 1 with gradient accumulation over eight batches, giving an effective batch size of 8. Each epoch comprised 1,500 accepted training batches. The objective combined ignore-aware cross-entropy and soft macro-Dice with weights of 0.8 and 0.2. Cross-entropy weights were derived from target-task pixel counts using inverse frequency raised to the power 0.25 and normalised to a mean of one. Stage-2 training continued until global epoch 44, and the checkpoint with the highest validation macro Dice was retained.

All configurations were trained using seeds 1337, 2027 and 31415. The training seed was the replicated experimental unit for comparisons among stage-2 conditions.

\subsection{Experimental augmentation interventions}

Four stage-2 conditions were evaluated: unaugmented baseline, real-artefact augmentation, random erasing and matched clean-tissue replacement. Augmentation was applied to accepted fitting tiles with probability 0.5. For each augmented tile, target modified-area fraction was drawn uniformly from 0.05 to 0.35. Components were added until realised alpha-mask coverage reached the target or 60 placement attempts had been made. The tumour--stroma reference mask remained unchanged.

For each sampled component, horizontal flipping was applied with probability 0.5, rotation was sampled uniformly from 0° to 360°, and isotropic scale was sampled uniformly from 0.6 to 1.6. Components larger than the host image were reduced to fit. The effective area contributed by one component was capped at approximately one-third of the target coverage to limit overshoot. Placement coordinates were sampled uniformly from valid origins. Alpha masks were resized bilinearly and smoothed using a Gaussian filter with radius 1.5 pixels before compositing.

Real-artefact augmentation inserted the sampled RGB artefact component through its transformed alpha mask. Random erasing used the same sampled masks, geometry, placement and target coverage but filled the region with the mean RGB colour of the host image plus Gaussian noise with standard deviation 6 intensity units. Noise used an independent deterministic random-number stream so that generating the fill did not alter geometry or placement draws.

Matched clean-tissue replacement inserted artefact-free tissue through the transformed mask. When the sampled artefact entry lacked an associated clean crop, the experimental fallback remapped the event to an entry in the available clean-tissue subset without consuming additional draws from the main augmentation random-number stream. Consequently, clean-tissue replacement achieved similar mean realised coverage but drew on a subset of smaller source components and was not paired to identical mask geometry for every real-artefact augmentation event.

A dedicated augmentation random-number generator was initialised from each training seed independently of the main data-loading and optimisation streams. Experimental and public-release configurations are distinguished in Supplementary Data 4, Table 22. In particular, the public release provides deterministic behaviour based on seed, sample identifier and mode but is not represented as an exact reimplementation of every event in the experimental training pipeline.

\subsection{Held-out composited real-artefact robustness evaluation}

Robustness was evaluated by compositing held-out real histological artefacts onto evaluation tiles. Artefact components were sampled from an evaluation library constructed from the TCGA-common-source held-out partition, which was disjoint from the training artefact library at both ROI and WSI levels. S0 represented the uncorrupted image, whereas S1--S3 represented increasing perturbation severity with target artefact coverages of 5\%, 15\% and 30\%, respectively. Components were composited until the target coverage was reached or 60 placement attempts had been made. Geometric transformations and alpha-mask feathering followed the training augmentation procedure.

Corruptions were generated deterministically for each tile and severity. A random seed derived from the tile key, corruption identifier and severity using an MD5-based hash ensured that all model conditions received byte-identical corrupted versions of each evaluation tile.

\subsection{Robustness endpoint and frozen-backbone analysis}

Baseline artefact burden was measured on each uncorrupted evaluation tile using the default three-class dense QC predictor. The robustness analysis was restricted to the high-artefact stratum, defined as tiles with predicted artefact fraction ≥0.005 (≥0.5\%). This stratum comprised 288 unique tiles, defined from the uncorrupted images before robustness perturbation, and the same tiles were evaluated for every model condition. The complete validated evaluation cohort comprised 500 low-burden (\textless0.01\%), 264 moderate-burden (0.01\% to \textless0.5\%) and 288 high-burden (≥0.5\%) tiles. Stratification used unperturbed images and was held fixed across corruption conditions.

For each severity, tumour--stroma confusion matrices were accumulated after excluding reference pixels labelled 255. Confusion counts were pooled across all 288 tiles before tumour Dice, stroma Dice and macro Dice were calculated. Reported Dice values therefore reflected pooled pixel counts rather than the arithmetic mean of per-tile Dice values.

The principal reported robustness endpoint was pooled macro Dice in the high-artefact stratum under S3 held-out composited real-artefact corruption. Intervention effects were calculated within training seed relative to the corresponding unaugmented baseline and then summarised across the three seeds.

Backbone trainability and real-artefact augmentation were crossed in a 2 × 2 analysis comprising unaugmented/trainable, real-artefact/trainable, unaugmented/frozen and real-artefact/frozen conditions. Frozen and trainable runs were initialised from the corresponding seed- and condition-matched stage-1 checkpoints. The augmentation effect was calculated separately under frozen and trainable conditions. The interaction was the trainable-backbone augmentation effect minus the frozen-backbone augmentation effect. Analogous interactions were calculated for S0-to-S3 degradation and uncorrupted S0 performance. With only three training seeds, interactions were interpreted descriptively and not as tests of equivalence.

Backbone freezing was verified by comparing all stored CONCH v1.5 backbone tensors after stage 2 with their stage-1 initialisations. All 342 tensors were required to remain bitwise identical, with maximum absolute difference zero. Parameter and optimiser-state audits confirmed that only the 12.2-million-parameter segmentation head was trainable in frozen-backbone runs.

\subsection{Statistical analysis and reproducibility}

Experimental units and uncertainty procedures were defined separately by analysis.

For development-stage dense-predictor comparisons, the experimental units were the three patient-level folds. Values were reported descriptively as mean ± s.d., and no population-level significance tests were performed.

For held-out quality-control evaluation and comparator benchmarking, the ROI was the inferential unit. Fold-specific predictor outputs were averaged within ROI before cohort-level summarisation. Macro Dice and macro IoU were calculated using only classes present in the ROI reference mask, and class-specific metrics were calculated only for ROIs containing that class.

For tile-level policy fidelity, footprints were nested within ROIs. Confidence intervals were therefore obtained by resampling ROIs and retaining all footprints from a selected ROI together.

For the blinded pathologist audit, the crop was the observed unit, with clustering by ROI or TCIA case as specified. Uncertain usability calls were excluded from the primary binary analysis. Perfect observed agreement in the 100-crop presence audit produces degenerate nonparametric bootstrap intervals because resampling reproduces the same concordant observations. Such intervals do not quantify population certainty or performance below the sampled positive-burden range.

For CAMELYON16 classification, the slide was the inferential unit. U-policy effects were estimated using paired class-stratified slide-level bootstrap resampling of out-of-fold predictions. Fold--seed values were presented as descriptive measures of partition and fitting stochasticity and were not treated as independent biological observations. The 15 U-policy-versus-U0 contrasts used pointwise 95\% bootstrap intervals without adjustment for multiple comparisons. U4c was selected post hoc from the combined policies; subsequent U4c-focused analyses were exploratory.

For TCGA-LUSC, independently fitted models were the replicate units for descriptive comparisons among policies. Tiles were used to calculate performance metrics but were not treated as independent experimental replicates. The nine cells of the crossed sampling-seed × optimisation-seed design were retained as matched computational replicates and were not analysed as nine independent biological observations. No equivalence margin was prespecified.

For PathQCAug, training seeds were the replicated units. Intervention contrasts and factorial interactions were calculated within seed before summarisation. The cached evaluation tiles contributed to each seed-level endpoint but were not treated as independent replicates for comparisons among fitting conditions.

Unless otherwise stated, confidence intervals were percentile intervals from 10,000 bootstrap resamples. Values reported as mean ± s.d. refer to variation across the experimental unit specified for that analysis. Development ablations, TCGA-LUSC crossed-replicate contrasts and PathQCAug factorial interactions were interpreted descriptively.

Fold definitions, random seeds, matched-dataset manifests, run configurations, selected checkpoints, training histories and evaluation outputs were retained. Validation crop selection was deterministic. Held-out cohorts were not used for model fitting or hyperparameter optimisation. Evaluations used fixed trained checkpoints, and no model parameters were updated using held-out evaluation data. Exploratory and post hoc analyses are identified explicitly where applicable.

\section{Results}
\begin{figure*}[!tp]
\centering
\includegraphics[width=1.000\textwidth]{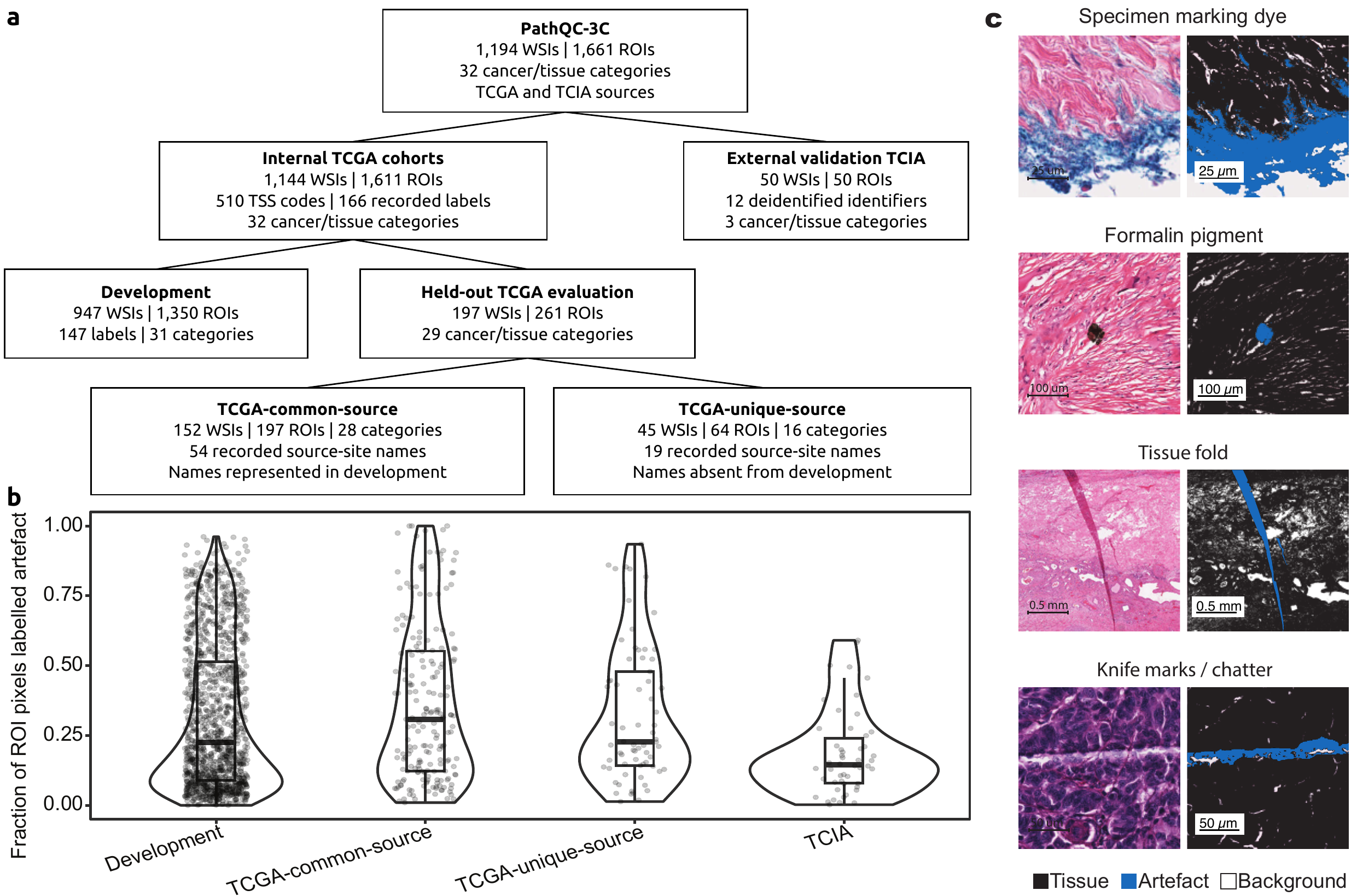}
\setcounter{figure}{1}\caption{\textbf{PathQC-3C cohort structure, artefact burden and representative dense annotations. a, Cohort composition and evaluation hierarchy. PathQC-3C comprises 1,194 H\&E WSIs and 1,661 ROIs covering 32 cancer/tissue categories. The TCGA component comprises 1,144 WSIs and 1,611 ROIs from 510 barcode TSS codes mapped to 166 distinct recorded GDC source-site names. Development contains 947 WSIs and 1,350 ROIs from 147 recorded names. Held-out TCGA evaluation contains 197 WSIs and 261 ROIs: TCGA-common-source comprises 152 WSIs and 197 ROIs from 54 names represented during development; TCGA-unique-source comprises 45 WSIs and 64 ROIs from 19 names absent from development. These units are registry names rather than verified independent institutions; common-source evaluation does not uniformly hold out source-site--category combinations. External validation used 50 TCIA WSIs and 50 ROIs with 12 separately reported deidentified source identifiers and three cancer/tissue categories.} b, Distribution of ROI-level artefact fraction across the development, TCGA-common-source, TCGA-unique-source and TCIA cohorts. Points represent ROIs; violin plots show the distributions and embedded boxplots indicate medians and interquartile ranges. c, Representative H\&E ROIs and corresponding dense three-class annotations showing specimen marking dye, formalin pigment, tissue fold and knife marks/chatter. Tissue is shown in black, artefact in blue and background in white.}
\label{fig:2}
\end{figure*}

\subsection{PathQC-3C defines a three-class QC representation}

To support artefact-aware quality-control policies, we assembled PathQC-3C from public haematoxylin and eosin whole-slide images in TCGA~\cite{r24} and TCIA/CPTAC~\cite{r25,r26,r27}. The resource comprises 1,661 densely annotated regions of interest (ROIs) from 1,194 whole-slide images covering 32 cancer/tissue categories (Fig. 2a). The TCGA component represents 166 distinct recorded GDC source-site names; TCIA contributes 12 separate deidentified source identifiers. Each ROI was labelled using three mutually exclusive classes: tissue, background and artefact. The artefact class intentionally aggregates heterogeneous technical defects into an operational category intended to support exclusion, ranking, down-weighting or review rather than fine-grained artefact diagnosis (Fig. 1b, Fig. 2c and Supplementary Fig. 1).

Artefact burden varied across the development and evaluation cohorts. Median annotated artefact fraction was 0.226 (IQR, 0.089--0.514) in the development cohort. Across the held-out cohorts, median annotated artefact fractions were 0.308 (0.122--0.552) in TCGA-common-source, 0.222 (0.146--0.490) in TCGA-unique-source and 0.145 (0.079--0.240) in TCIA (Fig. 2b). ROIs with \textgreater50\% artefact coverage accounted for 25.7\%, 29.9\%, 25.0\% and 6.0\% of the respective cohorts. PathQC-3C therefore combines dense 20×-equivalent annotation with source-site and cancer/tissue-category diversity in a common tissue/background/artefact representation (Table~\ref{tab:resources}).

\subsection{Artefact delineation guides predictor selection}

We first tested whether conventional foreground detection was sufficient for artefact-aware quality control. A two-class Swin--UPerNet-Base model, in which background and artefact were combined as non-tissue, achieved a macro Dice of 0.957 ± 0.003 (Supplementary Data 4, Table 1). Tissue Dice was similar between the binary and matched three-class models (0.972 versus 0.970), but the binary model could not distinguish empty background from artefact-affected tissue and therefore could not support different handling of these two conditions.

We next compared candidate predictors for the three PathQC-3C classes in a baseline architecture comparison using 512 × 512-pixel inputs and the Dice--IoU objective. Tissue and background were consistently well delineated across architectures, whereas artefact Dice ranged from 0.464 ± 0.164 for EfficientNet-B0 U-Net++~\cite{r28,r29} at 10× to 0.736 ± 0.058 for Swin--UPerNet-Base at 20× (Supplementary Data 4, Table 2 and Supplementary Figs. 2--4). Swin--UPerNet-Base also achieved the highest mean macro Dice (0.886 ± 0.016) while requiring less forward latency than Swin--UPerNet-Large and SegFormer-B5~\cite{r30} (47.92 versus 61.12 and 69.72 ms per batch, respectively; Supplementary Figs. 5 and 6). Complementary end-to-end validation-throughput measurements, incorporating data loading and evaluation overheads, are shown in Supplementary Fig. 7.

As an exploratory test of whether a general segmentation foundation model could serve as the dense quality-control predictor within TileNet, we adapted a frozen SAM3~\cite{r31} vision encoder to the same three-class tissue/background/artefact output space using a supervised semantic-segmentation decoder. Artefact delineation was lower and more variable than with the selected task-trained predictor (artefact Dice, 0.550 ± 0.179; Supplementary Data 4, Table 3). We therefore selected Swin--UPerNet-Base for subsequent optimisation.

Increasing spatial context produced the clearest improvement during configuration optimisation. A 1,024-pixel source crop, resized to the same 512 × 512-pixel network input, improved artefact Dice relative to 512- and 256-pixel source crops across the tested loss and foreground-sampling configurations (Supplementary Data 4, Table 4). The selected configuration used a 1,024-pixel source crop, unweighted Dice--cross-entropy and no minimum-foreground constraint, achieving macro Dice of 0.887 ± 0.009 and artefact Dice of 0.756 ± 0.034. Although a 256-pixel weighted-loss configuration achieved higher mean macro Dice (0.892 ± 0.033), the selected configuration provided the highest artefact Dice in the ablation grid with lower macro-Dice variability. Changes in class weighting, foreground sampling, batch size, per-epoch sampling budget and learning rate produced smaller or less consistent differences (Supplementary Data 4, Tables 4--7).

Because many artefacts combine fine local texture with larger-scale spatial structure, we also evaluated paired 10× and 20× multi-magnification inputs. This approach yielded a higher mean artefact Dice, reaching 0.795 ± 0.132 with late fusion of paired 10× and 20× inputs, but the gain was accompanied by substantially greater fold-to-fold variability, no consistent improvement in macro Dice and higher forward latency (65.12 versus 47.79 ms per batch; Supplementary Data 4, Table 8 and Supplementary Figs. 5 and 6). The single-scale 20× Swin--UPerNet-Base configuration with 1,024-pixel source crops was therefore retained as the default dense quality-control predictor within TileNet.

\begin{figure*}[!tp]
\centering
\includegraphics[width=0.886\textwidth]{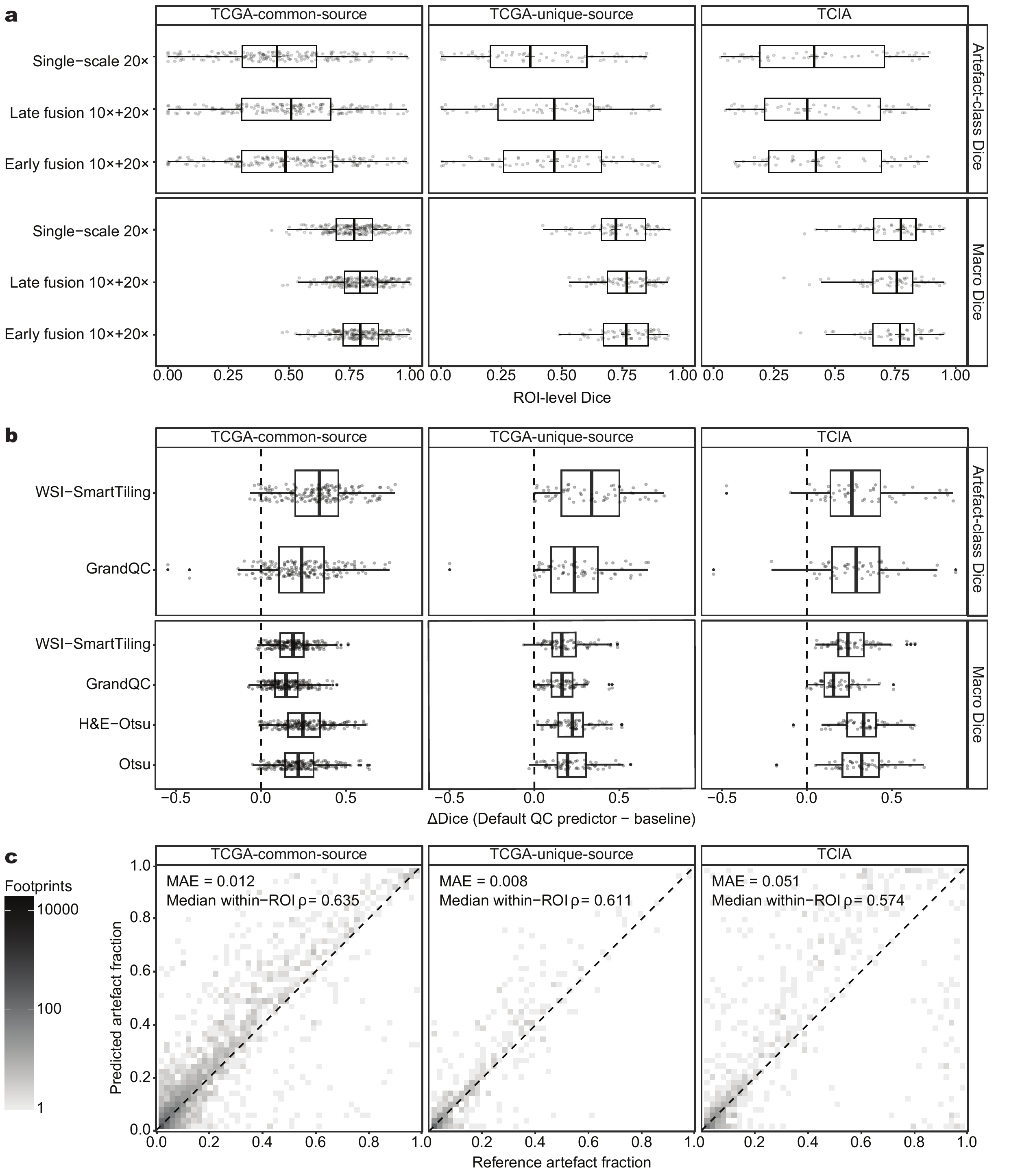}
\setcounter{figure}{2}\caption{\textbf{Held-out three-class quality-control performance, comparator benchmarking and tile-level policy fidelity.} a, ROI-level macro Dice and artefact Dice for the default single-scale 20× dense QC predictor and paired 10×+20× early- and late-fusion variants across TCGA-common-source, TCGA-unique-source and TCIA. Metrics from fold-specific checkpoints were averaged within ROI. Macro Dice is shown for all ROIs and artefact Dice for artefact-containing ROIs. Boxplots show medians, interquartile ranges and 1.5× interquartile-range whiskers; points represent ROIs. b, Paired ROI-level Dice differences between the default predictor and harmonised Otsu, H\&E-Otsu, WSI-SmartTiling and GrandQC outputs. Positive values favour the default predictor. Artefact Dice differences are shown only for artefact-containing ROIs and methods producing an explicit artefact class. c, Agreement between reference- and prediction-derived artefact fractions within non-overlapping 512 × 512-pixel footprints. Two-dimensional bins indicate footprint density and the dashed diagonal denotes perfect agreement. Panels report mean absolute error and median within-ROI Spearman correlation.}
\label{fig:3}
\end{figure*}

\begin{figure*}[!tp]
\centering
\includegraphics[width=0.778\textwidth]{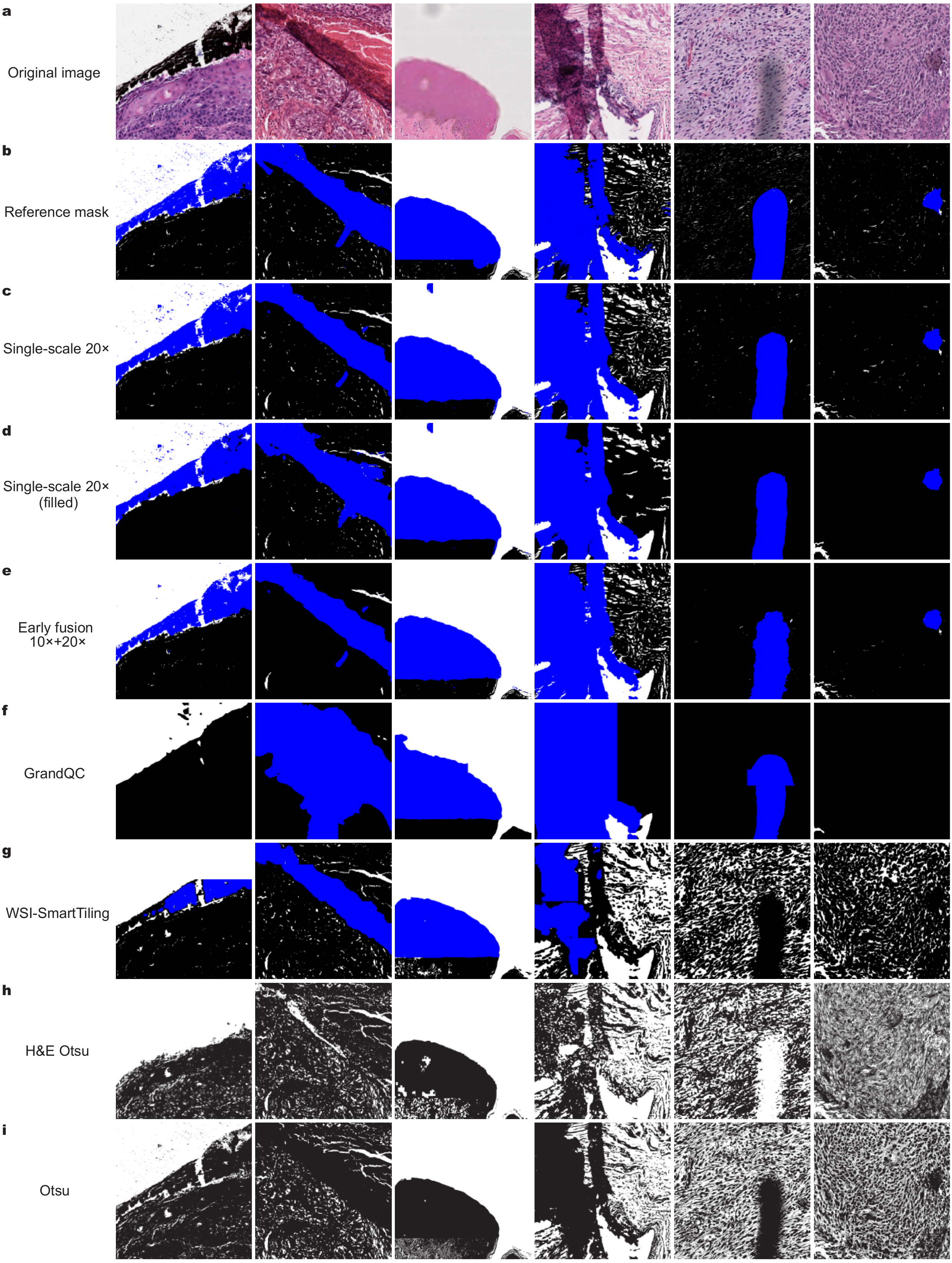}
\setcounter{figure}{3}\caption{\textbf{Qualitative examples of three-class quality-control prediction from the held-out cohorts.} Representative 20× ROIs from the held-out evaluation cohorts. a, Original H\&E images. b, Dense three-class reference masks. c--e, Dense QC predictor outputs from the default single-scale 20× model, the corresponding hole-filled post-processing variant and the early-fusion 10×+20× model. f--i, Comparator outputs from GrandQC, WSI-SmartTiling, H\&E-Otsu and Otsu. Masks show tissue in black, background in white and artefact in blue. Binary thresholding methods do not explicitly represent artefact, whereas the dense QC predictors provide separate tissue, background and artefact estimates suitable for downstream policy aggregation.}
\label{fig:4}
\end{figure*}

\subsection{Held-out quality-control performance and tile-policy agreement}

We evaluated the default single-scale 20× predictor across three held-out cohorts: TCGA-common-source, comprising held-out slides from recorded GDC source-site names represented during development; TCGA-unique-source, comprising recorded GDC source-site names absent from development; and a TCIA external-validation cohort assembled from separate CPTAC collections. Development-fold and held-out segmentation metrics were calculated using different aggregation procedures: development validation pooled predictions across deterministic crops, whereas held-out evaluation calculated metrics separately for complete ROIs before cohort-level averaging (Methods). These estimates are therefore not directly comparable. Across the three held-out cohorts, performance was similar, with macro Dice ranging from 0.730 to 0.768 and artefact Dice from 0.413 to 0.456, with artefact delineation remaining the limiting component of the three-class task (Fig. 3a and Supplementary Data 4, Tables 9--12).

To assess performance relative to existing quality-control approaches, we also benchmarked the default predictor against established threshold-based and learning-based methods on the same held-out ROIs. Under a harmonised three-class evaluation, the default predictor outperformed all evaluated comparators (Fig. 3b, Fig. 4, Supplementary Figs. 8 and 9 and Supplementary Data 4, Tables 10--12). Relative to GrandQC, the strongest learning-based comparator, mean paired ROI-level improvements ranged from 0.154 to 0.185 in macro Dice and from 0.243 to 0.286 in artefact Dice across the three held-out cohorts.

Since TileNet applies selection rules to the predicted fractions of tissue, background and artefact within tile footprints rather than directly to pixel-level segmentation boundaries, we tested how closely these predicted fractions agreed with those derived from the reference masks. Artefact fractions showed moderate rank agreement at 512 pixels and stronger agreement at 1,024 pixels, whereas tissue and background fractions showed stronger agreement at both scales (Fig. 3c and Supplementary Fig. 10).

We next examined whether differences in the estimated tissue and artefact fractions affected which tiles were retained. Under the combined rule (artefact fraction ≤0.01; tissue fraction ≥0.50), tile-selection decisions based on model predictions agreed with those based on the reference annotations for 93.9\%, 92.6\% and 88.8\% of 512-pixel footprints in TCGA-common-source, TCGA-unique-source and TCIA, respectively. The corresponding balanced accuracies were 0.938, 0.914 and 0.868, indicating that agreement was also high when accounting for differences in the frequencies of retained and rejected footprints. At 1,024 pixels, overall agreement was 93.7\%, 91.4\% and 87.8\%, respectively. Thus, improved continuous burden agreement at larger footprints did not translate into higher decision agreement. Most remaining disagreements were driven by the artefact criterion and were concentrated in a limited subset of ROIs (Supplementary Figs. 10 and 11). Prediction-derived fractions therefore reproduced most reference-derived tile-selection decisions across held-out cohorts despite imperfect pixel-level artefact delineation.

\subsection{Expert review separates artefact recognition from usability}

To examine how artefact burden relates to expert judgement, a board-certified pathologist from a separate institution, who had not participated in reference annotation, crop selection or model development, reviewed 100 blinded 512 × 512-pixel crops from 41 TCIA ROIs sampled across six reference-derived burden strata (Supplementary Data 1). The audit included 22 artefact-negative and 78 artefact-positive crops; the pathologist classified 43 as usable, 45 as rejected and 12 as uncertain (Supplementary Fig. 12 and Supplementary Data 4, Table 13). Artefact presence agreed with the reference annotations in all 100 crops (κ = 1.00), although artefact-positive crops were sampled above 5\% burden, limiting inference about smaller artefacts. Agreement between pathologist-rated and burden-derived severity yielded a quadratically weighted κ of 0.66.

Among the 88 definite usability assessments, the reference-derived combined rule (artefact fraction ≤0.01; tissue fraction ≥0.50) agreed with the pathologist in 73.9\% of cases (κ = 0.47; balanced accuracy, 0.733); in this sample, its decisions were identical to those of the reference-derived usability proxy (\textgreater5\% artefact or \textgreater70\% background). Disagreement was strongly asymmetric: the rule excluded 22 of 43 usable crops, 21 of them with artefact burdens between 5\% and 10\%, but retained only one of 45 rejected crops.

The prediction-derived rule achieved the same overall agreement (73.9\%; balanced accuracy, 0.735) while retaining 25 of 43 usable crops and excluding 40 of 45 rejected crops. Artefact burden discriminated rejected from usable crops with AUROCs of 0.904 and 0.846 using reference- and prediction-derived fractions, respectively (Supplementary Fig. 13 and Supplementary Data 4, Table 14). Thus, similar overall agreement concealed different exclusion and retention errors. This single-reader, burden-stratified audit does not establish a universal usability threshold or performance at routine prevalence.

We also evaluated optional hole filling for workflows requiring a more contiguous tissue representation. In an unblinded, illustrative assessment by a single expert histologist, filled outputs were favoured for tissue and background delineation, whereas artefact ratings were unchanged because the artefact channel was preserved (Supplementary Fig. 14). Filling nevertheless reduced pixel-level background Dice by deliberately reassigning selected reference-labelled background regions to tissue. We therefore retained the unmodified prediction as the default output and treated hole filling as an optional policy-support transformation.

Having evaluated held-out QC performance, agreement with reference-derived tile policies and associations with expert-assessed usability, we next tested how alternative QC policies affect downstream model behaviour.

\begin{figure*}[!tp]
\centering
\includegraphics[width=1.000\textwidth]{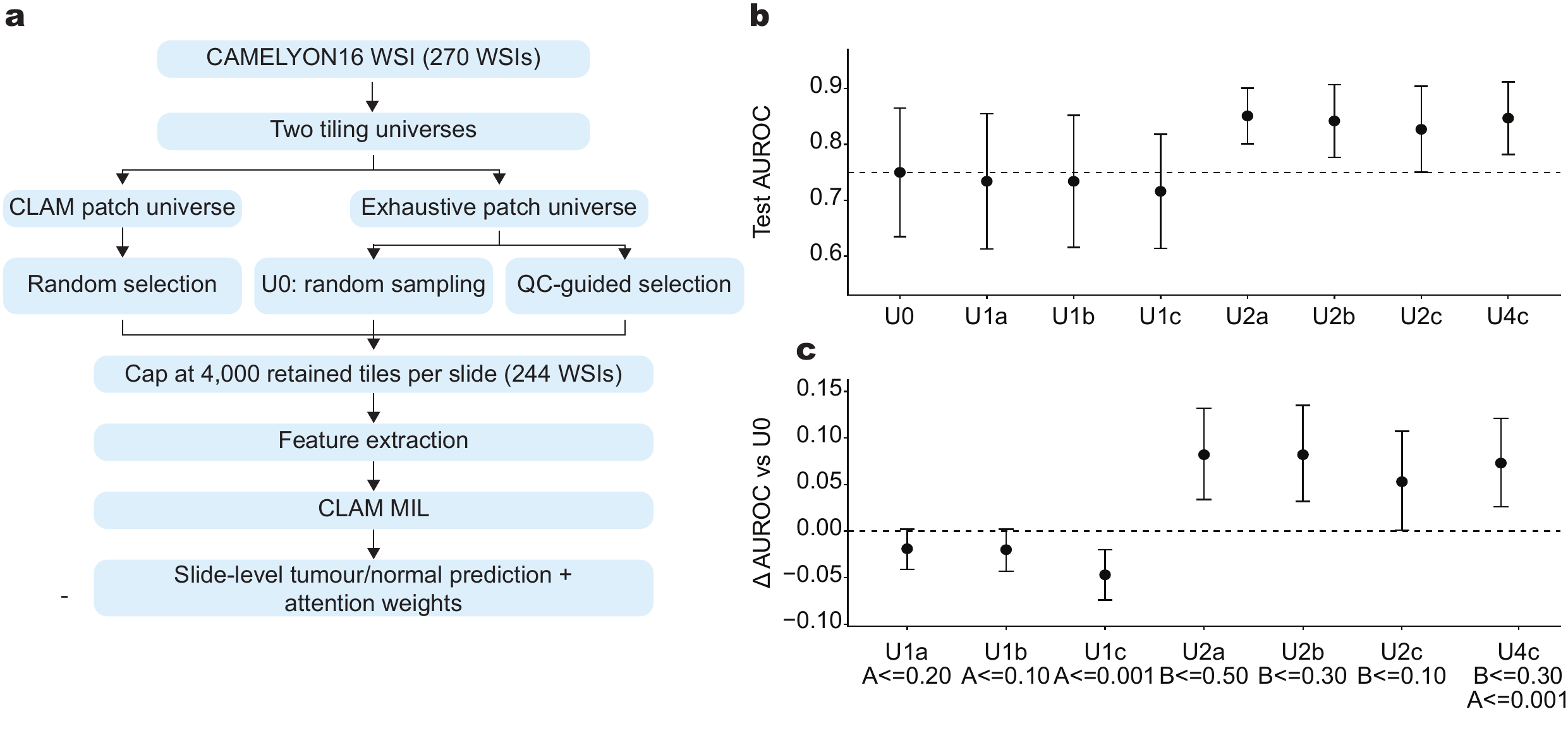}
\setcounter{figure}{4}\caption{\textbf{Tile-selection policy alters fixed-budget MIL performance in CAMELYON16.} a, Experimental design. Slides were represented using either the original CLAM candidate-patch universe (G0) or a common exhaustive candidate universe used for U0--U5. Analyses were restricted to 244 slides (134 normal, 110 tumour) that could supply the full 4,000-patch budget under all evaluated policies. U0 used reproducible random sampling from the exhaustive universe, whereas the remaining U policies applied QC-based eligibility rules before fixed-budget selection; all conditions used identical cross-validation partitions. b, Mean test AUROC ± s.d. across 15 fold--seed evaluations for U0, artefact-only policies (U1a--U1c), background-only policies (U2a--U2c) and the selected combined policy U4c. The dashed line denotes mean U0 AUROC. c, Paired out-of-fold AUROC differences relative to U0, with 95\% percentile confidence intervals from 10,000 stratified slide-level bootstrap resamples. Background-aware policies improved AUROC relative to U0, whereas artefact-only filtering did not. These channel-defined policies do not isolate physical artefact categories: scanner marks can be assigned to background and escape artefact-only exclusion. U4c was selected post hoc; intervals are pointwise and unadjusted for multiplicity.}
\label{fig:5}
\end{figure*}

\subsection{TileNet selection alters fixed-budget MIL predictions}

To test whether QC policies altered downstream multiple-instance learning, we evaluated CLAM~\cite{r3} for metastatic versus normal lymph-node classification in CAMELYON16~\cite{r22}. We distinguished two candidate-patch universes: an exhaustive tiling universe used to evaluate random and TileNet-guided selection policies (U0--U5), and the original CLAM foreground-selection procedure (G0) (Fig. 5a). To isolate the effect of tile-selection policy within the exhaustive universe, we fixed the computational input at 4,000 patches per slide. The common-support cohort comprised 244 slides (134 normal and 110 tumour) for which G0 and the most restrictive QC policy, U5c, could each supply the fixed 4,000-patch budget, thereby ensuring full-budget support for all evaluated conditions.

Within the exhaustive universe, U0 served as the matched reference, with 4,000 patches selected randomly from the candidate pool. U2a, which excluded patches with predicted background fraction \textgreater0.50, increased pooled out-of-fold AUROC from 0.758 to 0.839; the seed-averaged paired AUROC difference was +0.082 (95\% CI, 0.034--0.132), with a corresponding accuracy difference of +0.085 (0.041--0.130) (Fig. 5b,c and Supplementary Data 4, Table 15). Artefact-only exclusion did not reproduce this benefit. U1a (artefact ≤0.20) and U1b (artefact ≤0.10) showed no clear difference from U0, whereas U1c (artefact ≤0.001) reduced AUROC by 0.047 (95\% CI, −0.074 to −0.020). Combined background-and-artefact policies improved over U0 but did not exceed U2a in fold--seed mean AUROC. Among the combined policies, U4c (background ≤0.30; artefact ≤0.001) had the highest fold--seed mean AUROC (0.847 ± 0.065) and was selected post hoc for subsequent exploratory analyses. These policies act on predicted channels: scanner markings assigned to background can be removed by background-aware filtering while remaining eligible under artefact-only rules.

We next assessed tumour representation and attention localisation on all 110 tumour slides in the common-support cohort. U4c increased mean tumour-patch prevalence from 0.0113 under U0 to 0.0556. After accounting for tumour prevalence within each retained bag, U4c showed greater tumour-hit-fraction enrichment than U0 across the top 1--100 attended patches (paired differences, +0.153 to +0.172; all 95\% CIs excluded zero; Supplementary Fig. 15 and Supplementary Data 4, Table 16).

We also evaluated G0, which applied the same fixed-budget sampling principle to the original CLAM foreground-selected candidate universe. On the same 244-slide common-support cohort and with the same 4,000-patch budget, G0 achieved a mean test AUROC of 0.886 ± 0.072 and accuracy of 0.822 ± 0.072 across fold--seed evaluations. Because G0 and the U policies used different candidate universes, G0 serves as a descriptive reference rather than a matched comparator, and their performance differences cannot be attributed to QC-based selection alone. We therefore examined which regions the original CLAM foreground-selection procedure admitted and whether such regions could influence individual predictions.

Manual blinded review identified 40 scanner-mark regions outside biological tissue that had been admitted by the original CLAM foreground-selection procedure, across 27 of the 244 slides (Supplementary Fig. 16 and Supplementary Data 4, Table 17). Of these, 36 regions across 24 slides were assigned high confidence during blinded manual review and were retained for the primary QC-channel analysis. These regions were represented predominantly as non-tissue by the dense QC predictor, with a median tissue fraction of 0.0000 (IQR, 0.0000--0.0011) and a median combined background-plus-artefact fraction of approximately 1.0000. QC predictions were background-dominant in 21 regions and artefact-dominant in 15. Thus, scanner markings outside biological tissue were represented across both non-tissue channels. These marks remain physical technical artefacts irrespective of their predicted channel, and those assigned to background may escape artefact-only exclusion but be removed by background-aware policies. This distinction is consistent with the different filtering behaviour of U1 and U2, although its contribution to their performance differences was not isolated; a limited artefact-only effect therefore does not establish that technical artefacts are unimportant.

Inspection of the 100 highest-attention G0 patches further showed that highly attended instances could occur at these non-tissue locations (Supplementary Fig. 17a), whereas alternative tile-selection policies redistributed the spatial locations of highly attended patches (Supplementary Fig. 17b,c). To determine whether high-background patches could directly influence individual CLAM predictions, we mapped QC-predictor-derived background fractions onto the corresponding G0 bags and performed same-model counterfactual analyses. Patches with predicted background fraction greater than 0.50 were removed at inference, after which the unchanged out-of-fold G0 model was re-evaluated. This intervention altered the predicted class in at least one of the three independently trained models for 4 of the 27 slides containing manually identified scanner markings and in at least two models for one slide.

For tumor\_008, the scanner-mark-positive slide with the largest mean absolute counterfactual probability change (Supplementary Fig. 16a,b), patches with predicted background fraction \textgreater0.50 represented 19.5\% of the G0 bag but accounted for 94.9\% of CLAM attention and 96.4\% of the absolute tumour-versus-normal logit contribution, corresponding to approximately 4.9-fold enrichment relative to their abundance in the bag. Removing these patches reduced mean tumour probability from 0.959 to 0.153 and reversed the predicted class in all three independently trained G0 models. Because tumor\_008 is tumour-positive, this changed a correct positive prediction to a false negative in each model. The result demonstrates sensitivity to input removal without establishing a diagnostic benefit; because removal targeted all high-background patches, the effect cannot be attributed specifically to the manually delineated scanner mark.

Exploratory comparison of G0 and U4c identified 62 of 244 slides with discordant classifications in at least one seed, including 17 with disagreement across all three seeds (Supplementary Fig. 18 and Supplementary Data 4, Table 18). Among these 17 slides, G0 was correct across all three models for 14, U4c for two, and correctness was mixed for one. Removing patches with predicted background fraction \textgreater0.50 from the unchanged G0 models altered the classification in at least one seed for seven slides, at least two seeds for two slides and all three seeds for one slide. In tumor\_049, these patches represented 45.5\% of the bag but accounted for 97.5\% of attention and 97.8\% of absolute tumour-versus-normal logit contribution. Their removal reduced mean tumour probability from 0.928 to 0.084, reversing all three predictions. However, direct removal of high-background patches did not explain all reproducible G0--U4c disagreements.

Together, these analyses show that fixed-budget MIL performance depended on how predicted tissue, background and artefact fractions were translated into selection policies. Background-aware selection improved classification relative to matched random sampling within the exhaustive universe, whereas artefact-only exclusion did not reproduce this benefit. This contrast does not imply that technical artefacts are unimportant: scanner markings outside tissue were often assigned predominantly to background and could therefore be removed by background-aware policies while escaping artefact-only filtering. The channel analysis is consistent with this distinction but does not isolate the contribution of scanner marks to the cohort-level AUROC gain. Same-model interventions showed heterogeneous sensitivity to high-background patches and did not fully explain G0--U4c disagreement.

\begin{figure*}[!tp]
\centering
\includegraphics[width=1.000\textwidth]{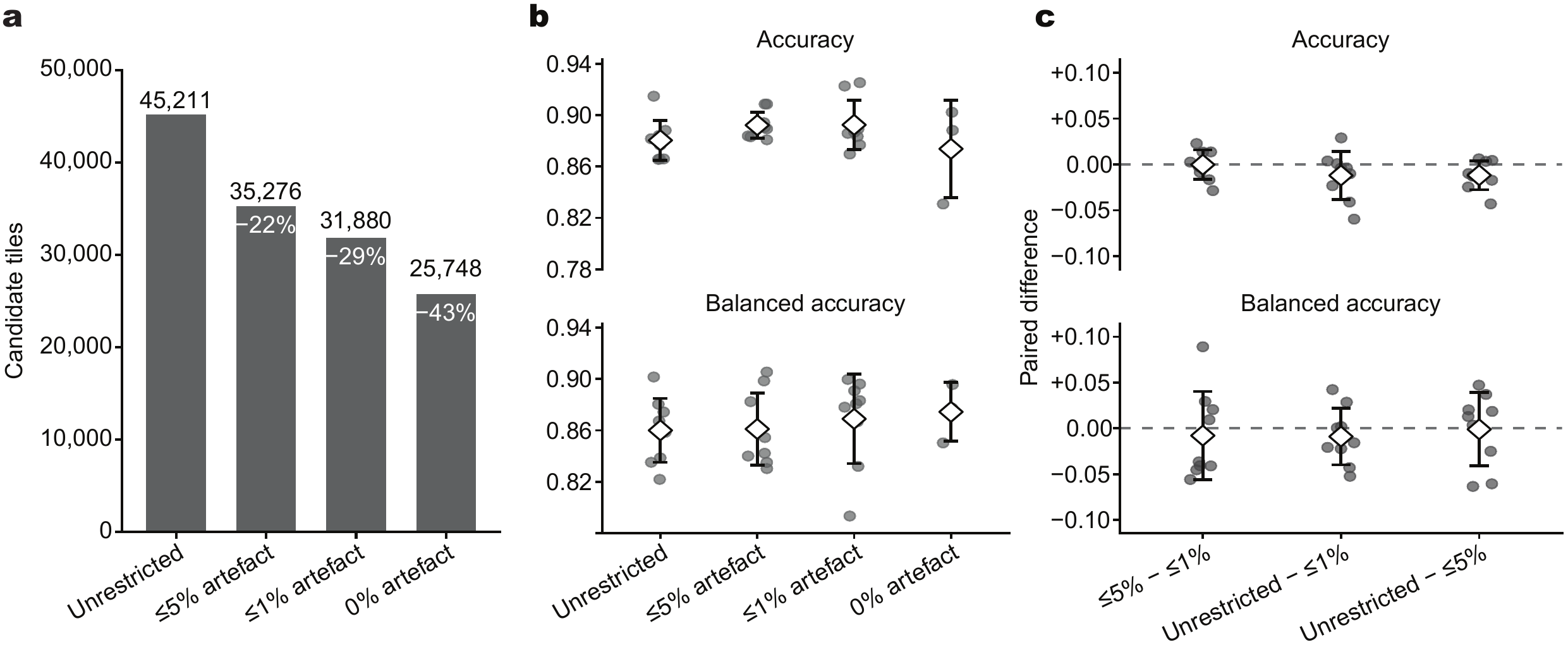}
\setcounter{figure}{5}\caption{\textbf{Artefact-aware filtering reduces candidate tile burden with modest matched performance differences.} a, TCGA-LUSC candidate tiles retained at a fixed \textless30\% predicted background threshold under unrestricted artefact inclusion, ≤5\% artefact, ≤1\% artefact and strict artefact exclusion (0\%). Counts precede training-set matching; percentages indicate reductions relative to unrestricted inclusion. b, Accuracy and balanced accuracy on the fixed artefact-excluded common test set after matching all training conditions to 18,183 tiles with identical class counts, patient identities and patient-by-class tile contributions. Points represent independently trained models; diamonds show means and error bars ±1 s.d. The ≤1\%, ≤5\% and unrestricted conditions each comprised three independently sampled matched training sets crossed with three optimisation seeds (n = 9 models), whereas strict exclusion used one deterministic matched set across three optimisation seeds (n = 3). c, Seed-matched pairwise differences in accuracy and balanced accuracy among the three artefact-permitting policies across the 3 × 3 tile-sampling-seed × optimisation-seed design. Positive values favour the first-named policy; dashed lines indicate no difference. The nine matched cells are shown descriptively and were not treated as independent experimental units.}
\label{fig:6}
\end{figure*}

\subsection{Artefact filtering shrinks the weak-label tile pool}

We next asked whether the artefact content of training tiles affected downstream tumour classification independently of training-set size and patient contribution. We evaluated a binary normal-versus-squamous-cell-carcinoma task in TCGA-LUSC, fixed the background criterion at \textless30\% for all conditions, and within that common threshold compared four artefact-retention policies: unrestricted artefact inclusion, maximum artefact fractions of 5\% and 1\%, and strict artefact exclusion (0\%) (Fig. 6a). To avoid confounding policy with training-set size or patient contribution, all conditions were matched to the strict-exclusion pool and contained 18,183 training tiles with identical class counts, patient identities and patient-by-class tile contributions. The three more permissive policies used three independently sampled matched sets crossed with three optimisation seeds; strict exclusion defined one deterministic set evaluated across three seeds.

Under strict exclusion, mean accuracy was 0.874 ± 0.038, 0.892 ± 0.019 at ≤1\% artefact, 0.892 ± 0.010 at ≤5\% and 0.880 ± 0.015 under unrestricted inclusion (Fig. 6b and Supplementary Data 4, Table 19). Class-specific metrics showed that these differing aggregate rankings reflected a redistribution of errors between recall and precision rather than a uniform change in classification performance. Under strict artefact exclusion, normal and SCC recall were nearly identical (0.876 ± 0.045 and 0.873 ± 0.061; absolute difference between the mean recalls, 0.003), whereas the artefact-permitting policies showed larger separations between mean class recalls (0.086--0.131), consistently favouring SCC recall. This greater recall symmetry under strict exclusion was accompanied by lower normal precision (0.721 ± 0.092) than under the ≤1\% and ≤5\% policies (0.792 ± 0.073 and 0.800 ± 0.049, respectively), while SCC precision was highest under strict exclusion (0.952 ± 0.014). Conversely, the ≤5\% policy combined the highest SCC recall (0.927 ± 0.028) with higher normal precision than strict exclusion. Thus, differences in accuracy, macro-F1 and balanced accuracy reflected policy-dependent precision--recall trade-offs, rather than a single artefact-retention policy consistently outperforming the others.

Seed-matched comparisons showed negligible differences in accuracy between the ≤1\% and ≤5\% artefact policies (mean paired difference, −0.0003). Unrestricted inclusion yielded approximately 0.012 lower accuracy than either moderate policy, although class-specific recall and balanced-accuracy differences were small and metric-dependent (Fig. 6c and Supplementary Data 4, Table 20). Detailed replicate-grouping sensitivity analyses and evaluation on a more permissive common test distribution are provided in Supplementary Results (Supplementary Fig. 19).

Artefact filtering nevertheless had a substantial effect on the volume of candidate data entering the pipeline (Fig. 6a). Before matching, unrestricted artefact inclusion retained 45,211 candidate tiles at the \textless30\% background threshold, compared with 35,276 for ≤5\% artefact, 31,880 for ≤1\% artefact and 25,748 following strict exclusion. Strict exclusion therefore reduced the candidate pool by approximately 43\% relative to unrestricted inclusion, and the moderate ≤1\% and ≤5\% policies by approximately 29\% and 22\%, respectively.

Together, these results show that stricter artefact filtering substantially reduced tile availability without consistently improving matched classification performance.

\begin{figure*}[!tp]
\centering
\includegraphics[width=1.000\textwidth]{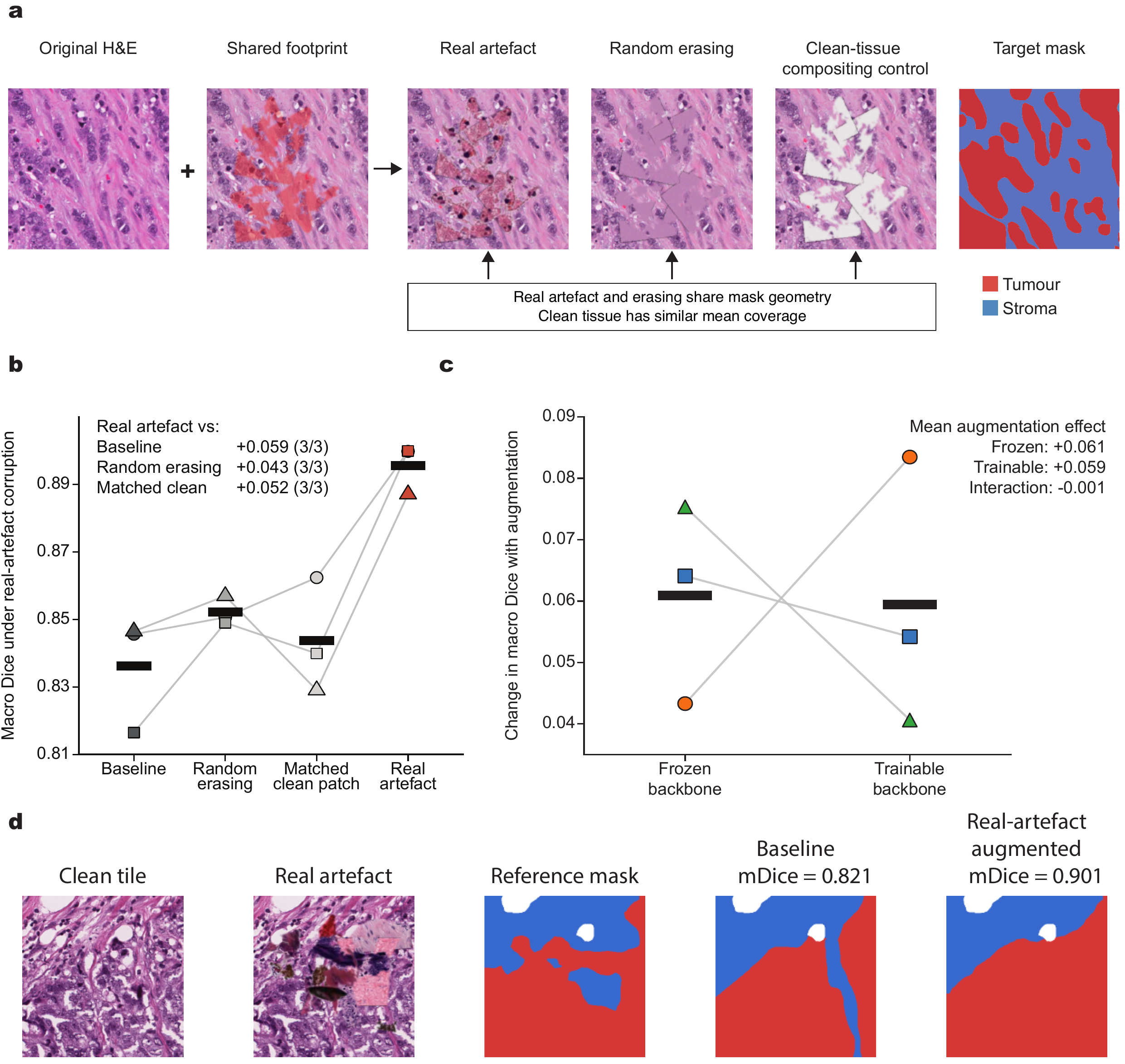}
\setcounter{figure}{6}\caption{\textbf{Real-artefact compositing improves robustness during foundation-model adaptation.} a, Experimental design comparing unaugmented training, random erasing, clean-tissue replacement and real-artefact compositing while preserving the tumour--stroma segmentation target. Random erasing matched the mask geometry, transformations and placement of the real-artefact intervention; clean-tissue replacement achieved similar mean total perturbation coverage but drew on a subset of smaller source components. b, Macro Dice under severe held-out composited real-artefact perturbation in the high-artefact stratum (288 tiles) across the four training conditions. Points represent independently trained models, lines connect seed-matched models and horizontal bars indicate means. Real-artefact compositing improved macro Dice over baseline by 0.0594 ± 0.0219 and exceeded both controls in all three seeds. c, Seed-matched effect of real-artefact compositing with frozen and trainable CONCH v1.5 backbones. The effect was positive in all three seeds under both conditions (frozen, +0.0609 ± 0.0163; trainable, +0.0594 ± 0.0219). d, Representative tumour--stroma segmentation before and after severe held-out composited real-artefact perturbation, showing predictions from the unaugmented and real-artefact-trained models.}
\label{fig:7}
\end{figure*}

\subsection{Real artefact compositing improves segmentation robustness}

Having used the three-class quality-control representation to guide tile selection, we next asked whether annotated artefacts could also be used constructively during downstream model training. We evaluated this in a tumour--stroma segmentation task using models initialised from the same CONCH v1.5 histopathology image encoder, accessed through the TITAN model~\cite{r32}. During training, masked real histological artefacts extracted from the PathQC-3C development partition were composited onto training tiles while the underlying tumour--stroma segmentation labels were left unchanged.

To assess whether the robustness gain could be reproduced by generic local perturbations, we compared real-artefact compositing with an unaugmented baseline and two controls. Random erasing used the same sampled mask geometry, transformations and placement as the real-artefact intervention, whereas clean-tissue replacement inserted structured artefact-free tissue at closely matched mean total perturbation coverage, drawing on a subset of smaller source components (Fig. 7a).

The principal reported robustness endpoint was macro Dice on the high-artefact-stratum tiles after severe held-out composited real-artefact perturbation. Real-artefact compositing increased mean macro Dice from 0.8362 ± 0.0171 in the unaugmented baseline to 0.8956 ± 0.0074, corresponding to a paired improvement of 0.0594 ± 0.0219 across three training seeds; the improvement was positive in all three seeds (Fig. 7b and Supplementary Data 4, Table 21).

Neither control reproduced this gain. Random erasing reached a mean macro Dice of 0.8522 ± 0.0042 and clean-tissue replacement 0.8438 ± 0.0170. Real-artefact compositing exceeded these conditions by 0.0434 ± 0.0116 and 0.0518 ± 0.0125, respectively, with both differences positive in all three seeds. Thus, the robustness gain was reproduced neither by geometry-matched random erasing nor by clean-tissue replacement at comparable mean total perturbation coverage.

We next asked whether this gain required adaptation of the model representation. Real-artefact compositing improved performance in all three seeds when the model backbone remained frozen, with a paired gain of 0.0609 ± 0.0163 compared with 0.0594 ± 0.0219 when the backbone was trainable (Fig. 7c). Absolute performance under composited real-artefact perturbation was lower with the frozen backbone, but the augmentation-associated gain was retained. Because only three training seeds constituted replicated experimental units, this comparison was interpreted descriptively rather than as evidence of equivalence between backbone conditions.

Macro Dice on the original, non-composited evaluation images remained similar (0.9138 without versus 0.9142 with real-artefact compositing), while mean degradation from uncorrupted images to S3 decreased from 0.0776 to 0.0186. A representative example illustrates the corresponding reduction in segmentation error under S3 composited real-artefact corruption (Fig. 7d).

To support reuse of this strategy, we subsequently implemented PathQCAug as a standalone augmentation module derived from the experimental pipeline. PathQCAug preserves the central design principle of compositing masked histological artefacts while leaving the downstream target unchanged; differences between the experimental implementation and released software are documented in Supplementary Data 4, Table 22.

\section{Discussion}
This study treats artefact-aware QC as an explicit, configurable policy layer rather than a fixed preprocessing filter. PathQC-3C provides dense tissue/background/artefact supervision, TileNet applies explicit keep, reject or ranking policies to aggregated quality estimates, and PathQCAug uses artefact information during model adaptation. The central design choice is to separate quality estimation from downstream policy: the predictor estimates image content, while task-specific rules determine its use.

The predictor experiments indicate that foreground detection alone is insufficient for this purpose. The binary control delineated tissue accurately but, by collapsing background and artefact into one non-tissue class, could not distinguish empty background from artefacts. Across candidate three-class models, artefact Dice accounted for most between-model separation, and artefact delineation remained less reliable than tissue delineation in the held-out cohorts. Gains over harmonised comparators, largest for artefact delineation, indicate closer alignment with the target tissue/background/artefact task but do not establish superiority in the comparators' native applications, given their different taxonomies, resolutions and intended uses.

Pixel-level accuracy alone does not establish operational value. Prediction-derived QC fractions reproduced most reference-derived tile-selection decisions across held-out cohorts, despite lower pixel-level agreement for artefact than for tissue and background. The relevant validation question therefore extends beyond reproduction of expert contours to whether aggregated QC outputs support stable decisions at the operating scale of the downstream workflow. Agreement nevertheless depended on threshold and footprint size, and larger threshold-crossing errors were concentrated within a limited subset of ROIs. Calibration and failure-mode auditing therefore remain necessary even when average decision agreement is high.

The blinded single-pathologist audit further separated artefact recognition from tile-level action. Within the burden-stratified sample, binary artefact presence was concordant, whereas agreement on tile usability depended on how continuous artefact burden was translated into an operational decision threshold. Prediction-derived burden remained associated with pathologist rejection, and applying the same combined rule to reference- and prediction-derived fractions produced the same overall agreement while shifting the balance between retaining usable crops and excluding rejected crops. The prediction-derived combined rule excluded 18 of 43 crops judged usable and retained five of 45 rejected crops. This discrepancy highlights the difference between mask-derived selection criteria and expert usability judgements, and the need to calibrate thresholds and tile footprints against task-specific costs of exclusion and retention. PathQC-3C should therefore be interpreted as an expert-derived representation of quality state rather than a universal definition of usability.

The downstream selection experiments show why this distinction matters. In CAMELYON16, background-aware QC policies improved classification over reproducible random sampling within a matched exhaustive candidate universe under a fixed tile budget. The selected combined policy also increased tumour representation and attention localisation relative to random sampling. However, the original CLAM foreground-selection procedure achieved a higher descriptive mean AUROC using a different candidate universe, so these results do not establish superiority over the original CLAM pipeline. Moreover, U4c was selected post hoc among the combined policies based on observed fold--seed mean AUROC, and its subsequent analyses should be interpreted as exploratory.

The scanner-mark analysis illustrates why predicted QC channels must be distinguished from physical artefact categories. Scanner marks outside biological tissue are technical artefacts, but the dense predictor assigned 21 of 36 high-confidence marked regions predominantly to background and 15 predominantly to artefact. Background-aware filtering can therefore remove technically undesirable markings that artefact-only filtering leaves eligible. The small or inconsistent artefact-only benefit in this experiment should not be interpreted as evidence that removing technical artefacts is unimportant. Removing high-background inputs from unchanged G0 models altered individual predictions, including a correct-to-incorrect reversal in tumor\_008. Because the intervention removed all patches above a predicted-background threshold rather than only the marked region, it demonstrates input sensitivity without establishing scanner-mark causality or diagnostic benefit.

The matched weak-label lung experiment provided a different stress test of the same policy principle. Under a fixed \textless30\% background criterion and identical training-data budget, increasingly stringent artefact filtering produced small, metric-dependent changes in performance: the intermediate artefact thresholds (≤1\% and ≤5\%) achieved the highest mean accuracy, with differences in macro-F1 reflecting class-specific precision--recall trade-offs, whereas strict exclusion yielded the highest balanced accuracy and nearly identical normal and SCC recalls. Strict exclusion also reduced the pre-matching candidate pool by approximately 43\%, while the preferred policy depended on the evaluation metric and on task-specific trade-offs between performance and data exposure.

PathQCAug provided a complementary use of artefact information during model adaptation rather than only before it, compositing masked real histological artefacts onto downstream training images while leaving the segmentation target unchanged. In tumour--stroma adaptation, real-artefact augmentation reduced the mean degradation from uncorrupted to severe held-out composited real-artefact perturbation from 0.0776 to 0.0186 macro Dice, while uncorrupted performance remained similar. Geometry-matched random erasing recovered only part of this benefit, suggesting a contribution from the inserted image content beyond mask geometry, and clean-tissue replacement likewise did not reproduce the gain. However, the clean-tissue control drew on a subset of library entries with smaller source masks, which limited isolation of artefact-appearance effects from component-size effects.

The separation of predictor and policy also makes TileNet largely agnostic to dense-predictor architecture and input configuration. A frozen SAM3 encoder could populate the same three-class interface, although artefact performance was lower and more variable than with Swin--UPerNet-Base, while multi-magnification input produced modest but inconsistent gains. These findings support the selected single-scale 20× predictor while leaving the policy interface open to alternative models that provide sufficiently reliable tissue, background and artefact estimates.

Several limitations constrain interpretation. PathQC-3C combines heterogeneous artefacts into one operational artefact class, which supports common policies but can obscure subtype- and severity-specific actions. Although the reference masks underwent expert consensus review, independent annotations were not collected to quantify inter-annotator agreement in dense tissue, background and artefact delineation. The blinded pathologist audit involved one pathologist and a deliberately stratified sample; the held-out ROIs were tissue-enriched; and the TCIA source-site identifiers cannot establish institutional disjointness from TCGA. TCGA source-site names were not canonicalised to physical institutions, and some patients or slides contributed multiple ROIs; ROI-level uncertainty does not account for all possible higher-level patient or slide dependence.

The downstream experiments were methodological benchmarks rather than demonstrations of clinical utility and used curated datasets that may underrepresent the range of image-quality problems encountered in routine practice. CAMELYON16 analyses were restricted to 244 common-support slides, G0 and the U policies used different candidate universes, U4c was selected post hoc and confidence intervals were unadjusted for multiple comparisons; the weak-label lung test included ten patients, and PathQCAug effects were evaluated across three training seeds. Although the training and evaluation artefact libraries were disjoint at ROI and WSI levels, PathQCAug robustness was assessed using held-out real artefacts composited through the same general procedure used during training, rather than naturally occurring artefacts alone. Threshold-based policies remain attractive because of their transparency, but their operating points require task-specific validation against retained-tile burden, false-exclusion risk and downstream performance. Prospective multi-rater and multi-site studies should therefore evaluate subtype-specific labels, calibrated burden estimates and task-specific operating points under deployment-relevant prevalence, image-quality variation and acquisition conditions. No prospective assessment of diagnostic decisions, patient outcomes, laboratory turnaround time or rescan burden was performed.

\section{Conclusion}
PathQC-3C and TileNet enable explicit, auditable quality-control policies governing image-region selection and tile-budget allocation, while PathQCAug extends artefact awareness to model adaptation. Across the evaluated tasks, quality-aware decisions affected data exposure, predictions and robustness without supporting a universal artefact-exclusion rule. Validation across independent institutions, multiple readers and naturally occurring artefacts remains necessary before clinical deployment.

\section*{Data and Code Availability}
Supplementary Data~1--4 (the blinded pathologist audit, the sources of the precomputed TileNet masks, the TCGA-LUSC source-region identifiers and cohort partitions, and numerical Tables~1--25) and the Source Data workbook (numerical source data for Figs.~2, 3 and 5--7 and the source-site mapping in worksheet Fig2\_site\_mapping) are provided as ancillary files with this preprint.

Upon journal publication, the study-generated resources will be archived in public repositories. These comprise the PathQC-3C dataset (redistributable ROIs, three-class tissue/background/artefact masks and held-out evaluation data), 19,000 precomputed TileNet quality-control masks from 43 source collections, the blinded pathologist audit materials, and the artefact library supporting the real-artefact compositing experiments. Where redistribution of source-derived image content is not permitted, provenance manifests and deterministic reconstruction procedures will be provided to regenerate the corresponding assets from the original source images. PathQC-3C and the artefact-library assets are derived in part from diagnostic slide images from The Cancer Genome Atlas obtained through the NCI Genomic Data Commons; no controlled-access genomic or clinical data are included in the resources designated for release.

The source whole-slide images remain available from their original repositories: TCGA through the Genomic Data Commons (\url{https://portal.gdc.cancer.gov/}), the CPTAC collections through The Cancer Imaging Archive, CAMELYON16 through the CAMELYON16 Grand Challenge (\url{https://camelyon16.grand-challenge.org/Data/}) and the GigaScience Database (\url{https://doi.org/10.5524/100439}), and BCSS through \url{https://github.com/PathologyDataScience/BCSS}. Original source-data access conditions and third-party data-use requirements apply.

TileNet and PathQCAug will be released as Python packages and command-line tools under the MIT License; pretrained quality-control checkpoints and redistributable artefact assets will be archived in public data or model repositories. Third-party models and dependencies, including CONCH~v1.5, remain subject to their respective licences and are not redistributed.

\section*{Use of AI Tools}
Claude (Anthropic) was used to assist with editorial revision and refinement of methodological descriptions throughout the manuscript and Supplementary Material.

\section*{Acknowledgements}
We acknowledge CSC -- IT Center for Science, Finland, for awarding this project access to the LUMI supercomputer, owned by the EuroHPC Joint Undertaking, hosted by CSC (Finland) and the LUMI consortium through the Finnish national LUMI resource allocation. The results reported in this manuscript are in part based on data generated by The Cancer Genome Atlas Research Network. Data used in this study were also generated by the National Cancer Institute Clinical Proteomic Tumor Analysis Consortium.

This work was co-funded by the European Union's Horizon Europe Research and Innovation Programme under the Marie Skłodowska-Curie Actions grant agreement No. 101126602 (Data4Healthcare) and by the University of Oulu.

Funded by the European Union. Views and opinions expressed are however those of the author(s) only and do not necessarily reflect those of the European Union or the European Research Executive Agency (REA). Neither the European Union nor the granting authority can be held responsible for them.

\section*{Author Contributions}
M.P.C.: Dataset annotation, study design, experimental planning, software development, model training, data analysis, figure generation and original manuscript drafting. Y.G.: Experimental planning, methodological input, and manuscript review and editing. P.M.P.: Blinded pathologist audit of 100 crops from the TCIA external-validation cohort, assessing artefact presence, severity, tile usability and dominant artefact type. S.P.: Annotation-protocol development, annotation review, prediction-mask review, unblinded expert assessment of hole-filling post-processing, CLAM candidate-universe review, experimental planning, manuscript review and editing, and project supervision. J.V.: Annotation-protocol development, annotation review, prediction-mask review, blinded manual review of CAMELYON16 scanner-mark regions, CLAM candidate-universe review, experimental planning, manuscript review and editing, and project supervision. All authors read and approved the final manuscript.

\section*{Competing Interests}
P.M.P. is a consultant for Cells AI. All other authors declare no competing interests.

\clearpage
\twocolumn[{\centering{\fontsize{18}{22}\selectfont Supplementary Material\par}\vspace{0.6em}{\normalsize Quality-Control Policies Shape Model Behaviour in Computational Pathology\par}\vspace{1.2em}}]
\setcounter{section}{0}\setcounter{figure}{0}
\renewcommand{\thesection}{S\arabic{section}}
\renewcommand{\theHsection}{S\arabic{section}}
\renewcommand{\thefigure}{\arabic{figure}}
\renewcommand{\theHfigure}{S\arabic{figure}}
\captionsetup[figure]{name=Supplementary Fig.,font=scriptsize}
\renewcommand{\dblfloatpagefraction}{0.4}
\setlength{\dblfloatsep}{10pt plus 2pt}
\renewcommand{\thesubsection}{\thesection.\arabic{subsection}}
\renewcommand{\theHsubsection}{\thesection.\arabic{subsection}}
\section{Supplementary Methods}
\subsection{Data units, identifiers and leakage control}

All supervised dense quality-control predictor experiments operated on paired region-of-interest (ROI) images and dense pixel-wise reference masks. Each ROI image--mask pair constituted the atomic supervised-learning unit used for sampling, training and evaluation. ROI and patient identifiers were derived deterministically from file names to support reproducible partitioning and leakage control. ROI identifiers retained the coordinate suffix corresponding to the \_r\#\_c\# component so that distinct ROIs from the same whole-slide image (WSI) remained distinguishable. Patient identifiers were generated after removing magnification suffixes such as \_10x and \_20x and retaining the first three hyphen-delimited components of the remaining TCGA identifier, corresponding to the TCGA-XX-YYYY case format.

Ten cross-validation partitions were generated using GroupKFold, with patient identity supplied as the grouping variable. All ROIs belonging to the same patient were therefore assigned exclusively to either the fitting or validation partition within a fold. Fold integrity was checked directly by confirming that no patient identifier occurred in both partitions.

The complete partition specification was stored in patient\_folds.json and reused across architecture comparisons and ablation experiments. A stored partition file was accepted only when both the requested number of folds and the ordered list of ROI identifiers matched the current ROI universe exactly. A mismatch in either field was treated as an error requiring regeneration of the partition file. Architecture benchmarking and model-selection experiments were executed on folds 0--2. This three-fold subset was fixed before model training and was used unchanged for every reported development configuration. Because only three development folds were executed, comparisons among predictor configurations were treated descriptively rather than as population-level estimates of effect.

\subsection{Region-of-interest extraction and tissue-constrained sampling}

Candidate ROIs were sampled from a non-overlapping spatial grid defined in level-0 WSI coordinates. Candidate order was generated using a pseudo-random number generator seeded deterministically from the slide identifier. Slide-reported level-0 microns-per-pixel metadata were used to calculate the native level-0 dimensions corresponding to a 6,000 × 6,000-pixel field of view at 0.5 µm per pixel.

Tissue-constrained sampling used a low-resolution tissue mask derived from the WSI thumbnail. The thumbnail was transformed to HSV colour space, and fixed saturation and value thresholds were applied to obtain a binary tissue estimate. For each candidate ROI, its level-0 coordinates were mapped to thumbnail coordinates using the relevant downsampling factor, and estimated tissue fraction was calculated as the mean of the binary thumbnail mask within the mapped window. Up to ten candidate locations were evaluated per ROI-selection round. For each ROI-selection round, the first candidate with an estimated tissue fraction ≥0.90 was retained; when none met this criterion, the candidate with the highest estimated tissue fraction was selected. Approximately 98\% of the final ROIs met the ≥90\% thumbnail-derived tissue-content criterion. To improve representation of tissue categories and source sites with limited WSI availability, additional spatially distinct ROIs were retained from selected TCGA slides. This increased annotation coverage without increasing the number of independent source slides. Overall, 229 of the 1,144 TCGA WSIs contributed multiple ROIs, with between one and 26 ROIs retained per slide. In contrast, the external TCIA cohort contributed one ROI per WSI.

Selected regions were read from level 0 using OpenSlide. If a requested region intersected the WSI boundary and was smaller than the intended native dimensions, it was padded with white background before resizing. RGB regions were resized to 6,000 × 6,000 pixels using Lanczos interpolation. For paired 10× experiments, the corresponding 20× ROI was downsampled to 3,000 × 3,000 pixels. RGB images were downsampled using Lanczos interpolation, whereas indexed reference masks were resampled using nearest-neighbour interpolation. Level-0 offsets, native extraction dimensions, microns-per-pixel values and scale factors were retained to permit deterministic re-extraction and coordinate mapping.

\subsection{Label encoding and supervision integrity}

Reference masks encoded three mutually exclusive semantic classes: tissue, background and artefact. Tissue was assigned class 0, background class 1 and artefact class 2. Masks stored as RGB images were converted to indexed class maps using a fixed colour-to-class mapping. Pixels that did not map to a recognised class were assigned the ignore label 255.

Ignore pixels were retained throughout preprocessing and excluded from all loss and metric calculations. Cross-entropy losses used ignore\_index = 255, and overlap-based losses and metrics were calculated only over valid labelled pixels. Label maps were always resampled using nearest-neighbour interpolation to avoid introducing intermediate class values.

The artefact class was used as an operational category encompassing technical alterations that obscured, distorted or degraded the biological image. It was not intended to distinguish individual preparation, staining, mounting or scanning artefact subtypes. The same class definitions and ignore handling were applied across development, held-out inference, the SAM3 backbone-swap experiment and tile-level policy analyses.

\subsection{Training-plan construction and crop generation}

Each training epoch used a fixed crop-sampling budget, target\_per\_epoch, stored in the run configuration. For each fold, the training plan comprised exactly target\_per\_epoch pairs of (roi\_id, crop\_idx). ROI identifiers were sampled with replacement from the fitting partition using a fold-dependent random seed. Sampling with replacement permitted the crop budget to vary without changing the underlying patient partition or ROI universe. The crop\_idx value provided a stable sample identifier for deterministic bookkeeping; spatial crop coordinates were generated when the sample was accessed.

Validation plans were constructed independently by enumerating a fixed number of samples per validation ROI. The experiments reported here used one fixed deterministic crop per validation ROI. Validation crop coordinates were generated from the (roi\_id, crop\_idx) pair and were invariant across repeated runs and predictor configurations. Validation did not use stochastic crop resampling or image augmentation.

For single-magnification training, each ROI image and aligned mask were loaded and enlarged when necessary to support the requested source-crop dimensions. Training coordinates were sampled uniformly from all valid crop origins. When a minimum-foreground constraint was enabled, the fraction of tissue pixels in each candidate reference-mask crop was calculated before resizing. Candidate coordinates were resampled up to a fixed maximum number of attempts until the requested tissue fraction was attained. If the criterion could not be satisfied, the final sampled crop was accepted to ensure that data loading progressed. Minimum-foreground resampling was applied during training only.

For multi-magnification experiments, paired 10×- and 20×-equivalent images represented the same physical field of view. The 20× image served as the reference canvas, and the 10× image was represented at half the linear resolution. Crop coordinates were sampled in the 10× coordinate system and mapped to the 20× image using the fixed scale factor. The corresponding mask crop was taken from the aligned 20× annotation. The 10× crop, 20× crop and mask were then transformed to a common 512 × 512-pixel grid for model input and supervision. Any minimum-foreground criterion was evaluated on the mask before resizing.

\subsection{Input processing, augmentation and model fusion}

Input normalisation and preprocessing followed the requirements of each architecture family. Convolutional models used ImageNet mean and standard deviation normalisation. Transformer-based models used the corresponding image processor loaded from the locally cached model snapshot.

Image augmentation was applied only during training. Colour jitter altered brightness, contrast, saturation and hue. Gaussian blur was applied to each sampled training crop with probability 0.90, using a randomly selected radius of 1.5 or 2.5 source pixels before resizing to the network input. For paired 10×/20× multi-magnification inputs, the same blur decision and radius were applied to both aligned views. Although synthetic Gaussian blur has previously been used to generate out-of-focus artefacts for model training, including in GrandQC~\cite{r13}, here it was used to introduce variation in image sharpness without modifying the reference masks. Its effect on detecting focus-related artefacts was not evaluated separately. All augmentations were disabled during validation and held-out inference.

Early multi-magnification fusion concatenated aligned 10× and 20× RGB crops to form a six-channel tensor processed by one Swin--UPerNet model with an adapted input projection. Late fusion processed the aligned 10× and 20× crops using separate Swin encoder streams and combined the resulting representations within a custom fusion decoder. Because the early- and late-fusion implementations differed in encoder count, parameterisation and decoder structure, their comparison did not isolate fusion stage independently of architecture.

\subsection{Optimisation, class weighting and checkpoint selection}

Training was implemented using the Hugging Face Trainer. Optimisation used Adam with cosine learning-rate decay and bfloat16 mixed precision where supported. Unless specified otherwise in Supplementary Data 4, Tables 23 and 24, the initial learning rate was 2.0 × 10⁻⁵, the per-device batch size was 2 and gradient accumulation was performed over six steps. Training continued for a maximum of 64 epochs with early-stopping patience of five validation evaluations. Within each fold, the checkpoint attaining the highest validation macro Dice was retained.

When cross-entropy contributed to the objective, class weighting was controlled by the run configuration. Unweighted runs used equal class contributions. Manually weighted runs used weights specified in the run configuration. In the automatic mode, weights were estimated separately for each fold using only masks from that fold's fitting partition. A fixed number of crops was sampled from each fitting mask; class-specific pixel counts were accumulated over valid pixels, transformed to inverse-frequency weights, normalised to unit mean, subjected to a stabilising power transformation and clipped to a bounded range. Pixel counts, class proportions and final weights were cached and written to the fold-level training log.

During development validation, logits were resized to the reference-mask dimensions before argmax classification. Class-specific Dice was calculated after pooling valid predictions and reference labels across the deterministic validation crops within a fold, and validation macro Dice was the unweighted mean of the three class-specific Dice values. This pooled-crop development metric differs from the ROI-level metric used for held-out evaluation.

\subsection{Fold-level learning-trajectory analysis}

Epoch-level validation metrics were read from the persistent fold logs for each architecture and configuration. Mean trajectories and standard deviations were calculated across the three executed development folds only for epochs at which all three folds contributed observations. Because early stopping terminated folds at different epochs, later epochs with incomplete fold coverage were excluded from the across-fold trajectory summaries.

Fold-specific curves were additionally displayed without averaging to show variation in optimisation dynamics and stopping time. Each fold's reported selected-checkpoint metric was taken from the epoch attaining the highest validation macro Dice in that fold. Consequently, values on the epoch-wise mean trajectories were not expected to equal the mean of the independently selected fold-specific checkpoints.

\subsection{Exploratory SAM3 backbone-swap experiment}

The SAM3 backbone-swap experiment used a frozen SAM3 vision encoder as a supervised dense feature extractor rather than as a prompt-driven segmentation model. The encoder was initialised from the facebook/sam3 checkpoint hosted on Hugging Face and loaded from a locally cached copy using the official SAM3 implementation. No point, box, mask, text or concept prompts were supplied.

The supervised decoder comprised a 1 × 1 convolution projecting the 1,024-channel SAM3 representation to 256 channels, followed by batch normalisation and ReLU activation; a 3 × 3 convolution with 256 output channels, batch normalisation and ReLU activation; and a final 1 × 1 convolution producing tissue, background and artefact logits. Encoder tokens were reshaped into a spatial feature map before decoding, and output logits were bilinearly upsampled to the supervision grid.

Training used the same three executed patient-level development folds and the same three-class label definitions as the baseline dense-predictor experiments. Source crops measured 512 × 512 pixels. Because the SAM3 encoder required 1,008 × 1,008-pixel inputs, each crop was resized internally to 1,008 × 1,008 pixels before encoder inference, and decoder logits were subsequently upsampled to the original 512 × 512-pixel supervision grid.

The fitting partition used 15,000 crops sampled with replacement per epoch, and validation used one deterministic crop per ROI. Training augmentation comprised colour jitter and Gaussian blur, whereas validation was unaugmented. Optimisation used Adam with an initial learning rate of 1 × 10⁻⁴, weight decay of 0, per-device batch size 2, gradient accumulation over six steps, cosine learning-rate decay, no warm-up, gradient clipping at 1.0 and bfloat16 mixed precision. The loss combined soft Dice and cross-entropy with equal weights and without cross-entropy class weighting. Training continued for up to 64 epochs with early-stopping patience of five epochs. The checkpoint with the highest validation macro Dice was selected separately in each fold. The SAM3 experiment assessed compatibility with the common dense three-class interface and was not used to select or validate downstream policies.

\subsection{Runtime and memory profiling}

Computational performance was evaluated using two complementary procedures.

For end-to-end validation profiling, runtime and throughput were extracted from the Trainer evaluation logs. Data-loader settings were held constant, including eight workers, persistent workers and pinned memory. Each validation sample represented one deterministic crop; for multi-magnification models, one sample comprised the paired 10× and 20× inputs processed jointly. Validation throughput was taken from the logged eval\_samples\_per\_second field. To reduce one-time initialisation effects, the first validation evaluation was excluded and the remaining evaluations were averaged within fold. Fold-level throughput estimates were then summarised across the three development folds. These measurements included data loading, source-image processing, host-to-device transfer, neural-network inference and evaluation overhead.

Architecture-level neural-network cost was evaluated separately using a standardised forward-only benchmark. Each architecture was reinstantiated and loaded with its stored fold-specific checkpoint. Measurements were performed on an AMD Instinct MI250X GPU using batch size 2, bfloat16 autocast and synthetic GPU-resident tensors matching the model's required 512 × 512-pixel input geometry. Early fusion received a six-channel tensor, whereas late fusion received separate three-channel 10× and 20× tensors.

Each benchmark comprised 20 untimed warm-up iterations followed by 100 timed forward passes. Warm-up permitted ROCm kernel initialisation and MIOpen solution selection to complete before measurement. The device was synchronised immediately before and after each timed pass, and latency was measured using time.perf\_counter(). Image loading, source-image processing, host-to-device transfer, loss calculation and metric computation were excluded.

PyTorch peak allocated and reserved memory were recorded after resetting the corresponding peak-memory statistics. Peak allocated memory was the primary reported memory endpoint. It included parameters, model inputs and forward allocations tracked by the PyTorch caching allocator, but not total device memory or allocations outside the PyTorch endpoint. Parameter counts included all instantiated encoder and decoder parameters.

The forward benchmark was repeated for the checkpoints from each of the three development folds, and latency and memory were summarised as mean ± s.d. across folds. Both the 512- and 1,024-pixel source-crop configurations were represented by 512 × 512-pixel network tensors in this benchmark; the benchmark therefore compared architecture-level cost rather than native source-image extraction or resampling cost. Profiling used PyTorch 2.8.0 with ROCm 7.0.2, Transformers 4.45.2 and MIOpen 3.5.0.70002.

\subsection{Hole-filling post-processing and expert assessment}

Optional post-processing was applied to class maps generated by the default single-scale predictor. The operation could reassign selected background pixels to tissue but never altered pixels predicted as artefact.

Foreground was defined as the union of predicted tissue and artefact. Morphological closing with a disk structuring element of radius 1 pixel was applied to bridge single-pixel discontinuities, after which internal holes in the foreground mask were filled. Candidate pixels for reassignment were restricted to pixels originally predicted as background. Background components that became foreground after filling, were not connected to the image border and occupied ≤500 pixels were reassigned to tissue. Remaining isolated background components occupying \textless400 pixels were also reassigned to tissue. Artefact pixels were preserved unchanged.

Post-processing was performed at 0.5 µm per pixel. The 500- and 400-pixel thresholds therefore corresponded to approximately 125 µm² and 100 µm², respectively. These values were configurable policy-support parameters rather than learned properties of the dense predictor.

One expert histologist assessed paired filled and unfilled predictions for 20 held-out 6,000 × 6,000-pixel ROIs: ten from TCIA and ten from TCGA-unique-source. Each pair was reviewed with the corresponding H\&E image and dense reference mask. Tissue, background and artefact delineation were scored as Poor, Fair, Good or Excellent. A dominant error mode and free-text observations were recorded where applicable, together with an overall acceptability judgement. This was an illustrative single-rater assessment rather than a formal observer study, and no inter-rater analysis was performed. Filled and unfilled outputs were viewed together, and the histologist was unblinded to the displayed condition.

\subsection{CAMELYON16 feature extraction and CLAM training}

Patch embeddings were extracted using the default resnet50\_trunc encoder in the CLAM feature-extraction pipeline. This path instantiated the resnet50.tv\_in1k model from timm with ImageNet-1K pretrained weights and features\_only=True, using out\_indices=(3,). The selected feature map was reduced using adaptive average pooling, yielding one 1,024-dimensional vector per patch. Source patches were transformed to 224 × 224 pixels and normalised using ImageNet mean and standard deviation. The same encoder, pretrained weights and preprocessing were used for all G0 and U bags.~\cite{r34}

Slide-level classification used single-branch CLAM with the small model configuration and 1,024-dimensional inputs. Training used Adam, cross-entropy bag loss, cross-entropy instance loss, a bag-loss weight of 0.7, eight sampled instances for the instance-level objective, dropout 0.25, learning rate 2 × 10⁻⁴, L2 regularisation of 1 × 10⁻⁵ and weighted sampling. All available fitting slides were used within each fold. Training continued for up to 200 epochs with the CLAM early-stopping procedure. All hyperparameters were held constant across arms, folds and seeds.

For every arm, fold and seed, predictions were obtained only for the corresponding outer test partition. Slide-level AUROC was the primary classification endpoint and accuracy was secondary. Fold--seed summaries were reported as mean ± s.d. across the 15 fold--seed combinations but were not treated as 15 independent biological observations.

For primary pairwise inference, held-out predictions from the five test folds were concatenated separately within each training seed, yielding one out-of-fold prediction per slide and arm for all 244 slides. Each U policy was compared with U0 using 10,000 paired class-stratified slide-level bootstrap resamples. Normal and tumour slides were resampled separately with replacement while preserving the original class counts. Identical resampled slide indices were applied to both arms and all three seeds. AUROC and accuracy were calculated within each seed, the arm-minus-U0 difference was calculated within seed, and the three seed-specific differences were averaged within each bootstrap replicate. Point estimates were calculated analogously using the complete cohort. Percentile 95\% confidence intervals were obtained from the distribution of the averaged paired differences.

\subsection{TCGA-LUSC replicate sensitivity analyses}

Comparisons among the ≤1\%, ≤5\% and unrestricted policies paired models sharing the same tile-sampling and optimisation seeds. Each contrast therefore comprised nine exact cells from the fully crossed 3 × 3 design. These cells were not treated as nine independent experimental units, and the contrasts were summarised descriptively.

The exact paired differences were marginalised in two complementary ways. First, values were averaged over optimisation seeds within each tile-sampling seed, yielding three sampling-seed-level contrasts. Second, they were averaged over tile-sampling seeds within each optimisation seed, yielding three optimisation-seed-level contrasts. Because the design was balanced and fully crossed, either marginalisation preserved the overall mean paired difference. Marginalisation was used to examine directional consistency and sensitivity to replicate grouping, not to generate an alternative effect estimate or narrower inferential interval.

Accuracy variability within each artefact-permitting policy was decomposed descriptively using sums of squares from the balanced 3 × 3 design. Tile-sampling seed and optimisation seed each contributed 2 degrees of freedom. The residual contributed 4 degrees of freedom and included their interaction and other cell-level variation that could not be separated because only one model was observed per sampling-seed × optimisation-seed cell. Observed fractions were compared descriptively with the 25\%/25\%/50\% partition implied by the relative degrees of freedom under exchangeable cell-level variation. No formal variance-component inference was performed.

The same 30 models were additionally evaluated on a secondary common test set permitting ≤90\% predicted background and unrestricted predicted artefact. This set comprised 11,167 tiles (4,568 normal and 6,599 SCC) from the same ten test patients. The same performance metrics and seed-matched descriptive comparisons were calculated as for the primary test set.

No equivalence margin was prespecified. Similar results among policies were therefore interpreted as absence of a consistent performance advantage within the evaluated design rather than as evidence of equivalence.

\subsection{Experimental artefact-library construction}

The training artefact library was constructed from paired RGB tiles and artefact masks from the PathQC-3C development partition. Non-zero mask pixels were treated as artefact. Connected artefact components were eligible when they occupied 1--60\% of the source tile and had bounding-box width and height of at least 8 pixels. Each eligible component was stored as an RGB crop and corresponding alpha mask.

Source files were processed in a deterministic sorted order, and the first 5,000 eligible components were retained. Random seed 0 was used for matched clean-crop sampling. For each artefact component, candidate artefact-free tissue crops with the same dimensions were sampled from the corresponding source tile. Crops containing reference artefact pixels were rejected. Up to 64 candidate locations were evaluated, and the crop with tissue fraction closest to that of the artefact-containing region was retained, with early acceptance when the absolute tissue-fraction difference was ≤0.15. Matched clean crops were available for 1,148 of the 5,000 artefact entries; 3,852 entries lacked a same-source clean counterpart.

The clean-crop search required an artefact-free candidate region of the same dimensions within the corresponding source tile, a constraint expected to reduce the availability of counterparts for larger components. Consistent with this, the 1,148 clean-eligible entries had smaller non-zero alpha-mask areas than the full library (median, 2,925 pixels {[}IQR, 1,891--5,963{]} versus 6,848 pixels {[}IQR, 2,587--29,345{]}; 90th percentile, 10,887 versus 64,491 pixels). By contrast, median mask occupancy within the enclosing crop was similar (0.481 versus 0.474).

A separate evaluation artefact library was constructed from the TCGA-common-source held-out partition. Training and evaluation libraries were disjoint at both ROI and WSI levels.

\subsection{PathQCAug software implementation}

The real-artefact compositing strategy was subsequently implemented as PathQCAug, a reusable Python module for artefact-aware augmentation of downstream histopathology tasks. PathQCAug accepts an RGB image, an optional downstream target and an artefact-library specification and returns the augmented image, unchanged downstream target and augmentation metadata. User-configurable parameters include augmentation probability, target-coverage range, placement attempts, scale and rotation ranges, horizontal flipping, alpha-mask smoothing and random seed. The same interface supports real-artefact insertion and corresponding control modes.

PathQCAug implements the general augmentation strategy evaluated in the tumour--stroma experiments but is not intended as a step-for-step reimplementation of every operation in the experimental training pipeline. Differences between the manuscript implementation and the public software release, including deterministic sampling and control-generation behaviour, are detailed in Supplementary Data 4, Table 22.

\section{Supplementary Results}
\subsection{Detailed optimisation of the dense quality-control predictor}

Source-crop size showed the most consistent association with artefact delineation across the evaluated optimisation grid. In all six matched combinations of loss formulation, cross-entropy weighting and minimum-foreground sampling, the 1,024-pixel source-crop condition achieved higher mean artefact Dice than both the corresponding 512- and 256-pixel conditions (Supplementary Data 4, Table 4). The relationship was not monotonic across the three source-crop sizes, however, because the 256-pixel condition exceeded the 512-pixel condition in four of the six matched comparisons. Source-crop size therefore altered performance through the sampled spatial context and resampling regime rather than through a simple monotonic increase in input size; all source crops were processed to the same 512 × 512-pixel network input.

Under the unweighted Dice--cross-entropy objective with no minimum-foreground constraint, artefact Dice was 0.731 ± 0.096, 0.733 ± 0.076 and 0.756 ± 0.034 for 256-, 512- and 1,024-pixel source crops, respectively. The corresponding macro-Dice values were 0.884 ± 0.030, 0.883 ± 0.021 and 0.887 ± 0.009, while background Dice decreased from 0.952 ± 0.014 to 0.947 ± 0.015 and 0.938 ± 0.015. The highest mean macro Dice across the 18 configurations in Supplementary Data 4, Table 4 was obtained using the 256-pixel weighted Dice--cross-entropy condition (0.892 ± 0.033), rather than the ultimately selected configuration. The selected 1,024-pixel unweighted condition nevertheless achieved the highest mean artefact Dice in the grid and the lowest fold-to-fold variability in macro IoU, macro Dice and artefact Dice, with standard deviations of 0.010, 0.009 and 0.034, respectively. Selection was therefore based on artefact performance and cross-fold stability rather than the maximum mean aggregate score.

Minimum-foreground sampling and cross-entropy weighting provided no consistent advantage. At the 512-pixel source-crop setting, increasing the minimum foreground fraction from 0.0 to 0.1 reduced artefact Dice from 0.733 ± 0.076 to 0.673 ± 0.140 under unweighted Dice--cross-entropy and from 0.722 ± 0.046 to 0.659 ± 0.154 under the weighted formulation. The corresponding differences at 1,024 pixels were smaller. Cross-entropy weighting improved artefact Dice for some 256-pixel conditions but reduced it at 512 and 1,024 pixels, indicating an interaction with sampled image context rather than a generally favourable effect.

Once backbone, source-crop size, objective and sampling criteria were fixed, batch size and per-epoch crop budget produced only modest differences. Batch sizes of 2, 4 and 6 yielded macro Dice of 0.887 ± 0.009, 0.885 ± 0.010 and 0.886 ± 0.015 and artefact Dice of 0.756 ± 0.034, 0.753 ± 0.041 and 0.756 ± 0.053, respectively (Supplementary Data 4, Table 5). Increasing the training budget from 15,000 to 30,000 sampled crops per epoch changed macro Dice from 0.887 ± 0.009 to 0.886 ± 0.009 and artefact Dice from 0.756 ± 0.034 to 0.754 ± 0.032 (Supplementary Data 4, Table 6).

Across learning rates of 1.2 × 10⁻⁵, 2.0 × 10⁻⁵ and 3.0 × 10⁻⁵, macro Dice was 0.884 ± 0.012, 0.887 ± 0.009 and 0.891 ± 0.018, respectively, and artefact Dice was 0.753 ± 0.046, 0.756 ± 0.034 and 0.773 ± 0.068. Although 3.0 × 10⁻⁵ produced the highest mean macro and artefact Dice, the standard deviation of artefact Dice was twice that at 2.0 × 10⁻⁵. The reference learning rate was therefore retained at 2.0 × 10⁻⁵ in favour of greater cross-fold stability. All development-stage comparisons used the same three patient-level folds and were interpreted descriptively rather than as formal estimates of effect size.

\subsection{Fold-level learning dynamics and multi-magnification variability}

Epoch-wise validation trajectories supported the selected-checkpoint comparisons but also demonstrated substantial variation in optimisation dynamics among development folds. Across the baseline architectures, macro-Dice trajectories separated the lower-performing EfficientNet-B0 U-Net++ configurations from the transformer-based models, whereas artefact-Dice trajectories showed greater between-model separation and variability. Tissue and background Dice remained comparatively high throughout training, while the artefact class accounted for most of the between-architecture separation (Supplementary Figs. 2 and 3). Individual fold trajectories terminated at fold-specific early-stopping epochs (Supplementary Fig. 4), and each selected-checkpoint value was taken from the epoch attaining the highest validation macro Dice in that fold. Consequently, epoch-wise means in the trajectory figures do not correspond directly to the aggregate selected-checkpoint values in Supplementary Data 4, Table 2.

Multi-magnification fusion increased mean artefact Dice but also increased fold-to-fold dispersion. With 512-pixel source crops, fold-specific artefact Dice was 0.693, 0.821 and 0.686 for single-scale input; 0.849, 0.817 and 0.613 for early fusion; and 0.910, 0.840 and 0.605 for late fusion. These values corresponded to ranges of 0.135, 0.236 and 0.305, respectively. With 1,024-pixel source crops, the corresponding values were 0.738, 0.795 and 0.735 for single-scale input; 0.866, 0.820 and 0.643 for early fusion; and 0.915, 0.815 and 0.654 for late fusion, giving ranges of 0.060, 0.223 and 0.261. Both fusion configurations exceeded single-scale input in folds 0 and 1 but were lower in fold 2 at the 1,024-pixel setting. For late fusion, which achieved the highest mean artefact Dice, the fold-specific differences from single-scale input were +0.177, +0.020 and −0.081. These patterns were treated as descriptive evidence of instability rather than as a characterised fold-specific failure mode (Supplementary Data 4, Table 8).

Early and late fusion also differed materially in architecture. The single-scale and early-fusion configurations contained 121.30 and 121.35 million parameters, respectively, whereas late fusion used separate 10× and 20× encoder streams and contained 179.47 million parameters. Under the standardised forward-only benchmark, mean latency was 47.79 ± 0.17 ms per batch for single-scale input, 48.32 ± 0.12 ms for early fusion and 65.12 ± 0.17 ms for late fusion. Peak PyTorch allocated memory during the same procedure was 1.850, 1.858 and 1.707 GB, respectively. These values quantify allocations tracked during the standardised forward procedure and do not include source-image extraction or paired-input preprocessing (Supplementary Fig. 5).

Larger source context produced the same directional change across the fusion comparison. Relative to the corresponding 512-pixel condition, 1,024-pixel source crops increased mean artefact Dice by 0.023 for single-scale input, 0.017 for early fusion and 0.010 for late fusion; seven of the nine architecture-by-fold contrasts favoured the larger crop. Because all conditions were processed to 512 × 512-pixel network inputs, these changes reflect sampled image context rather than an increase in network-input dimensions. The modest mean gains, greater fold-level dispersion and absence of an external TCIA improvement supported selection of the more stable single-scale configuration.

\subsection{Threshold and footprint dependence of tile-level policy fidelity}

Reference--prediction concordance depended on both the operating rule and the scale at which dense outputs were aggregated. For artefact-only rules evaluated over 512 × 512-pixel footprints, balanced accuracy was lowest under an exact-zero artefact criterion and increased at the 1\% and 5\% thresholds (Supplementary Fig. 10a). Adding the tissue criterion to the 1\% artefact rule increased balanced accuracy from 0.867 to 0.938 in TCGA-common-source, from 0.780 to 0.914 in TCGA-unique-source and from 0.827 to 0.868 in TCIA. Concordant rejection through the tissue gate masked 22.2\%, 16.5\% and 6.7\% of artefact-gate switches in the three cohorts, respectively. Thus, the combined rule partly buffered errors at the artefact threshold because the tissue criterion could independently determine rejection.

Increasing the aggregation footprint from 512 × 512 to 1,024 × 1,024 pixels improved continuous artefact-burden rank agreement but changed the distribution of footprints near the low operating thresholds. The fraction of exactly artefact-free reference footprints decreased from 0.727 to 0.558 in TCGA-common-source, from 0.685 to 0.504 in TCGA-unique-source and from 0.702 to 0.490 in TCIA. Conversely, the fraction with low-positive burden, defined as 0 \textless{} artefact fraction ≤0.01, increased from 0.134 to 0.258, from 0.180 to 0.313 and from 0.109 to 0.262, respectively. Improved continuous agreement at the larger footprint therefore did not translate into higher agreement for a fixed numerical rule, demonstrating that threshold meaning depends on aggregation scale (Supplementary Fig. 10b,c).

Residual disagreement under the combined rule was driven predominantly by the artefact gate. At 512 pixels, artefact-gate-only switches accounted for 4.9\%, 6.5\% and 6.0\% of footprints in TCGA-common-source, TCGA-unique-source and TCIA, whereas tissue-gate-only switches accounted for 0.7\%, 0.6\% and 2.5\%, respectively. A subset of artefact-gate switches reflected substantial underestimation rather than small perturbations around the 1\% boundary: among artefact-gate switches, reference artefact burden \textgreater5\% crossed below the predicted 1\% threshold in 12.8\%, 15.1\% and 18.5\% of cases across the three cohorts (Supplementary Fig. 11a).

These larger threshold-crossing errors were concentrated in relatively few ROIs. The five highest-contributing ROIs accounted for 54.5\% of the events in TCGA-common-source, 63.5\% in TCGA-unique-source and 81.0\% in TCIA. Residual policy disagreement was therefore neither evenly distributed among the component gates nor uniformly distributed across ROIs; it arose mainly from artefact-fraction disagreement concentrated in a limited number of difficult fields of view (Supplementary Fig. 11b).

\subsection{Slide-level heterogeneity in CAMELYON16 counterfactual analyses}

The scanner-mark audit first showed how these manually identified non-biological regions were represented within the three-class quality-control space. Among the 36 high-confidence scanner-mark regions across 24 slides used for the primary QC-channel analysis, the median predicted tissue fraction was 0.0000 (IQR, 0.0000--0.0011), whereas the median predicted background and artefact fractions were 0.5598 (0.1115--0.9387) and 0.4402 (0.0276--0.8885), respectively. The non-tissue prediction mass was distributed across both QC channels: 21 regions (58.3\%) were background-dominant and 15 (41.7\%) artefact-dominant. Scanner-mark-containing regions were therefore represented almost entirely as non-tissue by the dense predictor but were not confined to the predicted artefact channel (Supplementary Fig. 16 and Supplementary Data 4, Table 17). Scanner marks are physical technical artefacts regardless of their predicted channel. Background-aware filtering can remove marks assigned to background, whereas artefact-only filtering can leave them eligible. Thus, the limited artefact-only benefit does not demonstrate that technical artefacts are unimportant.

Same-model counterfactual effects nevertheless varied markedly among scanner-mark-positive slides. Removing G0 patches with predicted background fraction \textgreater0.50 changed the hard classification in at least one of the three independently trained models for four of the 27 scanner-mark-positive slides and in at least two models for one slide. The largest effect occurred in tumor\_008. High-background patches represented 19.5\% of its G0 bag but carried 94.9\% of the attention mass and 96.4\% of the absolute attention-weighted tumour-versus-normal logit contribution. Removing these patches reduced mean tumour probability from 0.959 to 0.153 and reversed the prediction in all three models. Because removal was defined using predicted background fraction rather than the manually delineated scanner-mark mask, this analysis demonstrates dependence on QC-defined high-background instances rather than a causal effect of the scanner mark itself. Because tumor\_008 is tumour-positive, its prediction reversal was a correct-positive to false-negative change in all three models, rather than a demonstrated improvement in diagnostic performance.

We next examined the 17 slides selected post hoc because G0 and U4c produced different hard classifications in all three independently trained models (Supplementary Fig. 18 and Supplementary Data 4, Table 18). Counterfactual sensitivity again varied substantially. In tumor\_049, high-background patches comprised 45.5\% of the G0 bag and showed attention and absolute-logit-contribution enrichments of 2.144 and 2.148, respectively. Their removal reduced mean tumour probability from 0.928 to 0.084 and reversed all three G0 predictions.

The direction and magnitude of these effects were not uniform. In normal\_079, removal increased mean tumour probability from 0.264 to 0.495 and changed two of the three classifications, indicating that the removed instances had provided net support for the normal class. Conversely, high attention or contribution enrichment did not necessarily imply a large probability effect. In tumor\_004, attention and absolute-logit-contribution enrichments were 2.892 and 2.763, respectively, but removal changed mean tumour probability by only 0.040 and produced no hard-classification change.

Several slides also showed reproducible G0--U4c disagreement despite negligible sensitivity to direct removal of high-background instances from the G0 bag. In tumor\_024, removal changed mean tumour probability from 0.983 to 0.986, an absolute change of 0.003, and in tumor\_006 it changed the probability from 0.945 to 0.959, an absolute change of 0.014; neither intervention changed a seed-level classification. Across all 17 reproducibly discordant slides, median enrichment of high-background patches was 1.30 for attention mass and 1.31 for absolute attention-weighted logit contribution.

These per-slide analyses therefore identify several distinct behaviours: high-background instances could dominate an individual prediction, support either class, or receive disproportionate attention and logit contribution without producing a large counterfactual probability change. Conversely, reproducible G0--U4c disagreement could persist even when direct removal of high-background instances had little effect on the original G0 prediction. Direct dependence on high-background patches therefore explains only a subset of the prediction differences associated with the two candidate-patch universes. The observed policy contrast is therefore consistent with how technical marks were distributed across predicted channels, but neither the counterfactual intervention nor the channel audit isolates their contribution to the cohort-level policy effect.

\subsection{Replicate-grouping sensitivity in the matched TCGA-LUSC experiment}

The deterministic strict artefact-exclusion training set contained 18,183 tiles, comprising 8,139 normal and 10,044 SCC tiles. All more permissive training conditions were matched to this set at the patient-by-class level and therefore reproduced these class counts and patient contributions. For the ≤1\%, ≤5\% and unrestricted artefact policies, three independently sampled matched training sets were crossed with three optimisation seeds, yielding a balanced 3 × 3 design. We used this structure to assess whether the observed policy differences depended materially on tile-sampling seed, optimisation seed or the way replicates were grouped.

Pairwise comparisons among the three artefact-permitting policies were first calculated for the nine exact tile-sampling-seed × optimisation-seed cells. Mean accuracy differed by −0.0003 ± 0.0161 for ≤5\% versus ≤1\% artefact, with an observed range of −0.0286 to +0.0228 and five positive and four negative contrasts. Unrestricted inclusion differed from ≤1\% artefact by −0.0121 ± 0.0261, with a range of −0.0597 to +0.0290 and three positive and six negative contrasts, and from ≤5\% artefact by −0.0118 ± 0.0158, with a range of −0.0430 to +0.0061 and three positive and six negative contrasts (Supplementary Data 4, Table 20). Thus, the two moderate artefact-retention policies were nearly indistinguishable in mean accuracy, whereas unrestricted inclusion was modestly lower on average but showed substantial replicate-level overlap.

The same contrasts were then marginalised by averaging over optimisation seeds within each tile-sampling seed and, separately, over tile-sampling seeds within each optimisation seed. Because the design was balanced and fully crossed, marginalisation did not alter the overall mean paired differences. For ≤5\% versus ≤1\% artefact, the three sampling-seed accuracy marginals ranged from −0.0075 to +0.0058 and the optimisation-seed marginals from −0.0056 to +0.0093. For unrestricted inclusion versus ≤1\% artefact, the corresponding ranges were −0.0239 to +0.0001 and −0.0230 to +0.0070. Directional consistency was strongest for unrestricted inclusion versus ≤5\% artefact: all three sampling-seed marginals were negative, ranging from −0.0182 to −0.0007, and all three optimisation-seed marginals were negative, ranging from −0.0175 to −0.0022. The narrower marginal ranges reflect averaging across three matched cells and were not interpreted as increased inferential precision.

This directional pattern did not extend uniformly across performance metrics. Under strict artefact exclusion, mean normal and SCC recall were 0.876 and 0.873, respectively, giving an absolute separation of only 0.003. By contrast, the three artefact-permitting policies showed separations of 0.086--0.131 between their mean class recalls, consistently favouring SCC recall. Within the unrestricted-versus-≤5\% comparison, normal-recall differences were positive in all three optimisation-seed marginals, ranging from +0.0018 to +0.0528, whereas SCC-recall differences were negative in all three, ranging from −0.0425 to −0.0066. Balanced-accuracy and patient-weighted-accuracy directions were less consistent, and the remaining policy contrasts likewise showed mixed marginal directions depending on metric and grouping factor. The apparently small differences in aggregate accuracy therefore reflected redistribution of class-specific errors rather than a uniform shift in classification quality.

Accuracy variability was additionally decomposed descriptively into tile-sampling, optimisation-seed and residual components. For the ≤1\% policy, these components accounted for 54.8\%, 15.2\% and 30.1\% of the total sums of squares, respectively; for the ≤5\% policy, 73.6\%, 0.3\% and 26.1\%; and for unrestricted inclusion, 28.5\%, 20.1\% and 51.3\%. The residual component included sampling-by-optimisation interaction and other cell-level variation that could not be separated because only one observation was available for each tile-sampling-seed × optimisation-seed cell. These proportions were compared descriptively with the 25\%/25\%/50\% partition implied by the corresponding degrees of freedom and were not used for formal variance-component inference.

Together, these sensitivity analyses show that the mean contrasts among artefact-retention policies were unchanged by alternative replicate groupings, although their directional consistency varied across metrics. The policies primarily altered the balance of class-specific errors, while both tile sampling and optimisation contributed to replicate-level variability.

As a secondary robustness analysis, the same matched classifiers were evaluated on a more permissive common test set allowing ≤90\% predicted background and unrestricted artefacts. Mean accuracy again differed only modestly among artefact-retention policies, ranging from 0.809 to 0.819, and seed-matched accuracy contrasts among the artefact-permitting policies remained close to zero (Supplementary Fig. 19). This secondary analysis therefore did not indicate a marked change in the relative behaviour of the training policies under the more permissive test distribution.

\subsection{Real-artefact augmentation improves robustness under matched controls and backbone freezing}

To assess whether the advantage of real-artefact augmentation could be attributed to perturbation coverage alone, we compared realised modified-area fractions across the three augmentation interventions. Mean realised coverage across the three training seeds was 0.213 for real-artefact augmentation, 0.213 for random erasing and 0.210 for clean-tissue replacement. Random erasing was matched exactly to the real-artefact condition in mask geometry, transformations and placement, whereas clean-tissue replacement was closely rather than exactly matched because library entries without an associated clean crop required remapping within the available clean-tissue subset. Despite this closely matched coverage, clean-tissue replacement improved high-stratum S3 macro Dice over the unaugmented baseline by only 0.0076 on average and in two of three seeds, compared with a mean improvement of 0.0594 for real-artefact augmentation (Supplementary Data 4, Table 21). The substantially larger gain with real-artefact augmentation was therefore not reproduced by clean-tissue replacement at similar mean total modified-image area, although differences in source-component size remained between these interventions.

We next examined whether the real-artefact augmentation gain depended on updating the foundation-model backbone. Under severe held-out composited real-artefact corruption, mean high-stratum S3 macro Dice was 0.8362 for the trainable-backbone baseline and 0.8231 for the frozen-backbone baseline, compared with 0.8956 and 0.8840 for the corresponding real-artefact-augmented models. Freezing therefore modestly reduced absolute corrupted performance in both augmentation conditions, whereas the augmentation gain remained similar with trainable and frozen backbones.

At the reported S3 severe-corruption endpoint, the trainable-minus-frozen interaction in the augmentation effect was −0.0015 ± 0.0382 across three training seeds (Fig. 7c and Source Data). This estimate was interpreted descriptively because the small number of training seeds did not support a conclusion of equivalence between frozen and trainable backbones. Together, these results indicate that the robustness gain associated with real-artefact augmentation was retained when adaptation was restricted to the segmentation head, while backbone fine-tuning was associated with modestly higher absolute performance under severe corruption.

\begin{figure*}[p]
\centering
\includegraphics[width=1.000\textwidth]{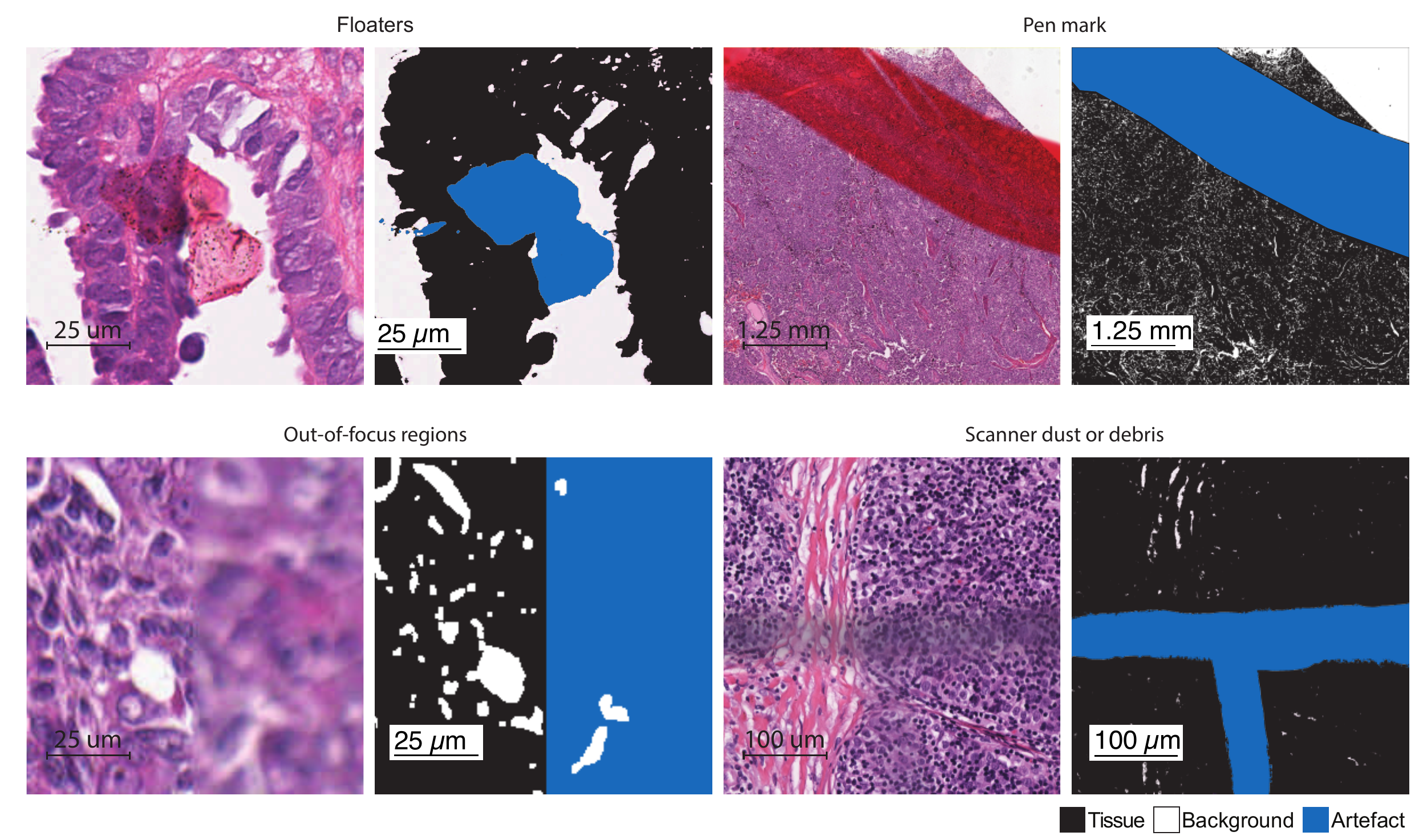}
\caption{\textbf{Representative PathQC-3C annotations across technical artefact types.} Representative H\&E ROIs and corresponding dense three-class reference masks illustrating the PathQC-3C annotation scheme. Masks show tissue in black, background in white and artefact in blue. Examples include floaters, pen marks, out-of-focus regions and scanner dust or debris arising during histopathology preparation and digitisation. Scale bars indicate physical dimensions.}
\end{figure*}
\begin{figure*}[p]
\centering
\includegraphics[width=1.000\textwidth]{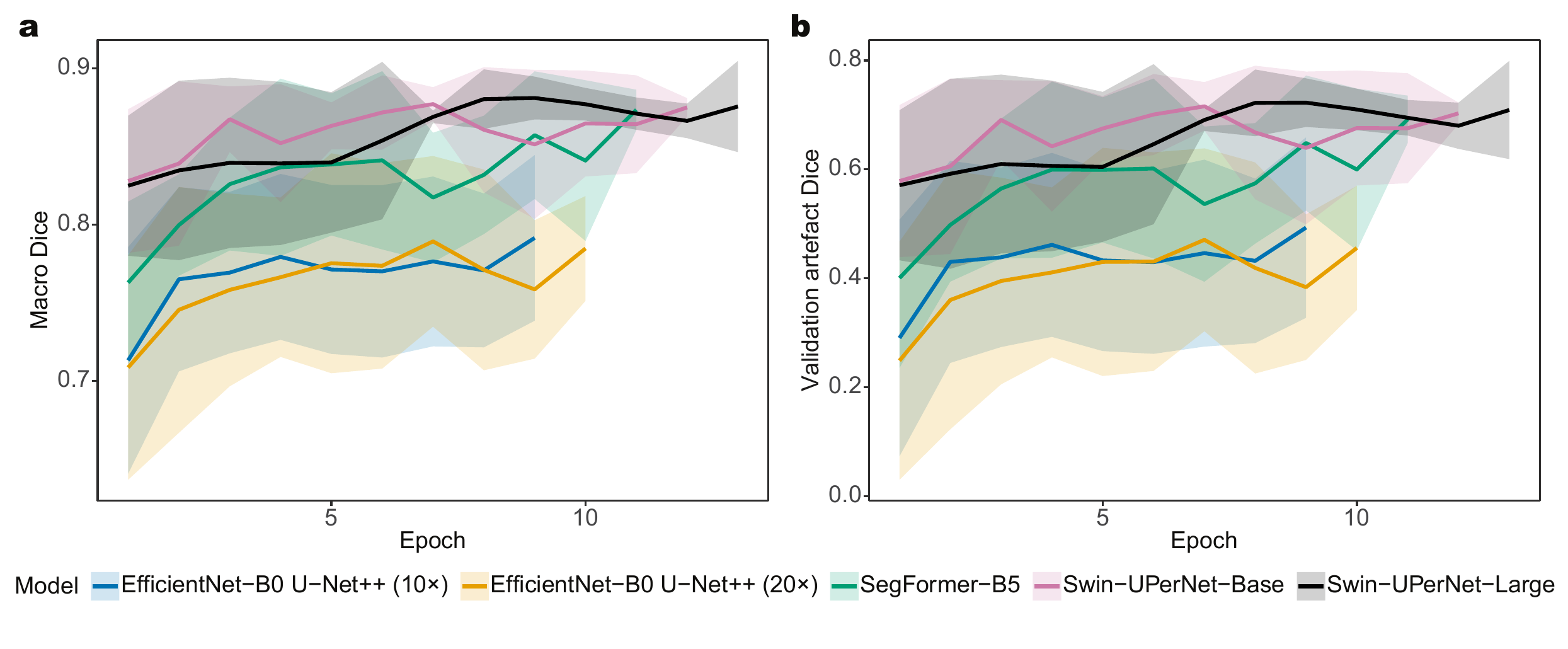}
\caption{\textbf{Learning dynamics of baseline dense quality-control predictors under matched training conditions.} \textbf{a,} Validation macro Dice across training epochs for five predictor configurations evaluated with 512 × 512-pixel inputs and the Dice--IoU objective. \textbf{b,} Validation artefact Dice for the same configurations. Models comprised EfficientNet-B0 U-Net++ at 10× and 20×, SegFormer-B5 at 20×, Swin--UPerNet-Base at 20× and Swin--UPerNet-Large at 20×. Lines show means across three patient-level cross-validation folds and shaded bands indicate ±1 s.d.; curves are shown only for epochs at which all three folds contributed. Macro Dice separated the lower-performing EfficientNet configurations from the transformer-based predictors, whereas artefact Dice showed greater between-model separation and variability. Trajectory values are epoch-wise summaries and therefore differ from the selected-checkpoint values in Supplementary Data 4, Table 2, which were calculated separately at each fold's best validation macro-Dice checkpoint.}
\end{figure*}
\begin{figure*}[p]
\centering
\includegraphics[width=1.000\textwidth]{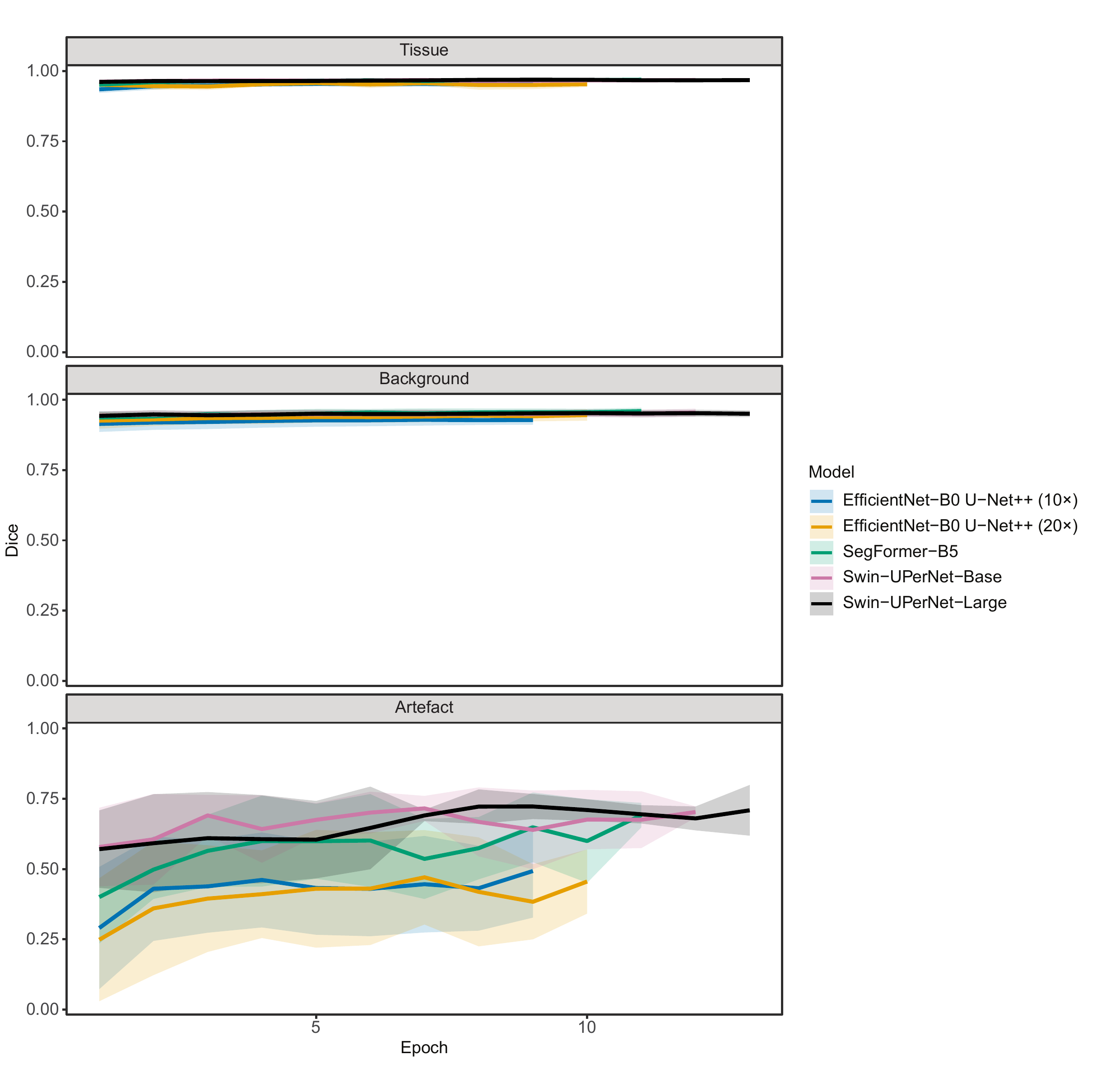}
\caption{\textbf{Class-wise learning dynamics of baseline dense quality-control predictors.} Validation Dice is shown across training epochs separately for tissue, background and artefact for five predictor configurations evaluated with 512 × 512-pixel inputs and the Dice--IoU objective. Models comprised EfficientNet-B0 U-Net++ at 10× and 20×, SegFormer-B5 at 20×, Swin--UPerNet-Base at 20× and Swin--UPerNet-Large at 20×. Lines show means across three patient-level cross-validation folds and shaded bands indicate ±1 s.d.; curves are shown only for epochs at which all three folds contributed. Tissue and background Dice remained consistently high across model families, whereas artefact Dice showed substantially greater between-model separation and variability during training. Trajectory values are epoch-wise summaries and therefore differ from the selected-checkpoint values in Supplementary Data 4, Table 2, which were calculated separately at each fold's best validation macro-Dice checkpoint.}
\end{figure*}
\begin{figure*}[p]
\centering
\includegraphics[width=0.965\textwidth]{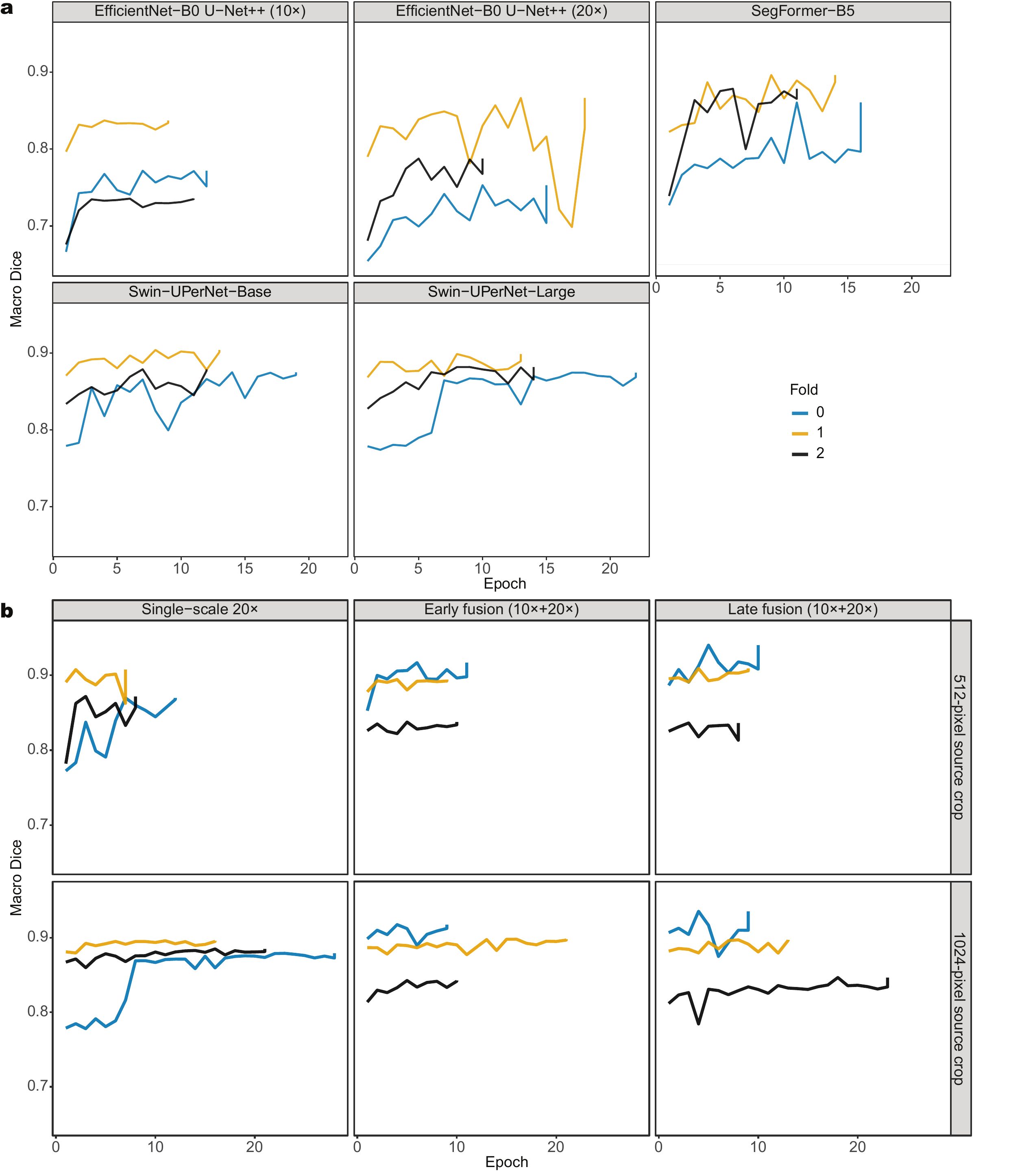}
\caption{\textbf{Fold-wise validation macro-Dice trajectories for baseline architectures and Swin--UPerNet-Base configurations.} a, Validation macro Dice across training epochs for five baseline dense QC predictor configurations evaluated with 512 × 512-pixel network inputs and the Dice--IoU objective: EfficientNet-B0 U-Net++ at 10× and 20×, SegFormer-B5 at 20×, Swin--UPerNet-Base at 20× and Swin--UPerNet-Large at 20×. b, Validation macro-Dice trajectories for six Swin--UPerNet-Base configurations combining single-scale 20×, early-fusion 10×+20× and late-fusion 10×+20× inputs with 512- or 1,024-pixel source crops, all trained with the unweighted Dice--cross-entropy objective. Source crops were resized so that all network inputs were 512 × 512 pixels. Curves represent the three patient-level development folds. Within each comparison, a common y-axis scale is used to facilitate assessment of optimisation dynamics and fold-to-fold variability. Trajectories terminate at each fold's final evaluation epoch under early stopping; the selected checkpoint was the epoch with the highest validation macro Dice within that fold. Fold 0 is shown in blue, fold 1 in orange and fold 2 in black.}
\end{figure*}
\begin{figure*}[p]
\centering
\includegraphics[width=1.000\textwidth]{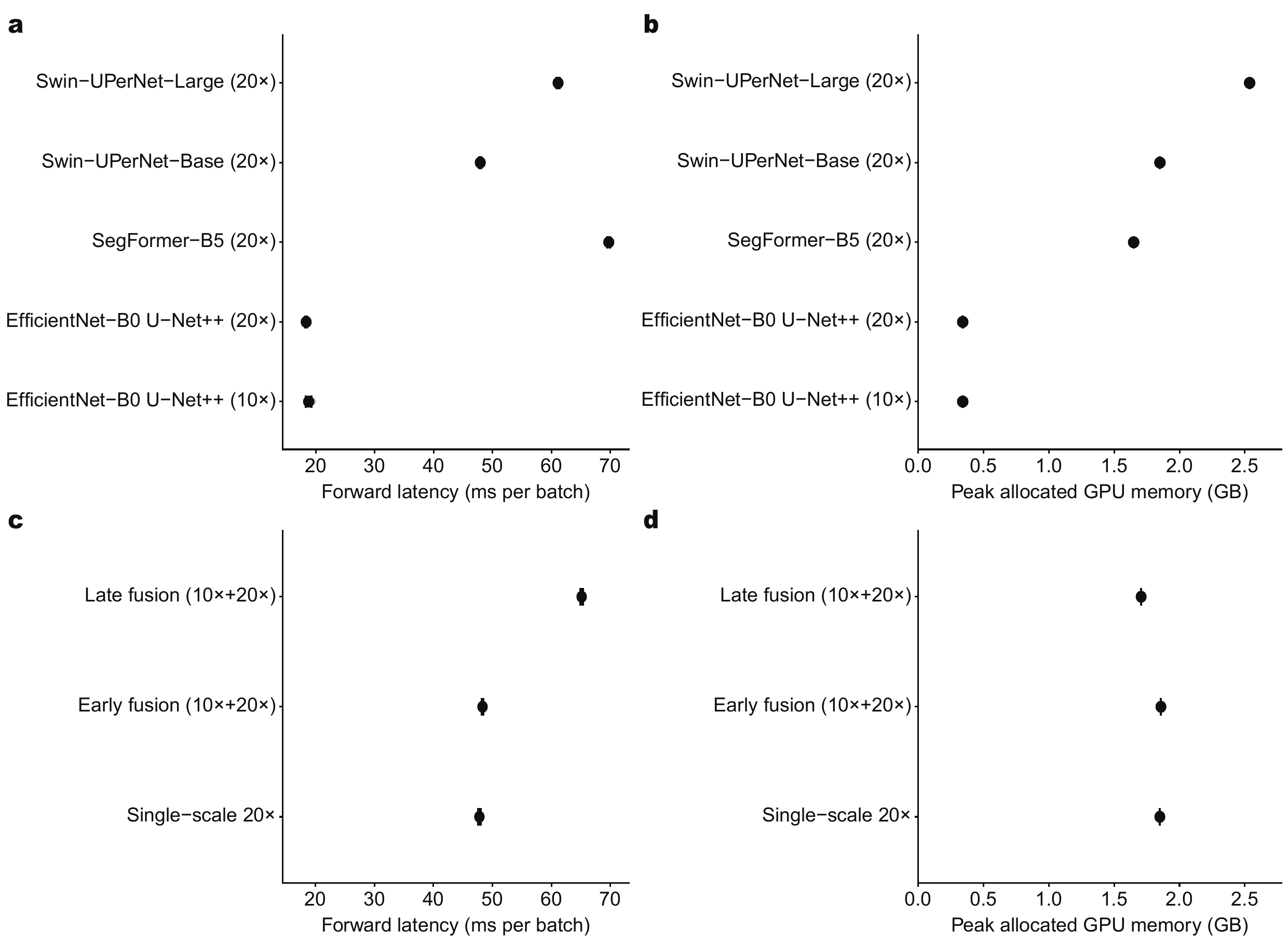}
\caption{\textbf{Standardised forward-only computational profiling of baseline architectures and Swin--UPerNet-Base configurations.} \textbf{a,} Mean forward latency for five baseline dense QC predictor configurations under a matched inference benchmark. \textbf{b,} Peak PyTorch allocated GPU memory for the same models. Configurations comprised EfficientNet-B0 U-Net++ at 10× and 20×, SegFormer-B5 at 20×, Swin--UPerNet-Base at 20× and Swin--UPerNet-Large at 20×. \textbf{c,} Mean forward latency for single-scale 20×, early-fusion 10×+20× and late-fusion 10×+20× Swin--UPerNet-Base configurations. \textbf{d,} Peak PyTorch allocated GPU memory for the same three configurations. All models were benchmarked on an AMD Instinct MI250X GPU using batch size 2, bf16 autocast and 512 × 512-pixel network inputs. Each run comprised 20 warm-up iterations followed by 100 synchronised timed forward passes using GPU-resident preprocessed tensors; image loading, preprocessing, host-to-device transfer, loss computation and metric evaluation were excluded. Points show individual development folds, and larger points with error bars indicate mean ± s.d. across three folds.}
\end{figure*}
\begin{figure*}[p]
\centering
\includegraphics[width=0.880\textwidth]{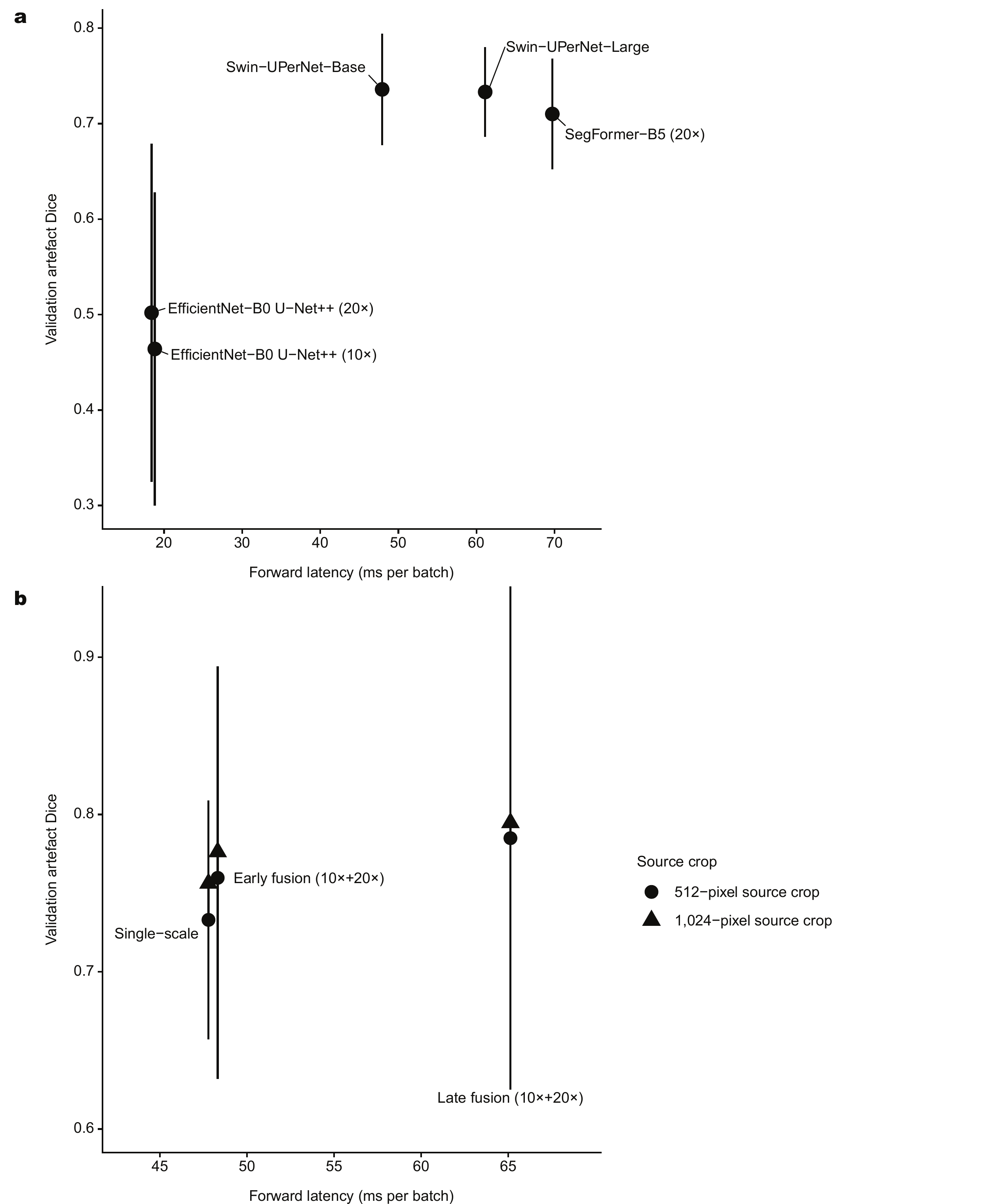}
\caption{\textbf{Artefact-segmentation performance relative to standardised forward-only computational cost.} a, Validation artefact Dice versus mean forward latency for five baseline dense QC predictor configurations: EfficientNet-B0 U-Net++ at 10× and 20×, SegFormer-B5 at 20×, Swin--UPerNet-Base at 20× and Swin--UPerNet-Large at 20×. Artefact Dice was taken from the checkpoint with the highest validation macro Dice within each patient-level development fold, and latency was measured using the standardised benchmark in Supplementary Fig. 5. Points indicate across-fold means; horizontal and vertical error bars show ±1 s.d. across three folds for latency and artefact Dice, respectively. b, Corresponding comparison for single-scale 20×, early-fusion 10×+20× and late-fusion 10×+20× Swin--UPerNet-Base configurations using 512- or 1,024-pixel source crops and the unweighted Dice--cross-entropy objective. The two source-crop conditions for each architecture share the same forward-latency estimate because both were resized to 512 × 512-pixel network inputs before inference; vertical separation therefore reflects differences in validation artefact Dice rather than neural-network forward cost. Benchmarks used an AMD Instinct MI250X GPU, batch size 2, bf16 autocast, 20 warm-up iterations and 100 synchronised timed forward passes with GPU-resident preprocessed tensors. Image loading, preprocessing, host-to-device transfer, loss computation and metric evaluation were excluded. These measurements quantify standardised model-forward cost and do not imply that native 1,024 × 1,024-pixel network inference would have the same computational cost as 512 × 512-pixel inference.}
\end{figure*}
\begin{figure*}[p]
\centering
\includegraphics[width=1.000\textwidth]{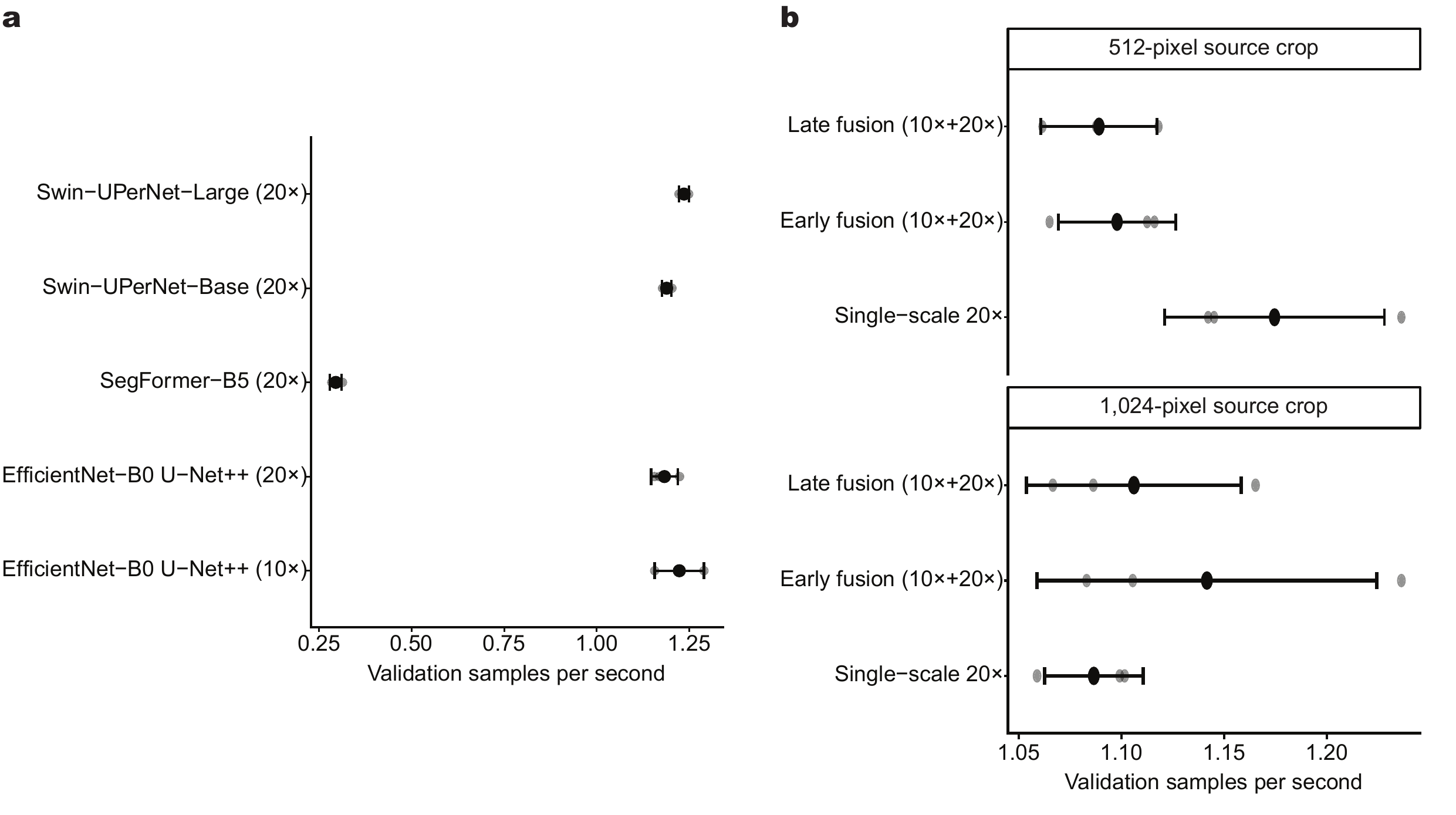}
\caption{\textbf{End-to-end validation throughput of baseline architectures and Swin--UPerNet-Base configurations.} a, Mean steady-state validation throughput for five baseline dense QC predictor configurations evaluated with 512 × 512-pixel network inputs: EfficientNet-B0 U-Net++ at 10× and 20×, SegFormer-B5 at 20×, Swin--UPerNet-Base at 20× and Swin--UPerNet-Large at 20×. b, Mean steady-state throughput for single-scale 20×, early-fusion 10×+20× and late-fusion 10×+20× Swin--UPerNet-Base configurations using 512- or 1,024-pixel source crops and the unweighted Dice--cross-entropy objective. In both source-crop conditions, inputs were resized to 512 × 512 pixels before network inference. Points show individual development folds; larger points and error bars indicate mean ± s.d. across three folds. Steady-state throughput was calculated after excluding the first evaluation epoch to reduce one-time initialisation overhead. Measurements include data loading, source-image processing, host-to-device transfer, model inference and evaluation overheads and are therefore distinct from the standardized forward-only benchmark in Supplementary Fig. 5.}
\end{figure*}
\begin{figure*}[p]
\centering
\includegraphics[width=1.000\textwidth]{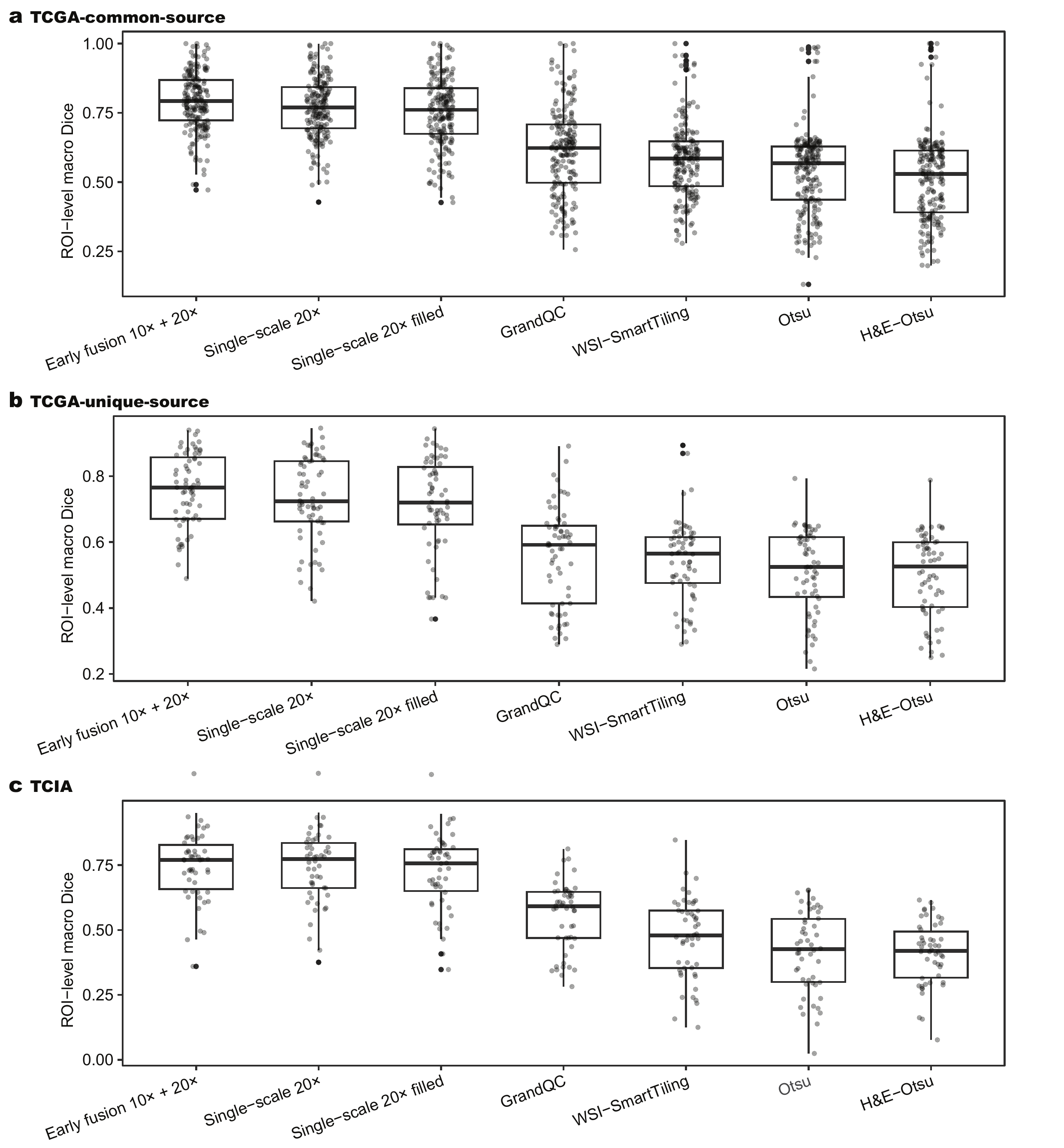}
\caption{\textbf{Held-out macro Dice comparison with quality-control baselines across cohorts.} ROI-level macro Dice is shown for a, TCGA-common-source, b, TCGA-unique-source and c, TCIA. For Swin--UPerNet-Base dense QC predictor configurations, metrics were calculated separately for each fold-specific prediction and averaged within ROI before plotting, yielding one value per ROI and configuration. Comparator methods were evaluated once per ROI. Macro Dice was calculated within each ROI as the mean over classes present in the reference mask. Points represent individual ROIs; boxplots show medians, interquartile ranges and 1.5× interquartile-range whiskers. Methods comprise early-fusion 10×+20×, single-scale 20×, single-scale 20× with hole filling, GrandQC, WSI-SmartTiling, Otsu and H\&E-Otsu.}
\end{figure*}
\begin{figure*}[p]
\centering
\includegraphics[width=1.000\textwidth]{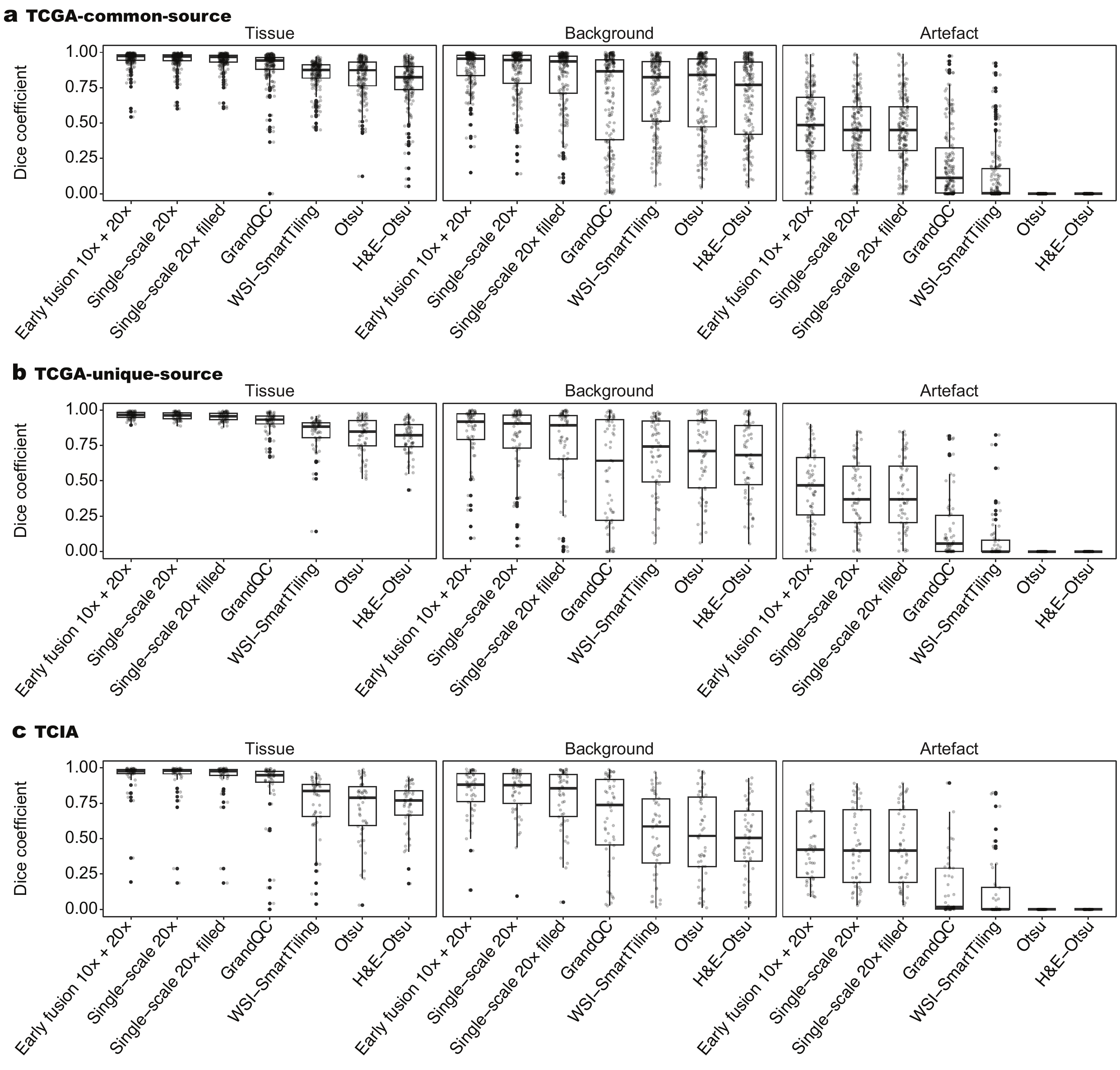}
\caption{\textbf{Held-out class-wise Dice comparison with quality-control baselines across cohorts.} ROI-level tissue, background and artefact Dice are shown for a, TCGA-common-source, b, TCGA-unique-source and c, TCIA. For Swin--UPerNet-Base dense QC predictor configurations, metrics were calculated separately for each fold-specific prediction and averaged within ROI before plotting, yielding one value per ROI and configuration. Comparator methods were evaluated once per ROI. Class-wise Dice was summarised only for ROIs in which the corresponding reference class was present; artefact Dice therefore includes only artefact-containing ROIs. Points represent individual ROIs; boxplots show medians, interquartile ranges and 1.5× interquartile-range whiskers. Methods comprise early-fusion 10×+20×, single-scale 20×, single-scale 20× with hole filling, GrandQC, WSI-SmartTiling, Otsu and H\&E-Otsu. Otsu and H\&E-Otsu are binary foreground--background methods and did not produce an artefact class after harmonisation to the common label space; their displayed artefact-Dice values are structural zeros and should not be interpreted as conventional artefact-detection estimates.}
\end{figure*}
\begin{figure*}[p]
\centering
\includegraphics[width=0.943\textwidth]{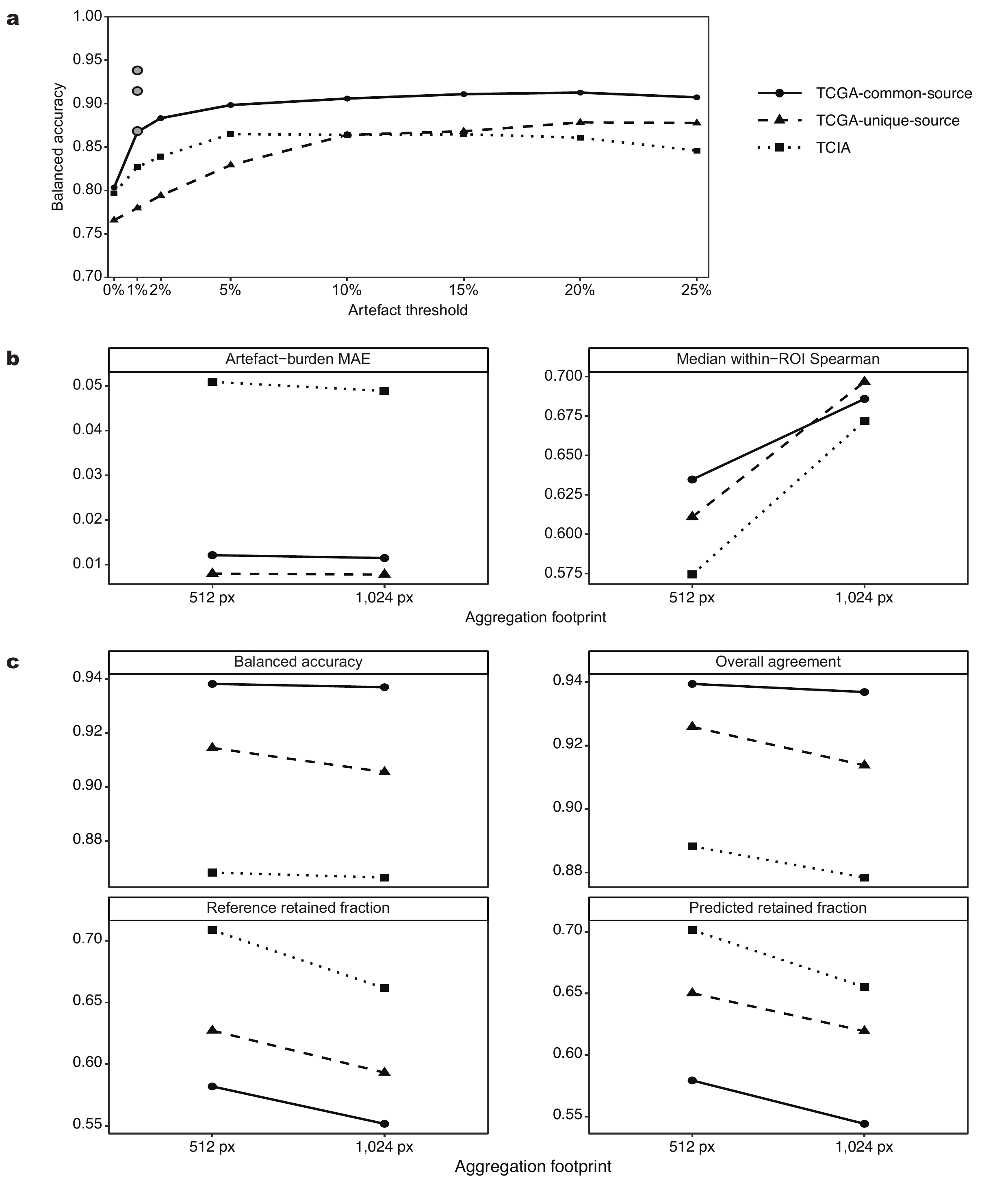}
\caption{\textbf{Aggregation-scale and threshold sensitivity of tile-level burden and policy fidelity.} a, Reference--prediction agreement for artefact-threshold tile-selection rules evaluated over non-overlapping 512 × 512-pixel footprints in TCGA-common-source, TCGA-unique-source and TCIA. Balanced accuracy is shown as a function of artefact-burden threshold, treating the reference-derived decision as the target and rejection as the positive class. Thresholds of 0\%, 1\%, 2\%, 5\%, 10\%, 15\%, 20\% and 25\% were evaluated as sensitivity operating points rather than as universally optimal cut-offs. Grey markers at the 1\% artefact threshold denote the combined rule requiring artefact fraction ≤0.01 and tissue fraction ≥0.50. b, Tile-level artefact-burden agreement at 512 × 512- and 1,024 × 1,024-pixel aggregation footprints, summarised by mean absolute error and median within-ROI Spearman correlation between reference-derived and prediction-derived artefact fractions. \textbf{c,} Reference--prediction agreement for the combined artefact-and-tissue rule at both footprint sizes. Balanced accuracy, overall agreement and reference- and prediction-derived retained fractions are shown for each cohort. These analyses assess reproducibility of specified operating rules using prediction-derived QC fractions and do not establish that any threshold is optimal for tile usability.}
\end{figure*}
\begin{figure*}[p]
\centering
\includegraphics[width=1.000\textwidth]{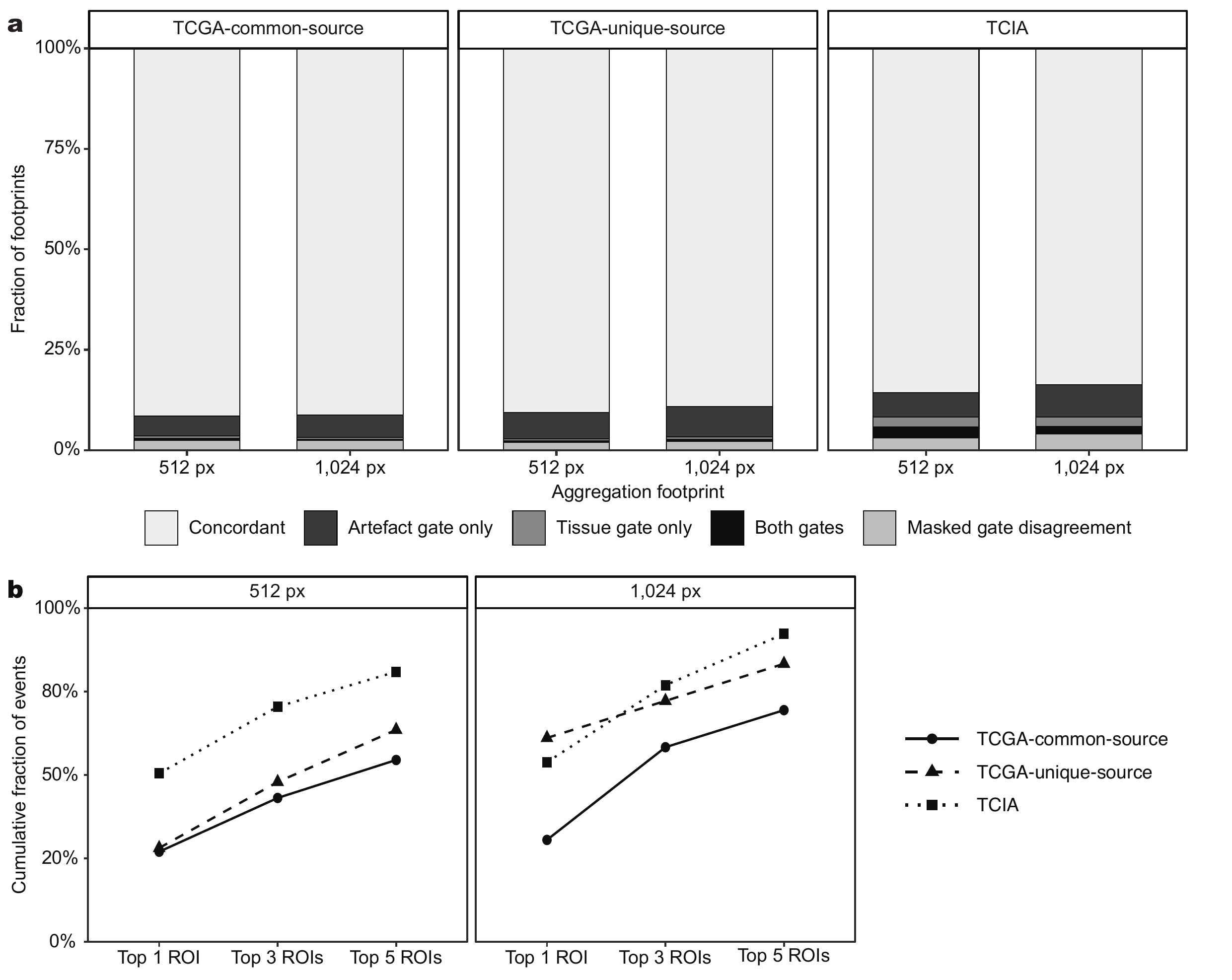}
\caption{\textbf{Mechanisms and concentration of QC-policy disagreement.} a, Decomposition of reference--prediction disagreement for the combined rule requiring artefact fraction ≤0.01 and tissue fraction ≥0.50. Footprints are classified as concordant, artefact-gate-only disagreement, tissue-gate-only disagreement, disagreement involving both gates, or masked gate disagreement, in which at least one component gate differed between reference and prediction but the final keep/reject decision remained concordant because the other gate already determined rejection. Results are shown for 512 × 512- and 1,024 × 1,024-pixel aggregation footprints across TCGA-common-source, TCGA-unique-source and TCIA. b, Concentration of larger artefact-underestimation events across ROIs, defined descriptively as reference-derived artefact fraction \textgreater0.05 with prediction-derived artefact fraction ≤0.01. Curves show the cumulative fraction of events accounted for by the one, three and five highest-contributing ROIs in each cohort at both footprint sizes. These descriptive analyses show that substantial threshold-crossing errors were concentrated within a limited subset of ROIs rather than distributed uniformly across the evaluated cohorts.}
\end{figure*}
\begin{figure*}[p]
\centering
\includegraphics[width=1.000\textwidth]{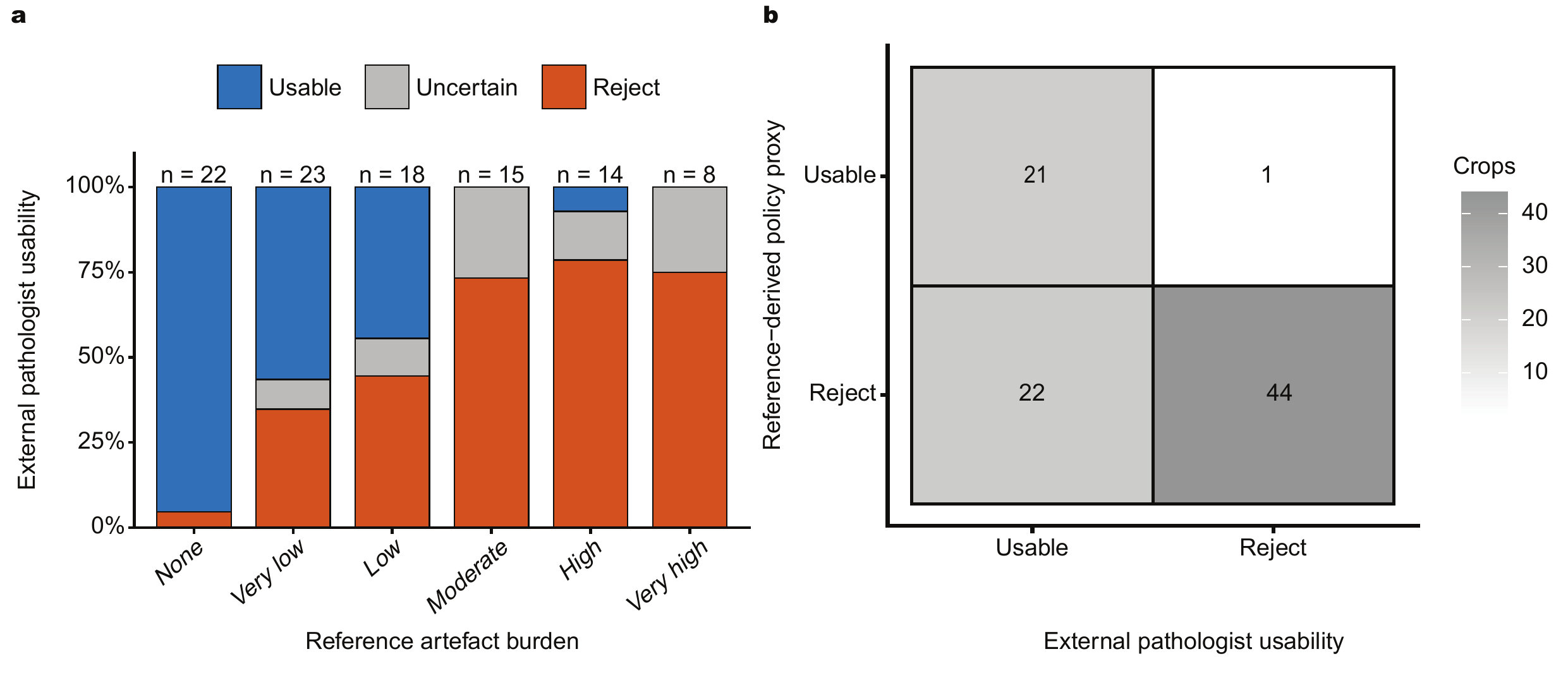}
\caption{\textbf{Artefact burden is associated with tile-level usability in the blinded pathologist audit.} a, Pathologist usability assessment across 100 crops from the TCIA external-validation cohort, stratified by reference-derived artefact burden. Artefact burden was defined as the fraction of valid pixels labelled artefact in the reference mask and grouped as none (0\%), very low (\textgreater5--\textless7\%), low (7--\textless10\%), moderate (10--\textless20\%), high (20--\textless50\%) or very high (≥50\%). Bars show the proportions rated usable, uncertain or reject; labels above bars indicate the number of crops in each stratum. b, Cross-tabulation of the reference-derived usability proxy and pathologist usability among the 88 crops with definite usable/reject calls; 12 uncertain calls were excluded. The proxy classified crops as reject when background exceeded 70\% or artefact exceeded 5\% of valid pixels and as usable otherwise. Numbers indicate crop counts.}
\end{figure*}
\begin{figure*}[p]
\centering
\includegraphics[width=1.000\textwidth]{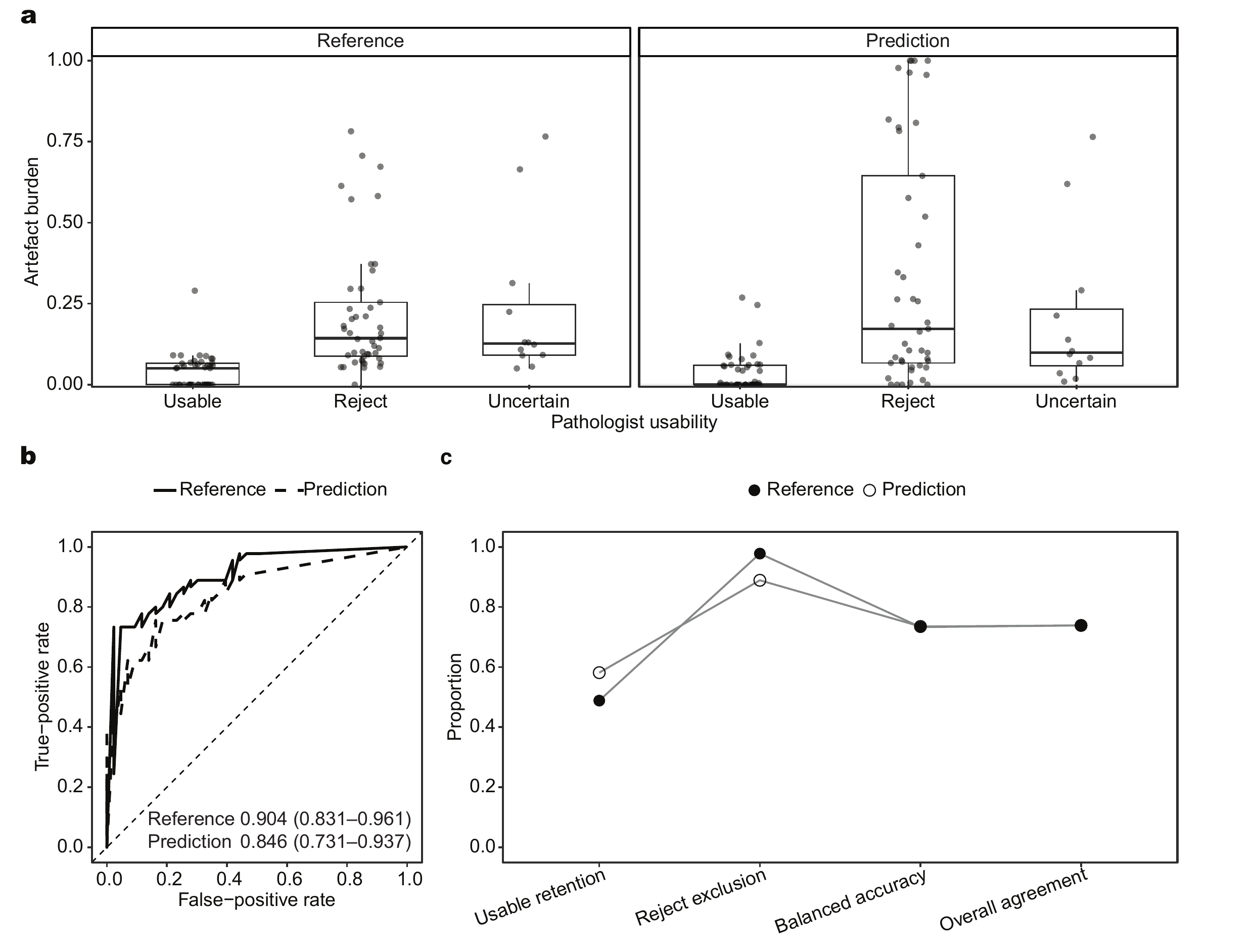}
\caption{\textbf{Prediction-derived burden and policy decisions in the blinded pathologist audit.} a, Distribution of reference-derived and prediction-derived artefact burden across crops rated usable, reject or uncertain by the pathologist from a separate institution. The primary discrimination analysis was restricted to the 88 crops with definite usability calls. b, Receiver-operating-characteristic curves for discrimination of pathologist-rejected versus usable crops using reference-derived and prediction-derived artefact burden. AUROC estimates and 95\% confidence intervals were obtained from 10,000 case-cluster bootstrap resamples preserving crops from the same TCIA case. \textbf{c,} Agreement with pathologist usability for the combined artefact-and-tissue rule requiring artefact fraction ≤0.01 and tissue fraction ≥0.50. Usable retention, reject exclusion, balanced accuracy and overall agreement are shown for reference-derived and prediction-derived decisions among the 88 crops with definite usability calls. Replacing reference-derived with prediction-derived fractions shifted the operating point toward greater retention while leaving overall agreement unchanged at 73.9\%.}
\end{figure*}
\begin{figure*}[p]
\centering
\includegraphics[width=1.000\textwidth]{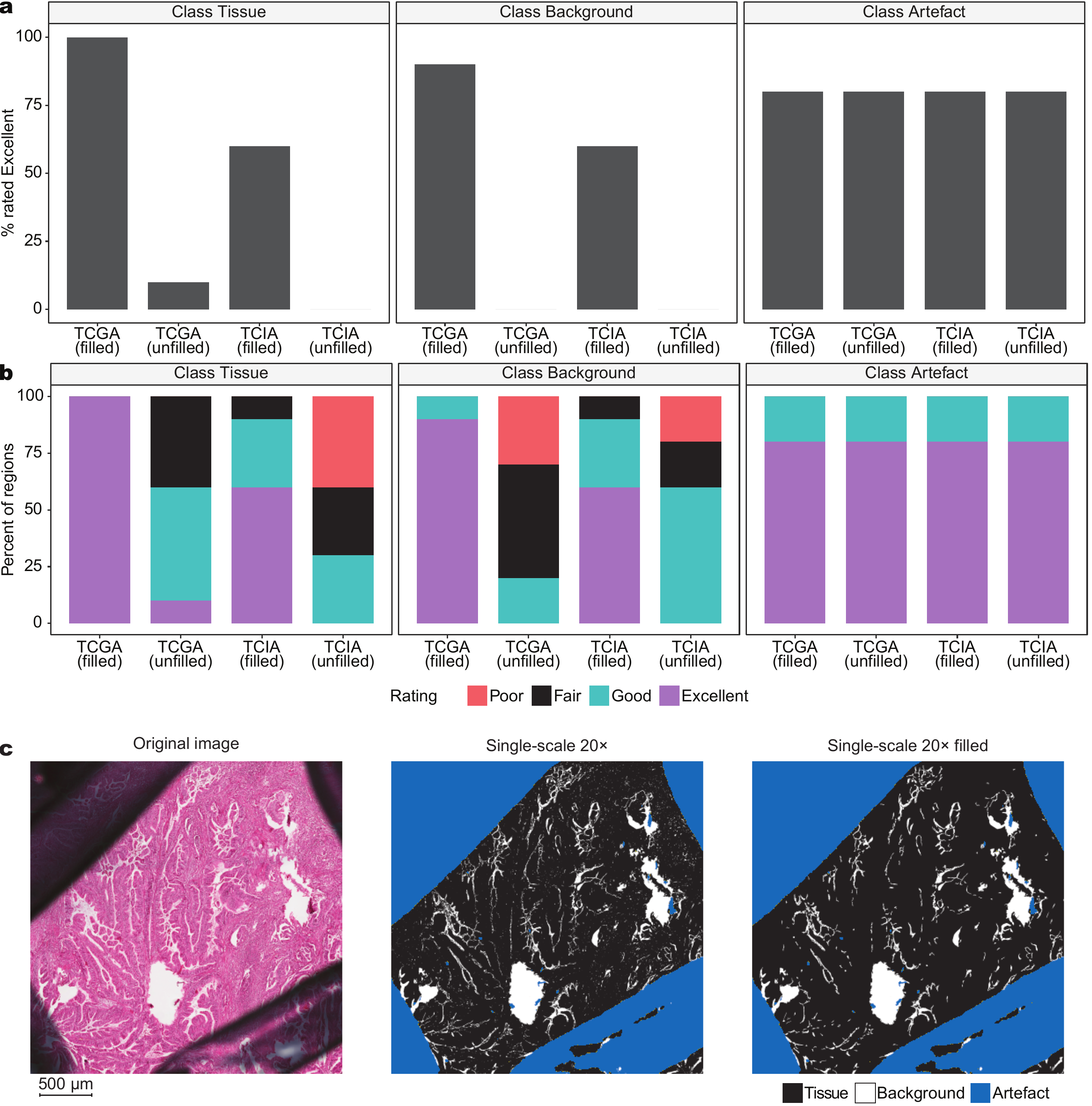}
\caption{\textbf{Expert visual assessment of hole-filling post-processing.} \textbf{a,b,} Expert histologist ratings comparing single-scale 20× predictions with and without hole filling across 20 held-out 20× ROIs (6,000 × 6,000 pixels; 10 from TCIA and 10 from the TCGA-unique-source cohort). One expert histologist scored tissue, background and artefact delineation using a four-level ordinal scale (Excellent, Good, Fair and Poor). Filled and unfilled variants were reviewed together in a fixed order and the assessment was therefore not blinded to post-processing condition. a, Proportion of ROIs rated Excellent for each class, cohort and post-processing condition. b, Full rating distributions shown as stacked bars. \textbf{c,} Representative example showing the effect of hole filling on the predicted mask. Shown are the original H\&E ROI, the single-scale 20× prediction and the corresponding hole-filled prediction. Masks show tissue in black, background in white and artefact in blue. Scale bar, 500 µm.}
\end{figure*}
\begin{figure*}[p]
\centering
\includegraphics[width=1.000\textwidth]{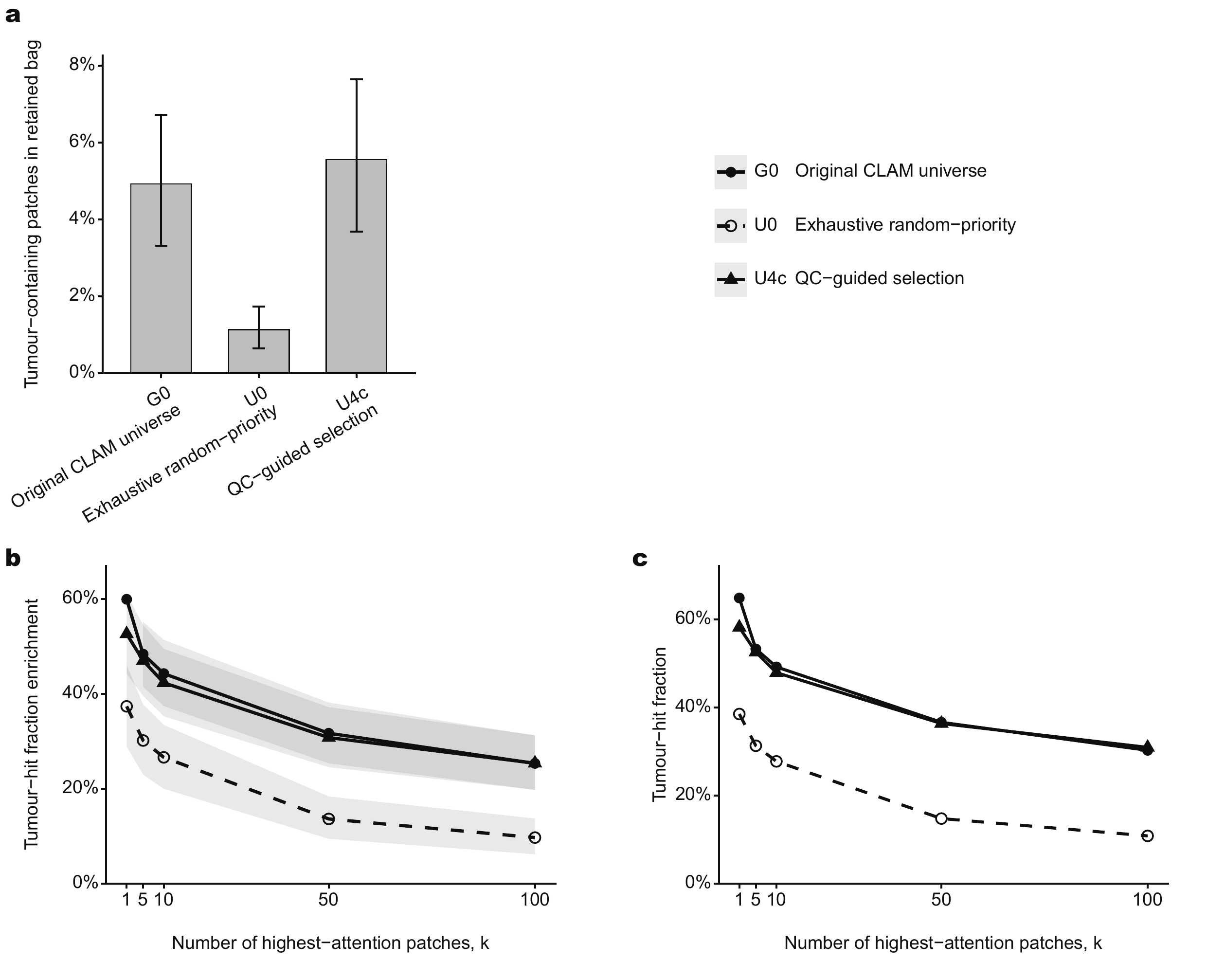}
\caption{\textbf{Tile-selection policy alters tumour representation and CLAM attention localisation.} a, Mean proportion of retained patches whose centres overlapped CAMELYON16 tumour annotations for the original CLAM candidate-patch universe (G0), reproducible random sampling from the exhaustive universe (U0) and QC-guided combined background-and-artefact selection (U4c). Analysis included all 110 tumour slides in the 244-slide common-support cohort. Error bars show pointwise 95\% slide-level bootstrap confidence intervals. b, Tumour-hit-fraction enrichment among the top-k highest-attention patches (k = 1, 5, 10, 50 and 100), calculated within slide as observed tumour-hit fraction minus tumour prevalence among all retained patches. Shaded bands show pointwise 95\% bootstrap confidence intervals. \textbf{c,} Mean fraction of top-k highest-attention patches whose centres overlapped annotated tumour. Attention was reconstructed from each of the three out-of-fold CLAM models, and localisation metrics were averaged across seeds within slide. U4c increased retained tumour prevalence relative to U0 (0.0556 versus 0.0113) and showed greater hit-fraction enrichment at all evaluated attention depths. No corresponding enrichment advantage over G0 was observed. G0 is therefore shown as a contextual cross-universe reference, whereas U4c versus U0 is a matched exploratory comparison because both operate on the same exhaustive candidate universe and fixed 4,000-patch budget. U4c was selected post hoc from the combined policies.}
\end{figure*}
\begin{figure*}[p]
\centering
\includegraphics[width=1.000\textwidth]{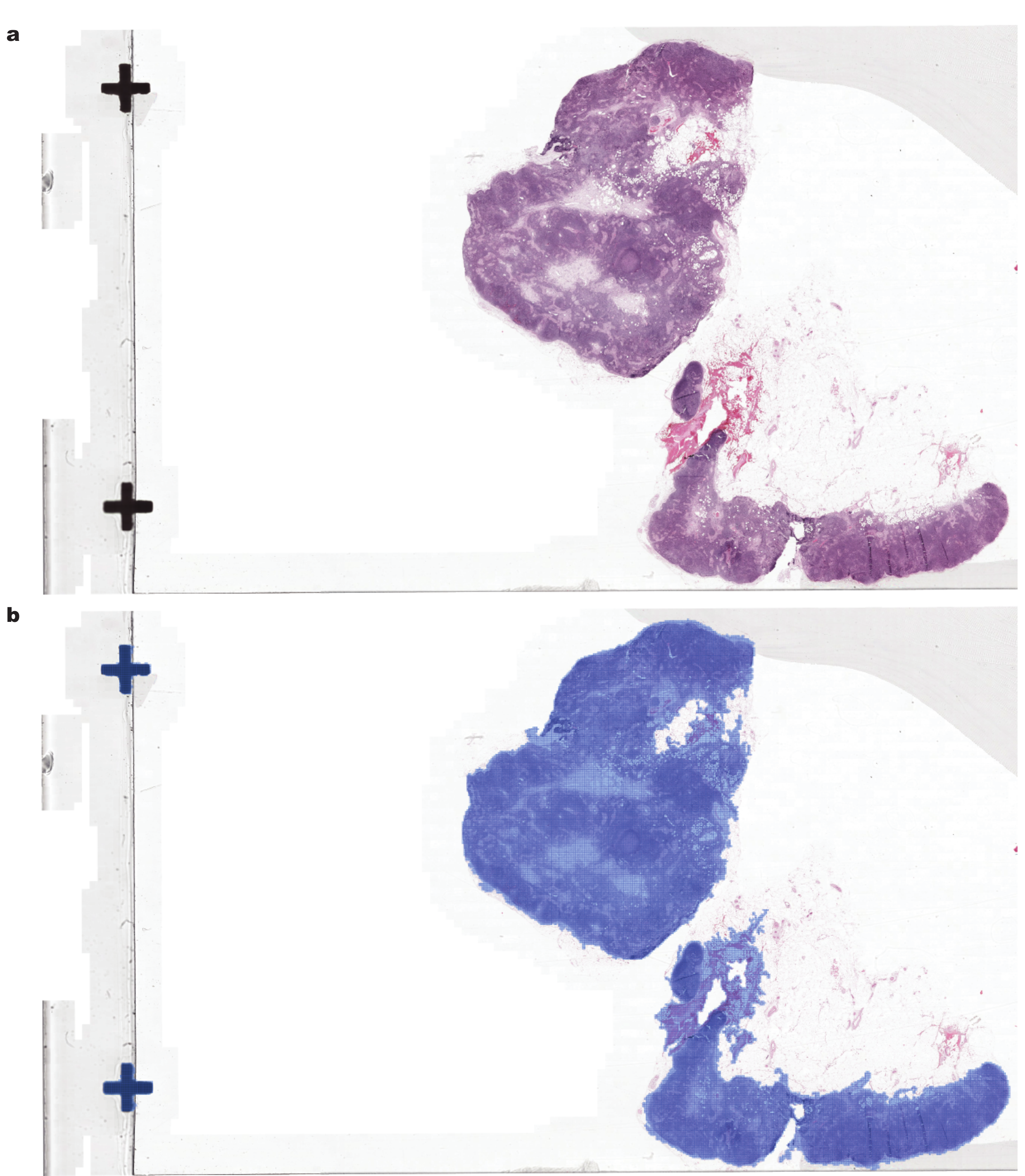}
\caption{\textbf{Scanner-mark inclusion by the original CLAM foreground-selection procedure in CAMELYON16.} a, Original CAMELYON16 whole-slide image tumor\_008, identified as scanner-mark-positive during manual review. b, Corresponding slide showing regions admitted by the original CLAM foreground-selection procedure (G0), including non-tissue scanner markings outside the biological tissue compartment. The blinded review identified 43 scanner-mark regions across 28 slides overall; after restriction to the 244-slide common-support cohort, 40 regions across 27 slides remained, of which 36 high-confidence regions across 24 slides formed the primary QC-channel analysis. The blue overlay indicates CLAM patch admission, not TileNet's predicted artefact channel; scanner marks were assigned across both predicted background and artefact channels.}
\end{figure*}
\begin{figure*}[p]
\centering
\includegraphics[width=0.909\textwidth]{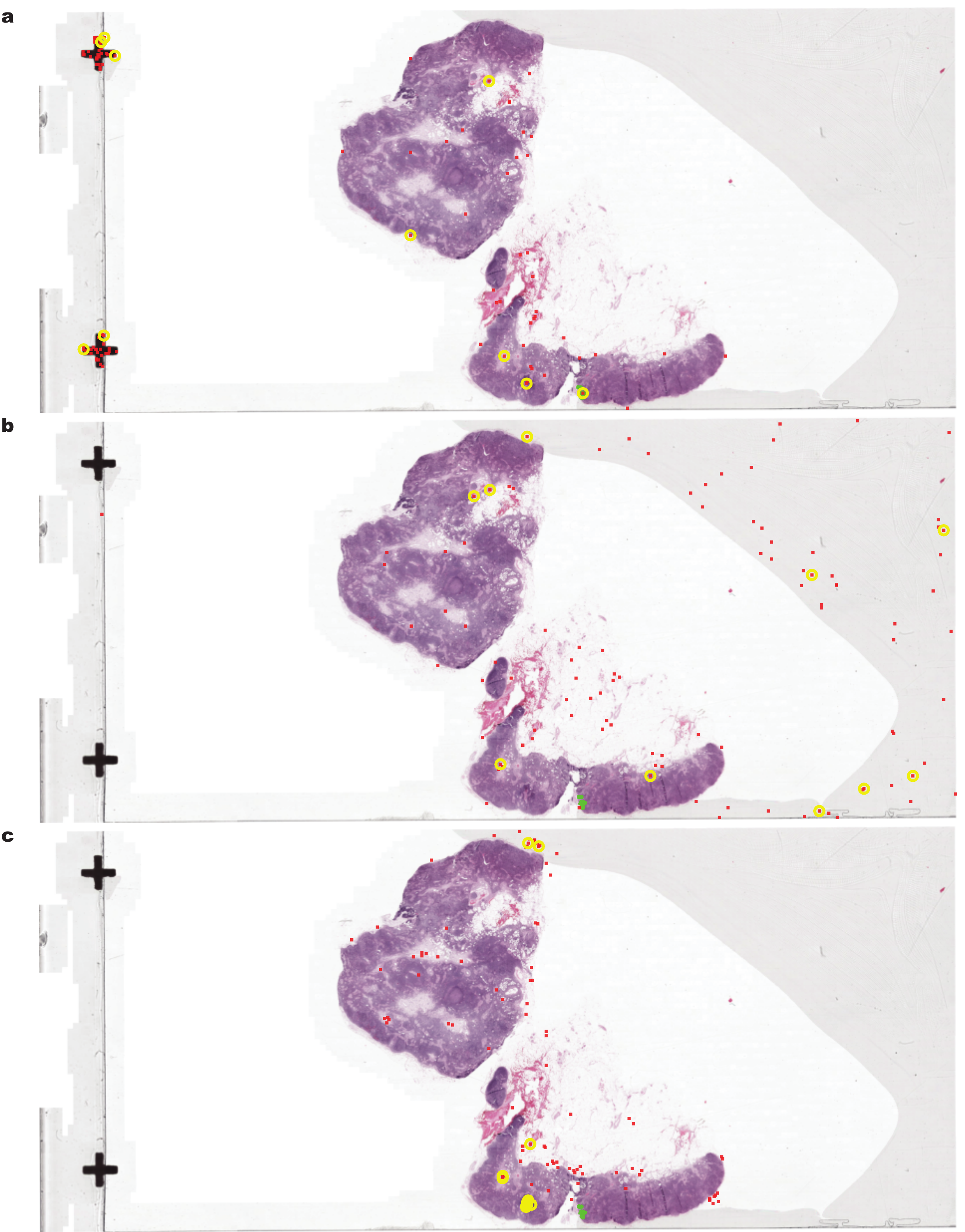}
\caption{\textbf{Spatial distribution of high-attention CLAM instances under alternative tile-selection policies.} Locations of the 100 patches receiving the highest CLAM attention weights are shown for the same representative CAMELYON16 slide under a, the original CLAM candidate-patch universe (G0), b, U0, reproducible random sampling without replacement from the exhaustive candidate universe, and \textbf{c,} QC-guided combined background-and-artefact selection (U4c). Red points indicate the top 100 highest-attention patches, and larger yellow-outlined markers identify the top 10. Under G0, highly attended patches include locations associated with non-tissue scanner markings outside the biological tissue compartment, whereas U0 and U4c produce different spatial distributions of high-attention instances. These maps illustrate how tile-selection policy alters both bag composition and the image regions receiving the greatest CLAM attention.}
\end{figure*}
\begin{figure*}[p]
\centering
\includegraphics[width=1.000\textwidth]{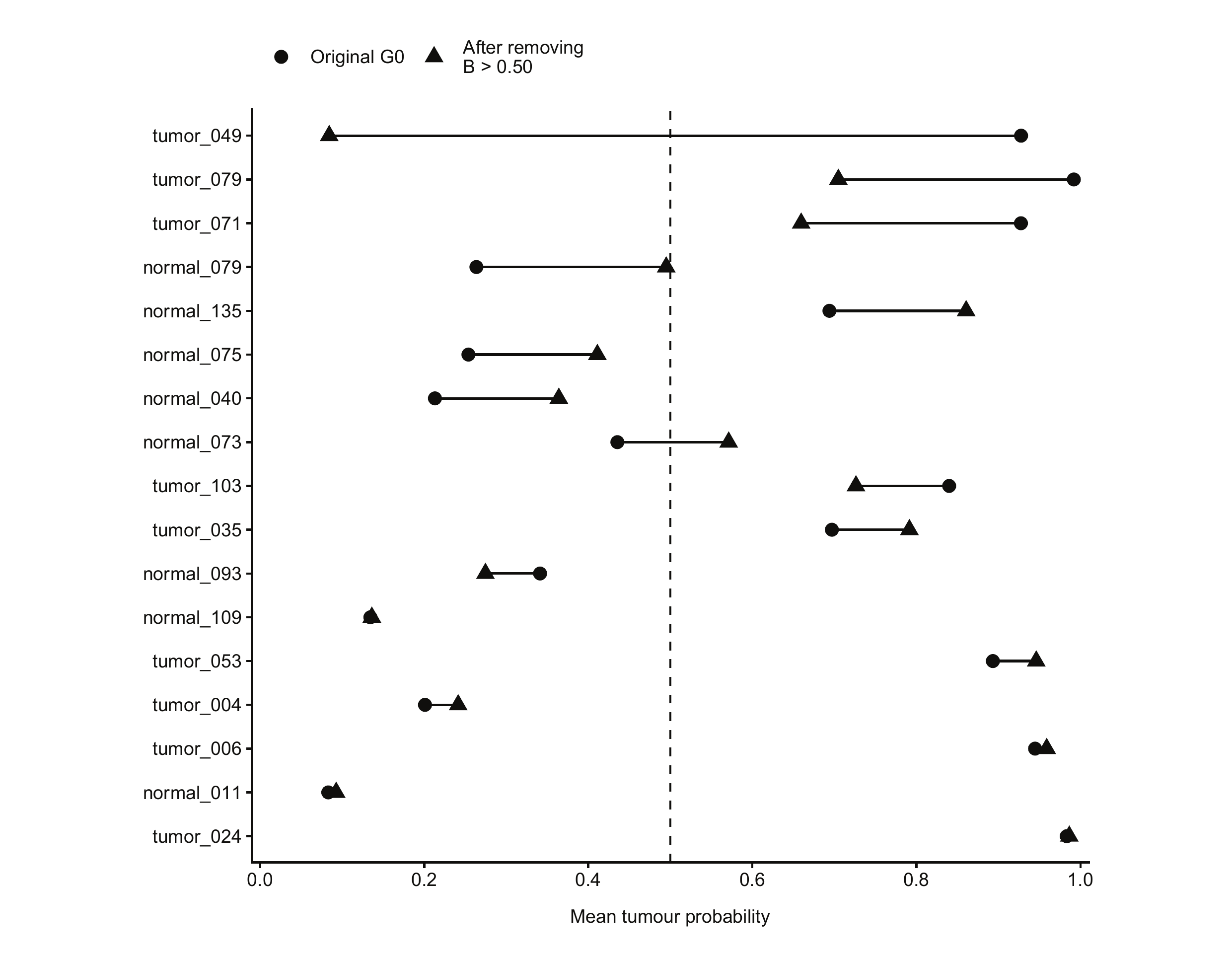}
\caption{\textbf{Same-model counterfactual analysis of reproducibly discordant G0--U4c predictions.} Mean tumour probabilities before and after removal of high-background patches are shown for the 17 CAMELYON16 slides on which G0 and U4c produced different hard classifications in all three trained models. This post hoc exploratory subset was defined by reproducible policy-dependent prediction disagreement. For each slide, G0 patches with predicted background fraction \textgreater0.50 were removed and the remaining bag was re-evaluated using the unchanged out-of-fold G0 model. Points show mean tumour probabilities across the three models before and after removal, with paired values connected within slides; slides are ordered by mean absolute probability change. The dashed vertical line denotes the 0.5 classification threshold. Removal changed the G0 classification in at least one model for 7 of 17 slides, in at least two models for 2 slides and in all three models for one slide; the median absolute probability change was 0.114. The heterogeneous responses, including minimal changes for some slides despite reproducible G0--U4c disagreement, indicate that direct dependence on QC-defined high-background instances explains only a subset of the cross-universe prediction differences.}
\end{figure*}
\begin{figure*}[p]
\centering
\includegraphics[width=1.000\textwidth]{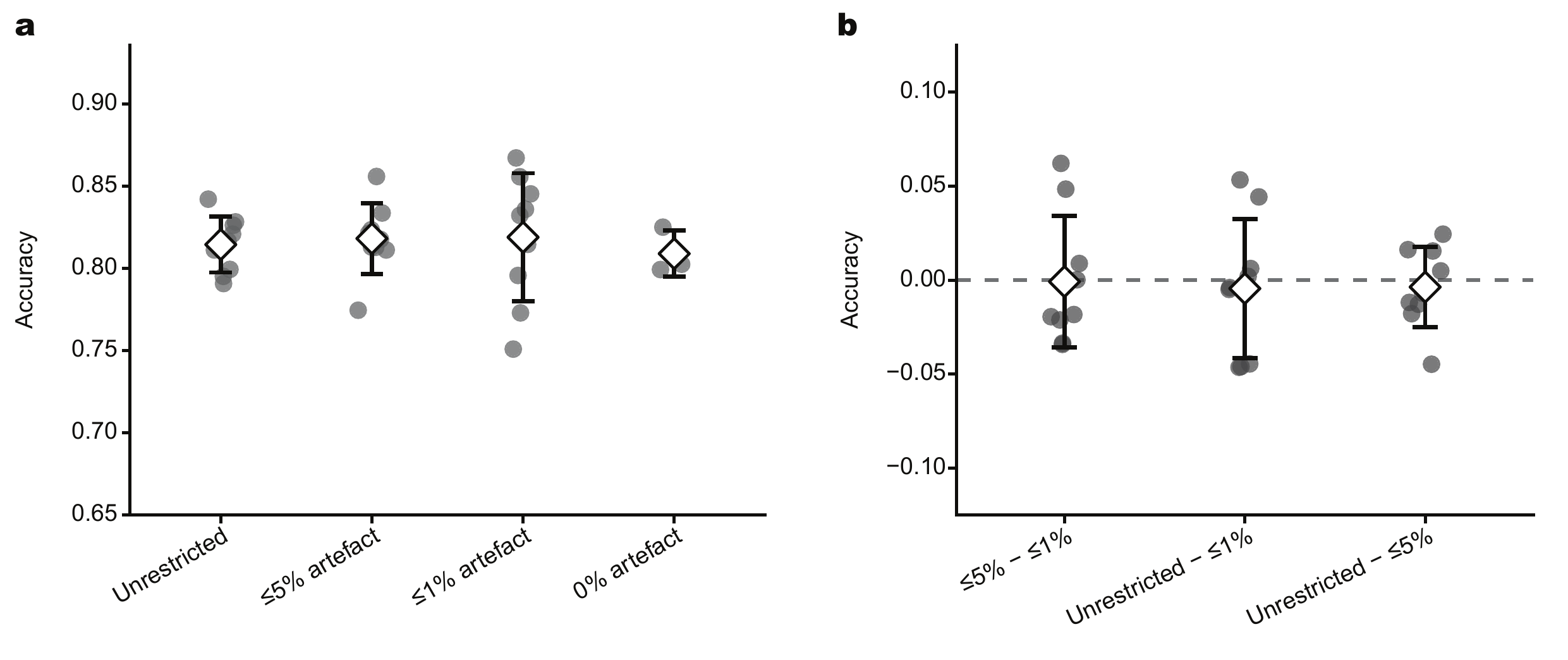}
\caption{\textbf{Matched TCGA-LUSC performance on a permissive quality-control test distribution.} a, Accuracy of the same 30 matched classifiers evaluated on a common secondary test set allowing ≤90\% predicted background and unrestricted artefacts. The set comprised 11,167 tiles from the same 10 test patients: 4,568 normal tiles (40.9\%) and 6,599 SCC tiles (59.1\%). Points denote separately trained models; diamonds indicate means and error bars show ±1 s.d. The ≤1\%, ≤5\% and unrestricted artefact policies each comprised three independently sampled matched training sets crossed with three optimisation seeds (n = 9 models per condition), whereas strict artefact exclusion used one deterministic matched training set evaluated across three optimisation seeds (n = 3). b, Seed-matched pairwise differences in accuracy among artefact-permitting policies for the nine cells of the fully crossed 3 × 3 tile-sampling-seed × optimisation-seed design. Positive values indicate higher accuracy for the first-named policy and the dashed horizontal line denotes no difference. The nine matched cells were analysed descriptively and were not treated as nine independent experimental units. This secondary analysis assessed robustness to a more permissive quality-control test distribution and was not part of the primary matched test-set comparison.}
\end{figure*}
\begin{figure*}[p]
\centering
\includegraphics[width=1.000\textwidth]{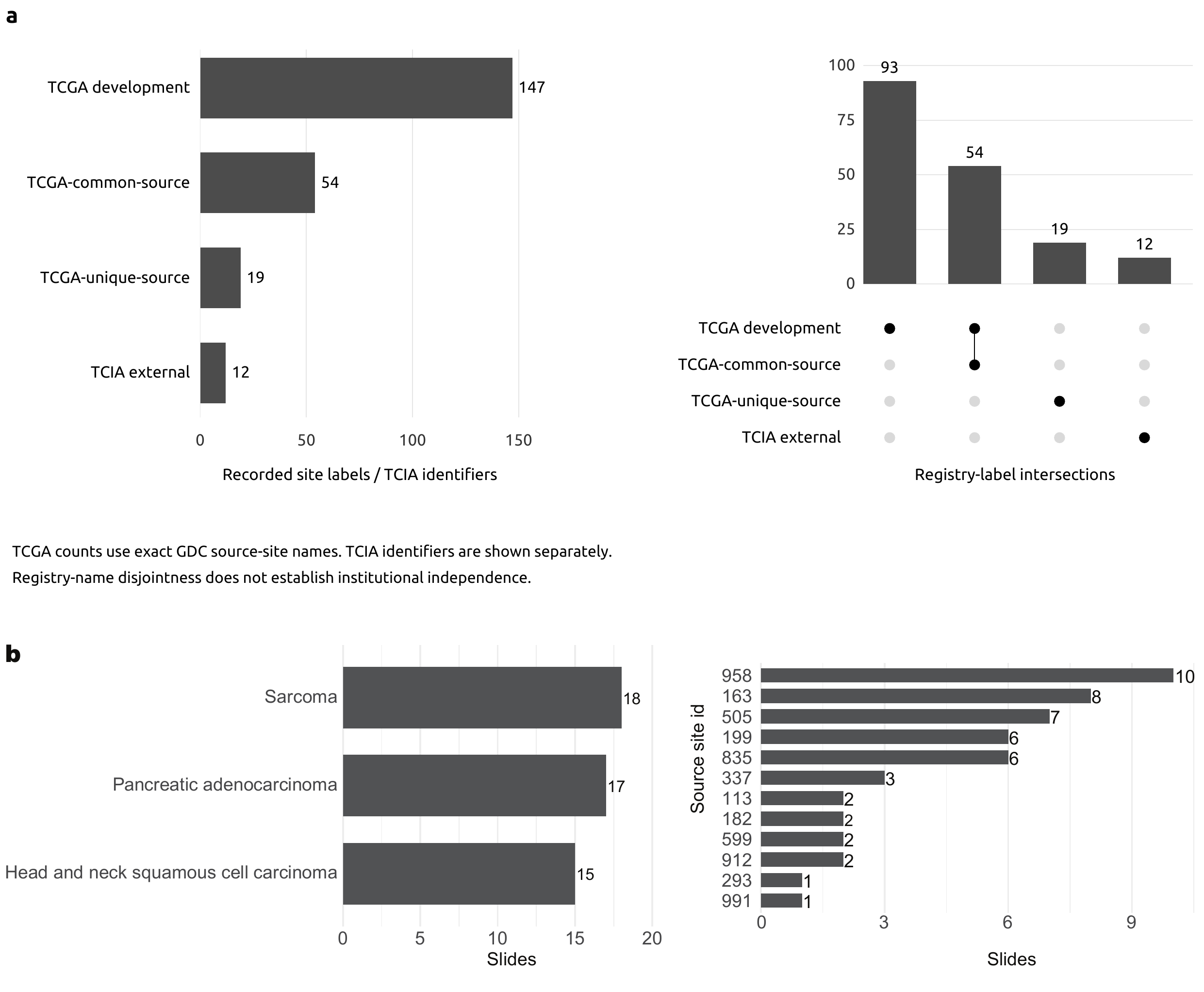}
\caption{\textbf{Recorded source-site labels and TCIA cohort composition. a, Exact recorded GDC source-site names across the TCGA cohorts: 147 in development, 54 in TCGA-common-source and 19 in TCGA-unique-source, yielding 166 distinct TCGA names overall. The 54 common-source names are represented in development. The corresponding barcode TSS counts are 395, 92 and 24, with 510 distinct codes overall because one code (FA) occurs in both development and TCGA-common-source (Source Data, worksheet Fig2\_site\_mapping). Multiple codes can map to the same recorded name. Names were not canonicalised to physical institutions, so the diagram establishes registry-name overlap rather than verified institutional overlap or independence. TCIA's 12 deidentified source identifiers are shown separately for composition; they cannot be cross-mapped reliably to the TCGA labels. Bars and membership points retain the original counts, including 93 development-only and 54 shared TCGA names.} b, Composition of the TCIA external-validation cohort. Left, slide counts across the three included diagnostic categories. Right, slide counts for each of the 12 deidentified TCIA source-site identifiers.}
\end{figure*}
\begin{figure*}[p]
\centering
\includegraphics[width=1.000\textwidth]{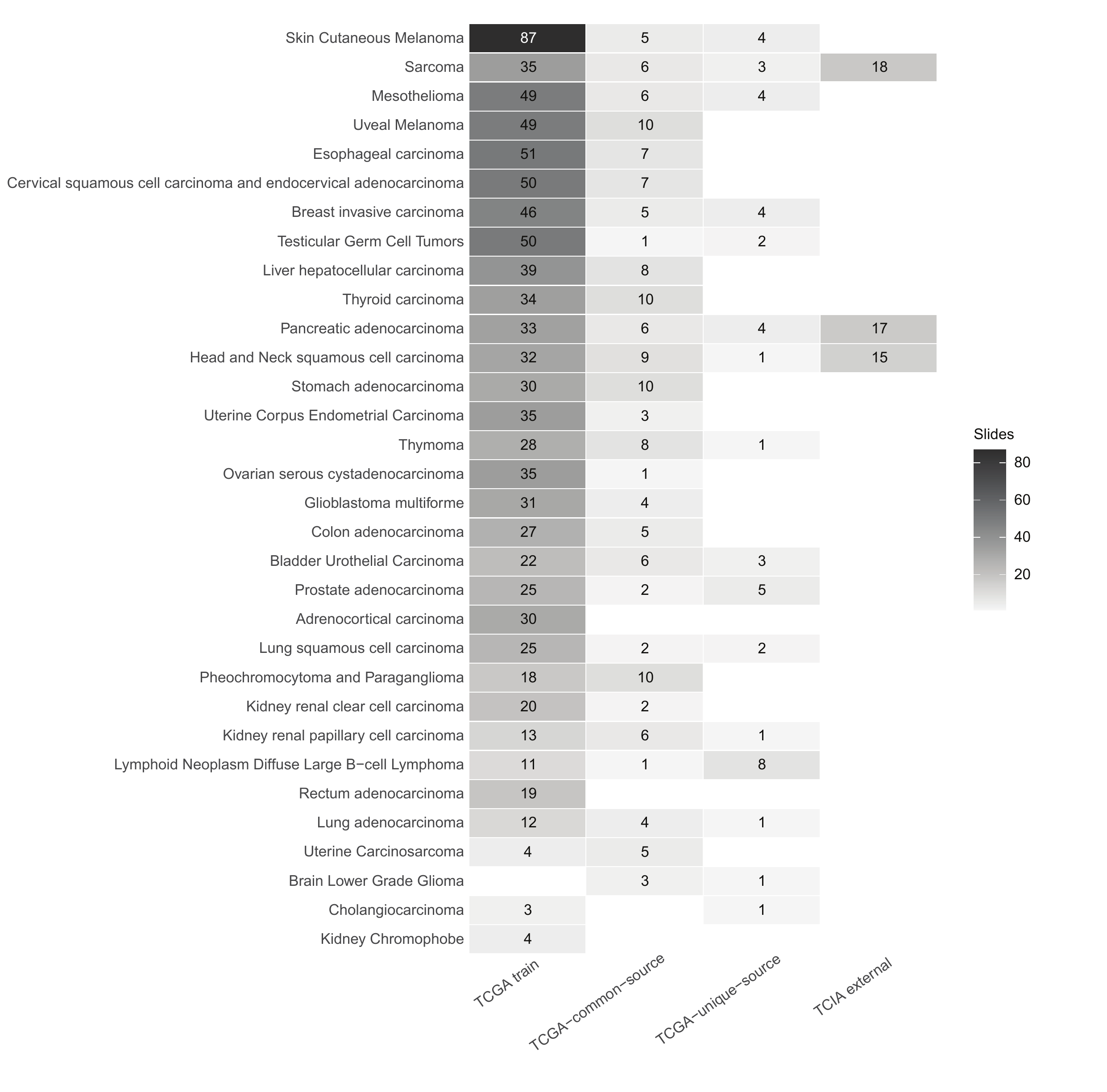}
\caption{\textbf{Slide-level cancer/tissue-category coverage across cohort partitions.} Heatmap showing WSI/slide counts by cancer/tissue category across the TCGA development, TCGA-common-source, TCGA-unique-source and TCIA external-validation cohorts. Cell values give slide counts, with shading encoding the same quantity. Cancer/tissue categories correspond to TCGA study/disease categories and TCIA diagnostic categories rather than strictly anatomical organ sites.}
\end{figure*}
\begin{figure*}[p]
\centering
\includegraphics[width=1.000\textwidth]{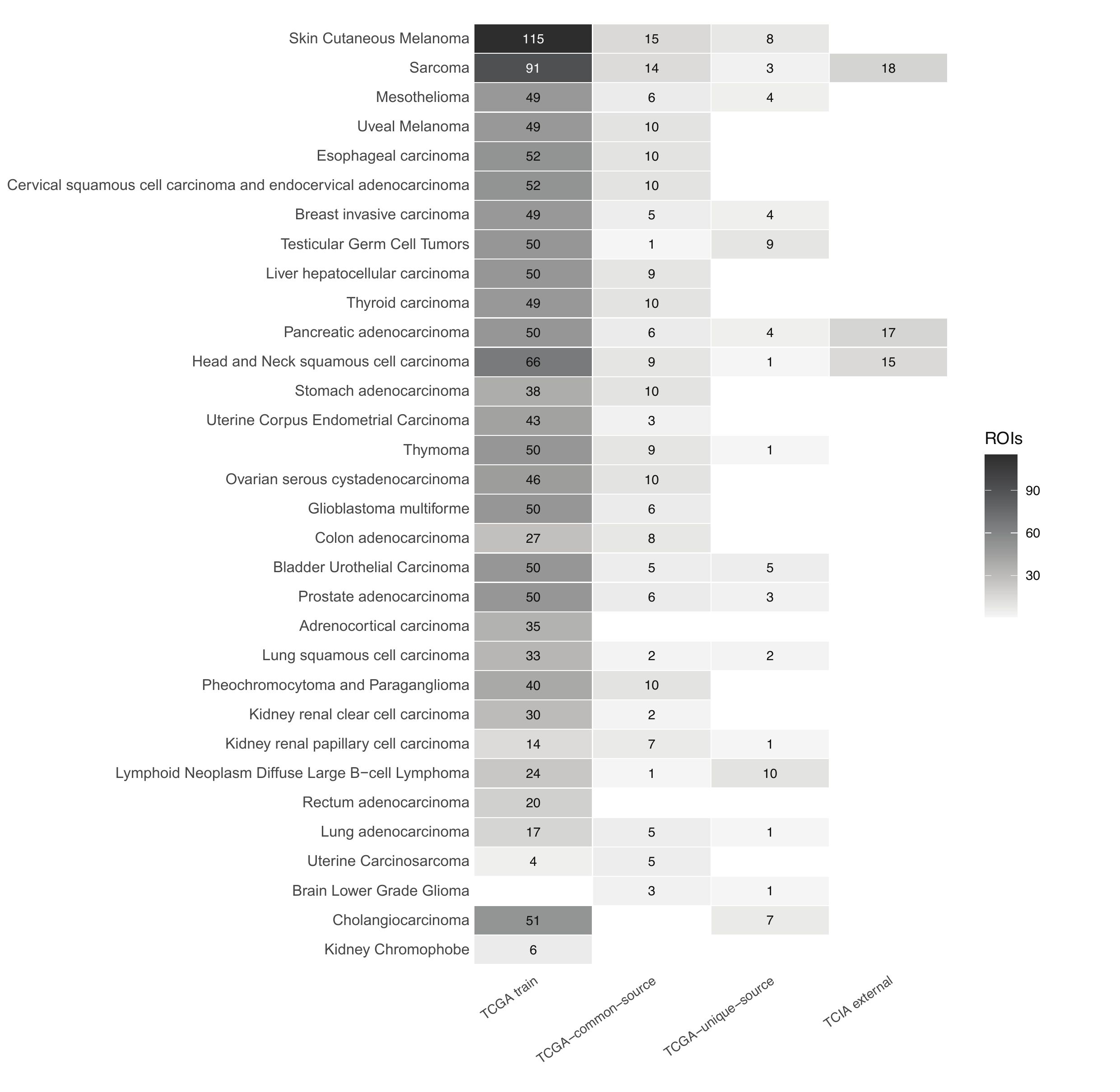}
\caption{\textbf{ROI-level cancer/tissue-category coverage across cohort partitions.} Heatmap showing annotated ROI counts by cancer/tissue category across the TCGA development, TCGA-common-source, TCGA-unique-source and TCIA external-validation cohorts. Cell values give ROI counts, with shading encoding the same quantity. Each TCIA slide contributed one ROI, whereas TCGA slides could contribute one or more ROIs; ROI-level coverage therefore complements the slide-level distribution shown in Supplementary Fig. 21.}
\end{figure*}

\clearpage

{\footnotesize
\begin{thebibliography}{99}
\bibitem{r1} A. Marcolini \emph{et al.}, ``histolab: A Python library for reproducible digital pathology preprocessing with automated testing,'' \emph{SoftwareX}, vol.~20, 2022, Art. no.~101237.
\bibitem{r2} J. M. Dolezal \emph{et al.}, ``Slideflow: Deep learning for digital histopathology with real-time whole-slide visualization,'' \emph{BMC Bioinformatics}, vol.~25, 2024, Art. no.~134.
\bibitem{r3} M. Y. Lu \emph{et al.}, ``Data-efficient and weakly supervised computational pathology on whole-slide images,'' \emph{Nat. Biomed. Eng.}, vol.~5, pp.~555--570, 2021.
\bibitem{r4} A. G. Berman, W. R. Orchard, M. Gehrung, and F. Markowetz, ``PathML: A unified framework for whole-slide image analysis with deep learning,'' medRxiv, 2021, doi: 10.1101/2021.07.07.21260138.
\bibitem{r5} A. Patil \emph{et al.}, ``Efficient quality control of whole slide pathology images with human-in-the-loop training,'' \emph{J. Pathol. Inform.}, vol.~14, 2023, Art. no.~100306.
\bibitem{r6} L. Browning \emph{et al.}, ``Artificial intelligence-based quality assessment of histopathology whole-slide images within a clinical workflow: Assessment of `PathProfiler' in a diagnostic pathology setting,'' \emph{Diagnostics}, vol.~14, 2024, Art. no.~990.
\bibitem{r7} B. A. Schreiber \emph{et al.}, ``Rapid artefact removal and H\&E-stained tissue segmentation,'' \emph{Sci. Rep.}, vol.~14, 2024, Art. no.~309.
\bibitem{r8} N. Kanwal \emph{et al.}, ``Are you sure it's an artifact? Artifact detection and uncertainty quantification in histological images,'' \emph{Comput. Med. Imag. Graph.}, vol.~112, 2024, Art. no.~102321.
\bibitem{r9} D. Tellez \emph{et al.}, ``Quantifying the effects of data augmentation and stain color normalization in convolutional neural networks for computational pathology,'' \emph{Med. Image Anal.}, vol.~58, 2019, Art. no.~101544.
\bibitem{r10} N. Otsu, ``A threshold selection method from gray-level histograms,'' \emph{IEEE Trans. Syst., Man, Cybern.}, vol.~9, no.~1, pp.~62--66, 1979.
\bibitem{r11} T. Kohlberger \emph{et al.}, ``Whole-slide image focus quality: Automatic assessment and impact on AI cancer detection,'' \emph{J. Pathol. Inform.}, vol.~10, 2019, Art. no.~39.
\bibitem{r12} K. Moutselos and I. Maglogiannis, ``The effect of tissue detection on false positives of diffusion-based artifact detection in histopathology,'' 2026, \emph{arXiv:2609.40083}.
\bibitem{r13} Z. Weng \emph{et al.}, ``GrandQC: A comprehensive solution to quality control problem in digital pathology,'' \emph{Nat. Commun.}, vol.~15, 2024, Art. no.~10685.
\bibitem{r14} F. Jabar \emph{et al.}, ``Fully automatic content-aware tiling pipeline for pathology whole slide images,'' \emph{Intell.-Based Med.}, vol.~12, 2025, Art. no.~100318.
\bibitem{r15} M. Scotto \emph{et al.}, ``SlideInspect: From pixel-level artifact detection to actionable quality metrics in digital pathology,'' \emph{Int. J. Imag. Syst. Technol.}, vol.~36, 2026, Art. no.~e70292.
\bibitem{r16} S. Kahaki \emph{et al.}, ``HistoART: Histopathology artifact detection based on large foundation model,'' \emph{Neural Comput. Appl.}, vol.~38, 2026, Art. no.~753.
\bibitem{r17} N. Kanwal \emph{et al.}, ``Equipping computational pathology systems with artifact processing pipelines: A showcase for computation and performance trade-offs,'' \emph{BMC Med. Inform. Decis. Mak.}, vol.~24, 2024, Art. no.~288.
\bibitem{r18} H. Xu \emph{et al.}, ``A whole-slide foundation model for digital pathology from real-world data,'' \emph{Nature}, vol.~630, pp.~181--188, 2024.
\bibitem{r19} Z. Liu \emph{et al.}, ``Swin Transformer: Hierarchical vision transformer using shifted windows,'' in \emph{Proc. IEEE/CVF Int. Conf. Comput. Vis. (ICCV)}, 2021, pp.~9992--10002.
\bibitem{r20} T. Xiao, Y. Liu, B. Zhou, Y. Jiang, and J. Sun, ``Unified perceptual parsing for scene understanding,'' in \emph{Proc. Eur. Conf. Comput. Vis. (ECCV)}, 2018, pp.~432--448.
\bibitem{r21} A. Jurgas \emph{et al.}, ``Improving quality control of whole slide images by explicit artifact augmentation,'' \emph{Sci. Rep.}, vol.~14, 2024, Art. no.~17847.
\bibitem{r22} G. Litjens \emph{et al.}, ``1399 H\&E-stained sentinel lymph node sections of breast cancer patients: The CAMELYON dataset,'' \emph{GigaScience}, vol.~7, 2018, Art. no.~giy065.
\bibitem{r23} M. Amgad \emph{et al.}, ``Structured crowdsourcing enables convolutional segmentation of histology images,'' \emph{Bioinformatics}, vol.~35, pp.~3461--3467, 2019.
\bibitem{r33} K. Clark \emph{et al.}, ``The Cancer Imaging Archive (TCIA): Maintaining and operating a public information repository,'' \emph{J. Digit. Imag.}, vol.~26, pp.~1045--1057, 2013.
\bibitem{r35} A. Dosovitskiy \emph{et al.}, ``An image is worth 16x16 words: Transformers for image recognition at scale,'' in \emph{Proc. Int. Conf. Learn. Represent. (ICLR)}, 2021.
\bibitem{r36} I. Loshchilov and F. Hutter, ``Decoupled weight decay regularization,'' in \emph{Proc. Int. Conf. Learn. Represent. (ICLR)}, 2019.
\bibitem{r24} J. N. Weinstein \emph{et al.}, ``The Cancer Genome Atlas Pan-Cancer analysis project,'' \emph{Nat. Genet.}, vol.~45, pp.~1113--1120, 2013.
\bibitem{r25} National Cancer Institute CPTAC, ``The Clinical Proteomic Tumor Analysis Consortium Sarcomas Collection (CPTAC-SAR) (Version 10),'' The Cancer Imaging Archive, 2019, doi: 10.7937/tcia.2019.9bt23r95.
\bibitem{r26} National Cancer Institute CPTAC, ``The Clinical Proteomic Tumor Analysis Consortium Head and Neck Squamous Cell Carcinoma Collection (CPTAC-HNSCC) (Version 19),'' The Cancer Imaging Archive, 2018, doi: 10.7937/k9/tcia.2018.uw45nh81.
\bibitem{r27} National Cancer Institute CPTAC, ``The Clinical Proteomic Tumor Analysis Consortium Pancreatic Ductal Adenocarcinoma Collection (CPTAC-PDA) (Version 15),'' The Cancer Imaging Archive, 2018, doi: 10.7937/k9/tcia.2018.sc20fo18.
\bibitem{r28} M. Tan and Q. Le, ``EfficientNet: Rethinking model scaling for convolutional neural networks,'' in \emph{Proc. 36th Int. Conf. Mach. Learn. (ICML)}, 2019, pp.~6105--6114.
\bibitem{r29} Z. Zhou, M. M. R. Siddiquee, N. Tajbakhsh, and J. Liang, ``UNet++: A nested U-Net architecture for medical image segmentation,'' in \emph{Deep Learning in Medical Image Analysis and Multimodal Learning for Clinical Decision Support} (Lecture Notes in Computer Science, vol.~11045). Cham, Switzerland: Springer, 2018, pp.~3--11.
\bibitem{r30} E. Xie \emph{et al.}, ``SegFormer: Simple and efficient design for semantic segmentation with transformers,'' in \emph{Proc. Adv. Neural Inf. Process. Syst. (NeurIPS)}, vol.~34, 2021, pp.~12077--12090.
\bibitem{r31} N. Carion \emph{et al.}, ``SAM 3: Segment anything with concepts,'' 2025, \emph{arXiv:2511.16719}.
\bibitem{r32} T. Ding \emph{et al.}, ``A multimodal whole-slide foundation model for pathology,'' \emph{Nat. Med.}, vol.~31, pp.~3749--3761, 2025.
\bibitem{r34} K. He, X. Zhang, S. Ren, and J. Sun, ``Deep residual learning for image recognition,'' in \emph{Proc. IEEE Conf. Comput. Vis. Pattern Recognit. (CVPR)}, 2016, pp.~770--778.
\end{thebibliography}
}
\end{document}